\documentclass[12pt,preprint]{aastex}

\slugcomment{}

\shorttitle{The Cool Subdwarf Investigation I}

\shortauthors{Jao et al.}

\received{2007 July 11}
\begin{document}

\title{Cool Subdwarf Investigations (CSI) I: New Thoughts for the Spectral
Types of K and M Subdwarfs}

\author{Wei-Chun Jao\altaffilmark{1}, Todd J. Henry\altaffilmark{1},
Thomas D. Beaulieu\altaffilmark{1}, John
P. Subasavage\altaffilmark{1}}

\affil{Department of Physics and Astronomy, Georgia State University,
Atlanta, GA 30302-4106} 

\email{jao, thenry, beaulieu, subasavage@chara.gsu.edu}

\altaffiltext{1}{Visiting Astronomer, Cerro Tololo Inter-American
Observatory.  CTIO is operated by AURA, Inc.\ under contract to the
National Science Foundation.}

\begin{abstract}

Using new spectra of 88 K and M-type subdwarfs, we consider novel
methods for assigning their spectral types and take steps toward
developing a comprehensive spectral sequence for subdwarf types K3.0
to M6.0.  The types are assigned based on the overall morphology of
spectra covering 6000\AA~to 9000\AA.  The types and sequence presented
link the spectral types of cool subdwarfs to their main sequence
counterparts, with emphasis on the relatively opacity-free region from
8200--9000\AA.  When available, supporting abundance, kinematic, and
trigonometric parallax information is used to provide more complete
portraits of the observed subdwarfs.  We find that the CaHn (n $=$
1--3) and TiO5 indices often used for subdwarf spectral typing are
affected in complicated ways by combinations of subdwarfs'
temperatures, metallicities, and gravities, and we use model grids to
evaluate the trends in all three parameters.  Because of the complex
interplay of these three characteristics, it is not possible to
identify a star as an ``extreme'' subdwarf simply based on very low
metallicity, and we suggest that the modifiers ``extreme'' or
``ultra'' only outline locations on spectroscopic indices plots, and
should not be used to imply low or very low metallicity stars.  In
addition, we propose that ``VI'' be used to identify a star as a
subdwarf, rather than the confusing ``sd'' prefix, which is also used
for hot O and B subdwarfs that are unrelated to the cool subdwarfs
discussed in this paper.

\end{abstract}


\section{Introduction}

The HR diagram is the most important map of stellar astronomy.  It
provides a relatively straightforward method for separating different
stellar luminosity classes, e.g.~supergiants, bright giants, giants,
subgiants, main sequence dwarfs, and white dwarfs, using their colors
and luminosities.  However, spectroscopic and trigonometric parallax
results have revealed a seventh distinct stellar luminosity class ---
the subdwarfs --- that lie below the main sequence dwarfs on the HR
diagram.

The realm of the subdwarfs has been previously explored by
\cite{Sandage1959}, \cite{Hartwick1984}, \cite{Monet1992},
\cite{Gizis1997}, and \cite{Jao2005}, to name a few.  Subdwarfs'
locations on the HR diagram are in part due to having metallicity
abundances lower than most field stars, which causes their opacities
to differ from those of regular dwarfs.  Subdwarfs are sometimes
called low metallicity halo stars or Galactic thick disk stars based
on their spectroscopic features, kinematics, and/or ages
\citep{Digby2003, Lepine2003, Burgasser2003, Reid2005, Monteiro2006}.
Regardless of how they are described, subdwarfs are fundamentally
different from their main sequence cousins.

Because of their generally high intrinsic velocities in the Galaxy,
many subdwarfs have been selected using high proper motion efforts,
such as the Lowell proper motion (\citealt{Giclas1971, Giclas1978}),
LHS (\citealt{LHS}), and L{\'e}pine-Shara Proper Motion-North (LSPM,
\citealt{Lepine2005a}) catalogs.  Recently, subdwarfs have been
selected by colors and spectroscopic observations via the Sloan
Digital Sky Survey (SDSS, \citealt{West2004}).  After initial flagging
as a potential subdwarf, spectroscopic and astrometric
(i.e.~trigonometric parallax work) followup efforts are typically
carried out to confirm or refute candidates as true subdwarfs.  Past
subdwarf identification efforts include \cite{Ryan1991},
\cite{Monet1992}, \cite{Carney1994}, \cite{Gizis1997},
\cite{Lepine2003}, \cite{West2004}, \cite{Reid2005} and
\cite{Burgasser2006}.  Most of these studies confirmed subdwarfs
spectroscopically, but only \cite{Monet1992} provides the crucial
trigonometric parallaxes that allow subdwarfs to be placed on the HR
diagram.

\cite{Gizis1997} presented a pioneering effort to assign numeric
subtypes for cool subdwarfs of spectral types K and M.  First, he used
the flux ratio of molecular band features, CaHn (n $=$ 1--3; we will
use ``CaH'' to indicate all three bands or index values throughout
this paper, unless otherwise specified) and TiO5 with pseudo-continuum
points to calculate spectroscopic indices
($f_{bands}$/$f_{continuum}$).  On TiO5 vs CaH1 or TiO5 vs CaH2$+$CaH3
plots, two high order polynomial lines can be used to separate a
continuous distribution of stars into three categories: regular
dwarfs, subdwarfs, and extreme subdwarfs.  Numerical subclasses were
then assigned using two independent linear fits for subdwarfs and
extreme subdwarfs.  For a decade, this methodology has been used to
assign spectral types for cool subdwarfs.

However, the current method of assigning subdwarf spectral types is
not directly linked to either their main sequence or giant
counterparts, as is typically (but not always) the case with normal
dwarfs and giants.  In addition, we have found that many subdwarfs are
assigned different subtypes even though their differences are limited
to CaH.  If these limitations can be overcome, a well-defined spectral
sequence would benefit many research areas beyond classification
efforts, including attempts to estimate effective temperatures and
distances, as well as providing insight into understanding Galactic
structure.

In this work, we first provide clarification and recommendations for
subdwarf terminology by addressing the usage of the confusing spectral
prefix ``sd''.  We then discuss our spectroscopic observations of 88
subdwarfs, generally targeting high proper motion stars in the
southern sky.  We next outline how synthetic spectra can assist us in
understanding cool subdwarf spectral features.  The bulk of this work
describes a detailed effort to provide a subdwarf spectral sequence
for stars having spectral types K3.0 through M6.0.  We then apply this
spectral typing method to those subdwarfs found in the SDSS database.
Once we have a detailed understanding of what makes a star a subdwarf,
we then discuss (1) why it is premature to assign precise parameters
to subdwarf spectra, (2) misunderstandings related to the terms
``extreme'' and ``ultra'' subdwarfs, (3) why the old method works for
dwarfs, but not subdwarfs, and finally (4) the recent set of subdwarf
spectral standards from \cite{Lepine2007}.

\section{``VI'' Subdwarfs are Different from ``sd'' Subdwarfs}
\label{sec.VI}

The first subluminous objects fainter than main sequence stars were
reported by \cite{Adams1922} when they were trying to determine the
luminosities of A-type stars.  In \cite{Adams1935} they called these
stars ``intermediate white dwarfs'' to separate them from typical
white dwarfs, and in the same paper reported the first six
``intermediate white dwarfs'' --- now known as LHS 405 (sdF3), LHS 540
(F8IV), LHS 1501 (A4p), LHS 2194 (sdF5), HD 132475 (F5/F6V), and HIP
68321 (A4).  All six have either A or F types (spectral types are from
SIMBAD) in the modern MK spectral classification system.  However, the
term ``subdwarfs'' was not suggested until \cite{Kuiper1939}.  He
expected three classes of objects to be found in his spectroscopic
survey of high proper motion stars--- white dwarfs, stars of large
spectroscopic parallaxes, and a class that was 2 or 3 magnitudes less
luminous than main sequence stars of the same color\footnote{Kuiper
assumed all of these high proper motion stars had $V_{tan}$ less than
474 km/sec.  Under this assumption, he could then make crude estimates
of their absolute magnitudes without having trigonometric parallaxes.
The results placed some stars three magnitudes below the main
sequence.}.  He suggested the name ``subdwarfs'' be used to represent
this final, independent, class of stars.  This name paralleled the use
of the term ``subgiants'' to describe stars that fall below the giants
on the HR diagram.  The name also eliminated the confusion with white
dwarfs, which are much less luminous than main sequence stars (and we
now know are a completely different type of object).  A year later,
\cite{Kuiper1940} reported the first three M-type subdwarfs ---
Kapteyn's star, LHS 20, and LHS 64 (\citealt{Gizis1997} confirmed that
all are M subdwarfs).  Although they were termed subdwarfs, he used
the same spectral classification as dwarfs (M0 and M2, see
\citealt{Kuiper1940} Table 1). The ``sd'' spectral classification
prefix for subdwarfs did not appear until \cite{Joy1947}, when he used
the strengthening of the Lindblad depression around 4226\AA.

In the late 1940's through 60's, the term ``subdwarfs'' also began
being used for a class of underluminous blue stars \citep{Humason1947,
Feige1958, Greenstein1966a, Greenstein1966b}.  The terminology was
based on the understanding that if these were high luminosity blue
stars, their distances would be outside the Milky Way, so it was
surmised that these stars should be underluminous, and therefore
closer.  Although their temperatures are similar to O and B dwarfs,
their spectral features are, in fact, quite different.  Generally,
O-type subdwarfs (sdO) and B-type subdwarfs (sdB) both have broad
Balmer absorption.  sdBs have weak or no He lines, while sdOs have
strong He II (4686\AA) or other He II lines \citep{Heber1992}.  Since
\cite{Feige1958}, such blue objects have been called sdO, sdB, or
sdOB-type stars.

\begin{figure}[h]
\centering
\includegraphics[angle=90, scale=0.7]{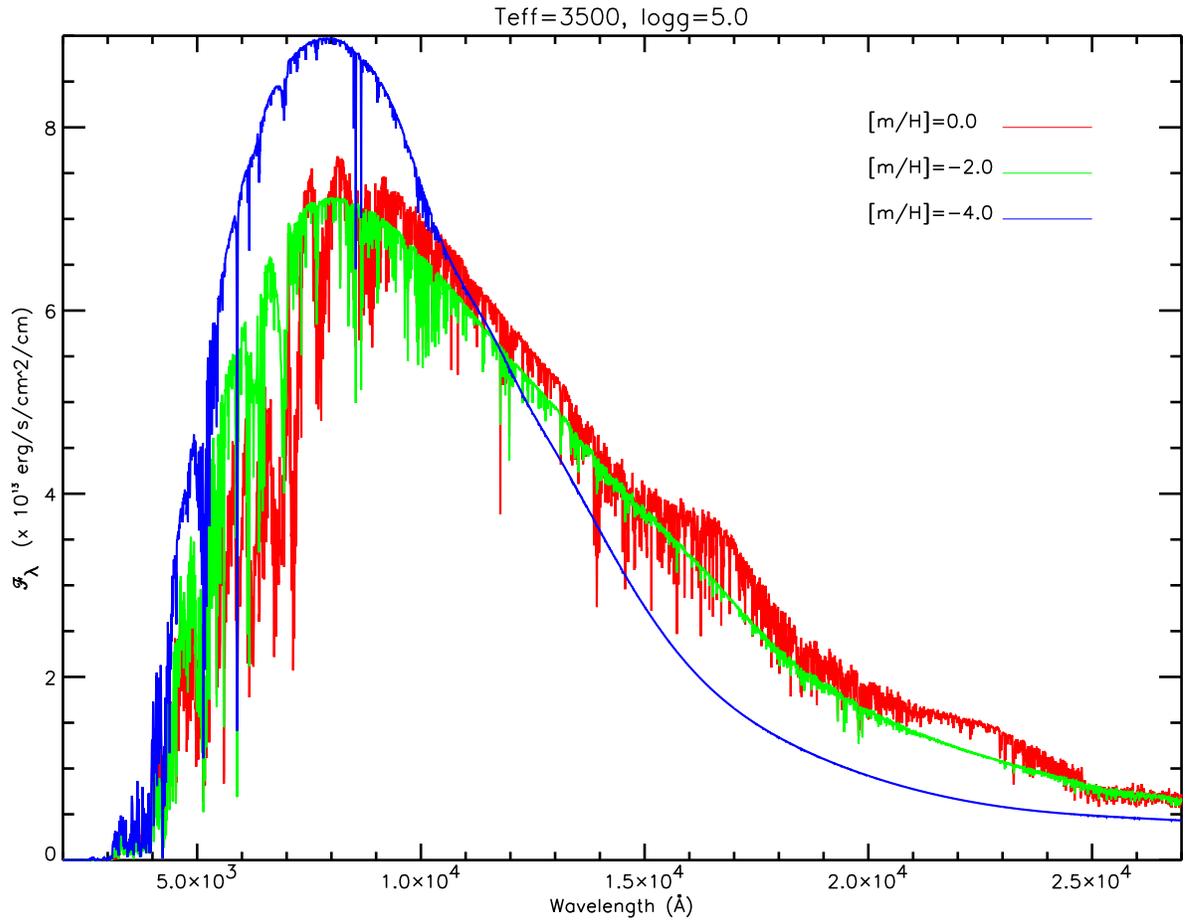}

\caption{Synthetic spectra for stars having $T_{eff}$ $=$ 3500K and
{\it log g} $=$ 5.0.  Red, green, and blue lines represent different
metallicities, 0.0, $-$2.0 and $-$4.0.  Note that the relative amounts
of blue and red fluxes trends toward bluer objects at lower
metallicities.}

\label{fig.continuum.model}
\end{figure}

Thus, we are left with the unfortunate situation that there are two
different classes of stars called ``subdwarfs.''  One is located at
the cool end of the HR diagram while the other is at the hot end.  The
two classes of stars are subluminous for completely different
astrophysical reasons but share the same ``sd'' spectral
classification prefix.  Cool subdwarfs usually have low metallicity
\citep{Chamberlain1951, Greenstein1966, Mould1976, Allard1995}, so
their opacties are different from dwarfs. For example,
\cite{Allard1995} discussed the possible opacity sources for a solar
type dwarf and a [m/H]=$-$2.5 subdwarfs.  TiO dominates the opacity
sources in the optical band. However, because of a decreasing
metallicity for subdwarfs, TiO opacity decreases dramatically. Hence,
this less blanketing from TiO bands causes more continuum flux
radiated from deeper and hotter layer of stellar atmosphere and their
spectrum falls closer overall to that of a blackbody, so these
subdwarfs appear bluer than dwarfs, as shown in
Figure~\ref{fig.continuum.model}\footnote{The details of these
synthetic spectra, the ``GAIA model grids'', are discussed in section
4.}. To the contrary, hot blue subdwarfs of types O and B are
progenitors of white dwarfs, and their subluminous nature is not
caused by metallicity at all.  Instead, hot subdwarfs represent a
stage in the stellar evolution cycle of an evolved star, and they
happen to be crossing the main sequence at the moment of observation.

\begin{figure}
\centering
\includegraphics[angle=90, scale=0.7]{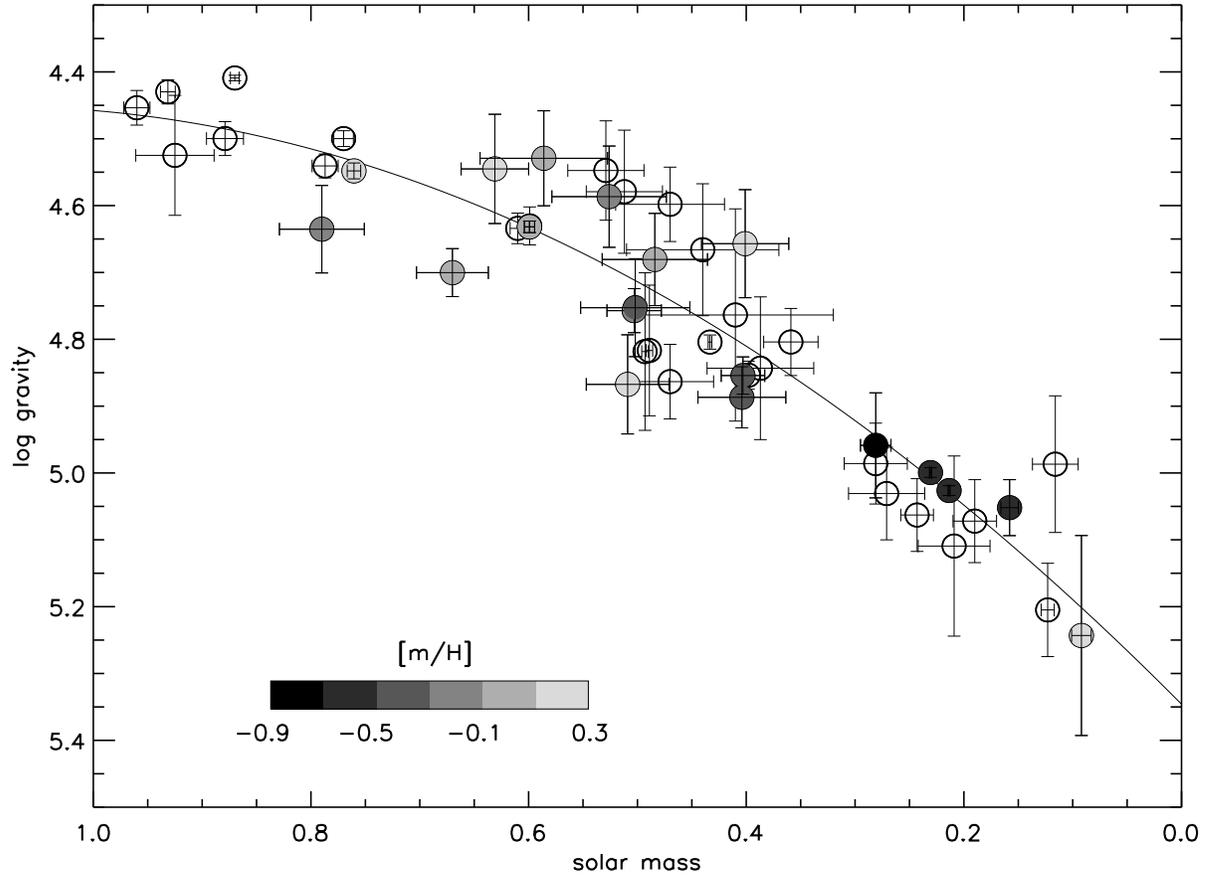}

\caption{The mass-gravity relation is shown using data from Table 1 of
\cite{Lopez2007}.  A solid line represents a polynomial fit to all of
the points, simply to be used as a guide.  Different shades for the
points represent different metallicity measurements.  Stars with
unknown metallicities are plotted as open circles.}
\label{fig.Lopez.Morales}
\end{figure}

Because they {\it are} different kinds of stellar objects, we suggest
that the two classes should not share the same spectral classification
notation, ``sd.''  \cite{Roman1955} argued that for types later than
G0, the spectral notation ``VI'' should be used for stars that are
$\sim$1--2 magnitude less luminous than main sequence stars.  Although
\cite{Jaschek1987} stated that ``this designation (VI) should
definitely be abandoned'', no specific reasons were actually given.
Here we propose that the luminosity class ``VI'' should be adopted for
cool subdwarfs, especially for K and M types.  The three primary
reasons are:

\begin{itemize}

\item Cool subdwarfs of types K and M lie clearly below main sequence
stars on the HR diagram, forming an additional class of objects.
Assigning their types as VI continues the progression outlined by
other classes.  Like the family of giants, which includes luminous
supergiants (I), bright giants (II), normal giants (III), and
subgiants (IV), the family of dwarfs includes main sequence dwarfs (V)
and their subdwarf (VI) counterparts.

\item Cool subdwarfs and OB subdwarfs are completely differently types
of objects with different origins.  The notation ``sd'' is suitable
for OB subdwarfs, which have not yet as a group been established to
have any sort of sequence on the HR diagram, at least when using using
parallaxes from the Yale Parallax Catalog, Hipparcos, and more recent
efforts.  On the other hand, there are dozens of cool subdwarfs with
parallax measurements, and they do form a coherent group below the
main sequence on the HR diagram, making a Roman numeral designation
reasonable.

\item Historically, the Roman numerals used for luminosity
categorization track with different gravities for the stars
classified, with higher gravities being assigned higher Roman
numerals.  We find that assigning ``VI'' for subdwarfs appropriately
continues this trend.  Figure~\ref{fig.Lopez.Morales} shows the
mass-gravity relation using data from Table 1 of \cite{Lopez2007}.  In
the best represented mass regime from 0.35$M_{\sun}$ to
0.70$M_{\sun}$, a crude trend indicates that lower metallicity stars
do indeed have higher gravities.  Although much more data are needed
to understand clearly how metallicity affects the mass-gravity
relation, current evidence supports using the ``VI'' designation for
low metallicity subdwarfs, which tend to exhibit higher gravities.

\end{itemize}

\noindent Thus, in order to separate the OB subdwarfs from the cool
subdwarfs, we suggest the `sd'' prefix should not be used for low
metallicity (and/or high gravity, as will be shown below) subdwarfs.

\section{Observations and Reductions}
\subsection{Observations}

Our subdwarf targets were selected from several different efforts,
including lists of spectroscopically identified subdwarfs
\citep{Gizis1997, Reid2005}, subdwarfs with parallax measurements
\citep{Jao2005, Costa2005, Costa2006}, candidates from proper motion
catalogs \citep{Deacon2005, Subasavage2005a, Subasavage2005b}, and
stars with metallicity measurements \citep{Carney1994, Cayrel2001,
Nordstrom2004}.  We define our subdwarfs of interest to be those with
$V-K_{s}\geqslant$ 2.0, [m/H] $\leqslant-$0.5, or having absolute
magnitudes at least one magnitude less luminous in $M_{Ks}$ than a fit
to main sequence stars of comparable color with trigonometric
parallaxes from the RECONS (Research Consortium on Nearby Stars) 10 pc
sample \citep{Henry2006}.

Spectroscopic observations were made with the 1.5-m and 4.0-m
telescopes at CTIO.  For the observations on the 1.5-m from 2002 to
2006, the R-C spectrograph with a Loral 1200$\times$800 CCD camera was
used with the \#32 grating (in first order) at tilt 15.1$^\circ$.  The
order-blocking filter OG570 was utilized to provide spectra covering
the range of 6000\AA~to 9500\AA~with a resolution of 8.6\AA.  The only
variation in observing parameters was that during the May and December
2006 observing runs a larger slit width of 6\arcsec~and 4\arcsec,
respectively, ~was used instead of the 2\arcsec~slit used in previous
runs, in order to minimize the differential color refraction (because
the slit orientation was not changed during observations).  For
observations on the 4.0-m in 2002, the R-C spectrograph with a Loral
3K$\times$1K CCD was used with the \#181 grating (in first order) at
tilt 58.8$^\circ$.  The order-blocking filter OG515 was utilized to
provide spectra covering the range from 5500\AA~to 10000\AA~with a
resolution of 6\AA.  Fringing at wavelengths longer than
$\sim$7000\AA~in the 4.0-m data was removed by customized IDL
routines.  Bias frames and dome flats (and sky flats at the 1.5-m)
were taken at the beginning of each night for calibration.  At least
two exposures were taken for each object to permit cosmic ray
rejection.  If stars were faint, additional observations were
sometimes made.  A 10 second Ne$+$He$+$Ar or Ne only arc lamp spectrum
was recorded after each target to permit wavelength calibration.
Several spectroscopic flux standard stars found in the $IRAF$
spectroscopy reduction packages were observed during each observing
run, usually nightly.  Reductions were carried out in the standard way
using IRAF reduction packages.  Wavelength and flux calibrations were
done using {\it onedspec.dispcor} and {\it onedspec.calibrate} within
IRAF, respectively.

Many of the subdwarfs discussed in this paper have new trigonometric
parallaxes and $VRI$ photometry acquired during our southern nearby
star program, CTIOPI (Cerro Tololo Inter-american Observatory Parallax
Investigation, see \citealt{Jao2005}).  In a future paper in this
subdwarf series, we will present the astrometric (particularly
trigonometric parallaxes) and photometric results.

\subsection{Identifying Subdwarfs}

During our five year spectroscopic campaign, we have acquired spectra
for more than 900 objects.  To glean subdwarfs from our spectroscopic
database, we calculated their spectroscopic indices listed in
Table~\ref{tbl.cahn.tio5.index} (targets listed alphabetically),
mimicking the methodology outlined by \cite{Gizis1997}.

\begin{figure}
\centering
\includegraphics[angle=90, scale=0.5]{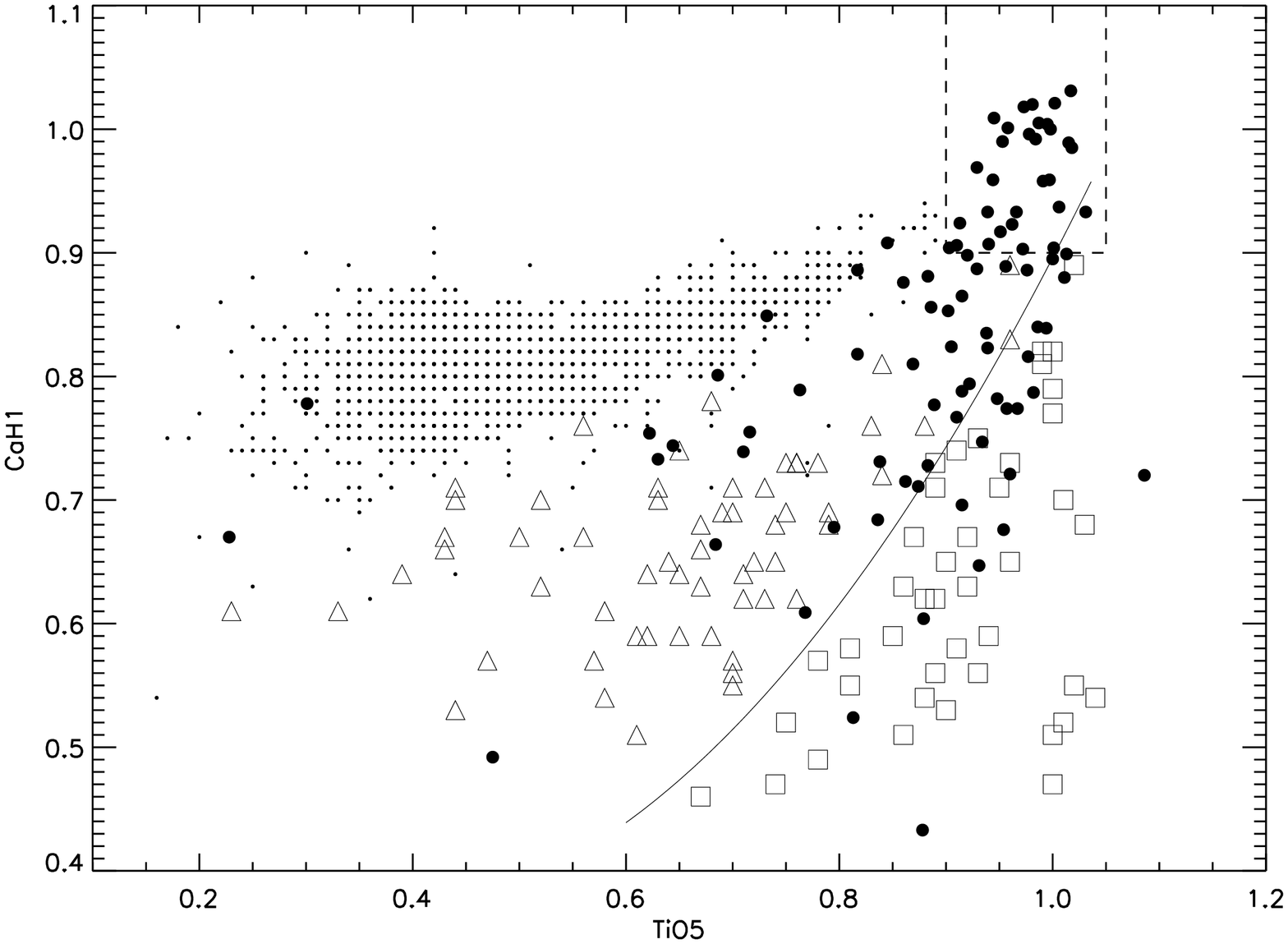}
\includegraphics[angle=90, scale=0.5]{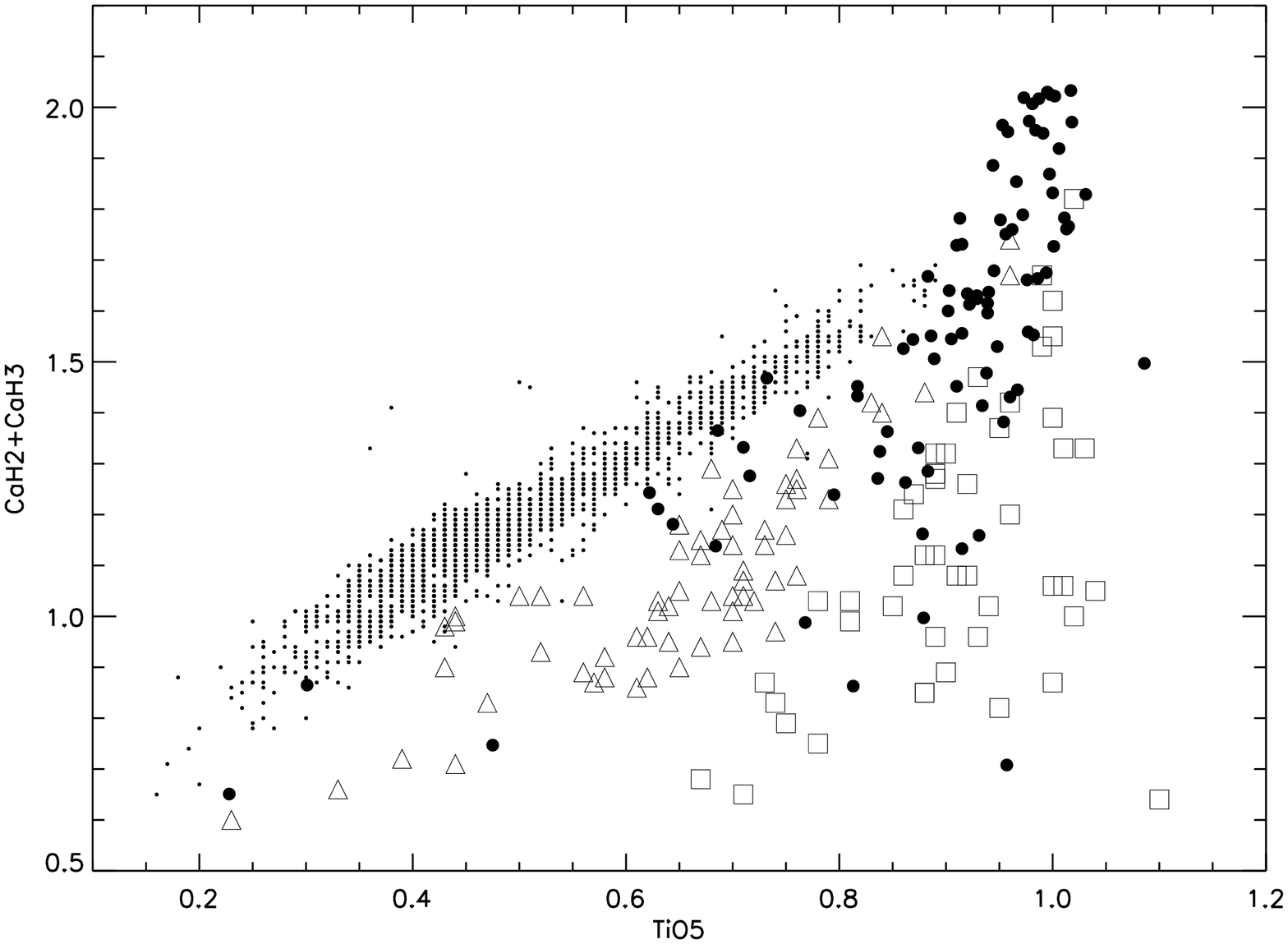}

\caption{The CaH1 and CaH2+CaH3 versus TiO5 indices are plotted for
our identified subdwarfs (solid circles).  For comparison, known cool
dwarfs (dots), subdwarfs (open triangles), and ``extreme'' subdwarfs
(open boxes) from \cite{Hawley1996}, \cite{Gizis1997}, and
\cite{Reid2005} are also shown.  The dashed box indicates subdwarfs
that do not exhibit strong spectroscopic indices, but which are either
below the main-sequence on the HR diagram or have published
metallicities [m/H] less than or equal to $-$0.5.  A solid line
indicates the separation between regular subdwarfs and extreme
subdwarfs adopted by \cite{Gizis1997}.}

\label{fig.cahn.plot}
\end{figure}

\begin{figure}
\centering
\includegraphics[scale=0.6, angle=90]{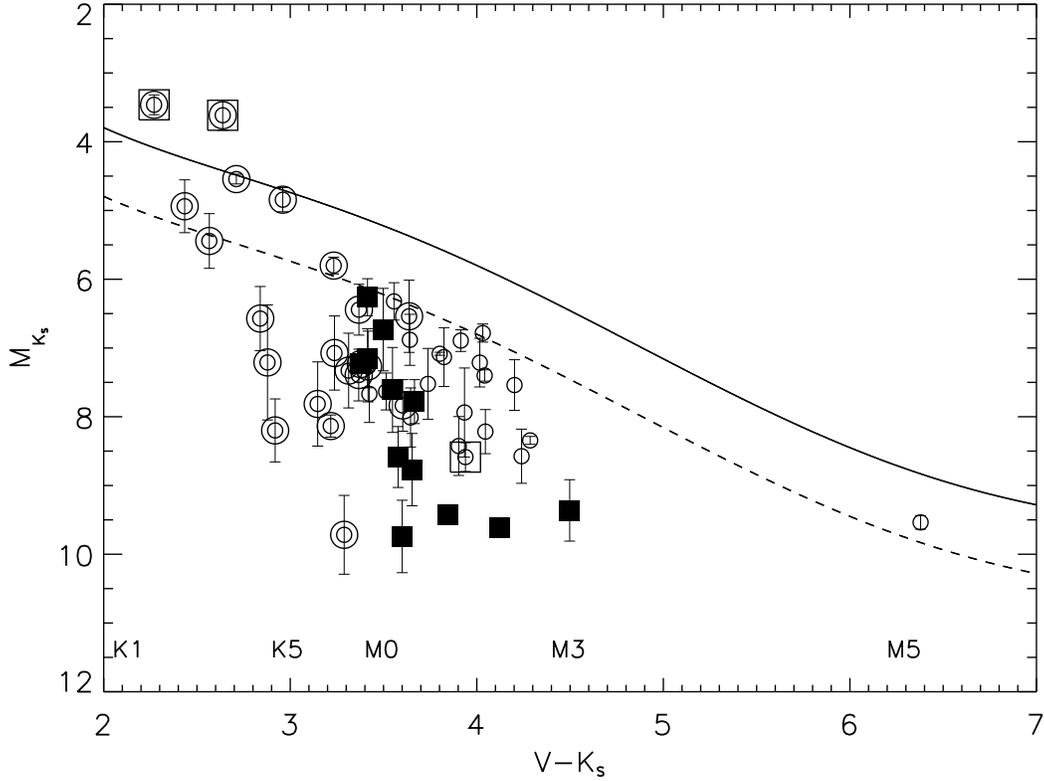}

\caption{HR diagram for subdwarfs listed in
Table~\ref{tbl.cahn.tio5.index} and shown in
Figure~\ref{fig.cahn.plot} that have trigonometric parallaxes.  Open
circles indicate subdwarfs and filled boxes indicate ``extreme''
subdwarfs based on spectroscopic indices.  Concentric circles indicate
stars with CaH1 index greater than 0.9 that are difficult to
distinguish from main sequence stars at our spectral resolution. Open
boxes indicate confirmed spectroscopic binaries.  A solid line
indicates a fit to main-sequence dwarfs, primarily from
\cite{Henry2004} with extra dwarf standard stars from \cite{Gray2003}.
The dashed line is one magnitude fainter than this solid line.  Note
that the K-type subdwarf sequence merges with the K dwarf sequence at
the blue end of this ($M_{K_{s}}$ vs $V-K_{s}$) plot.  The single
point at $V-K_{s}$ = 6.4 is LHS 2067A. The spectral types for dwarfs
are given at the bottom of the figure as references.}

\label{fig.hr.plot}
\end{figure}

Figure~\ref{fig.cahn.plot} shows TiO5 plotted against CaH1 and
CaH2$+$CaH3 for various samples of stars.  Our subdwarfs are shown
with solid circles.  For comparison, small dots indicate main sequence
stars from \cite{Hawley1996}, while open triangles and squares
represent subdwarfs and ``extreme'' subdwarfs from \cite{Gizis1997}
and \cite{Reid2005}.  Some stars (solid circles) having indices
located near or in the main sequence regions in these plots have been
manually checked to confirm that they are subdwarfs.  Using the HR
diagram in Figure~\ref{fig.hr.plot}, we confirm the low luminosities
of our spectroscopically selected subdwarfs ($V-K_{s}>$ 2.7) that have
accurate trigonometric parallaxes.

\begin{figure}
\hspace{-2cm}
\includegraphics[scale=0.8, angle=90]{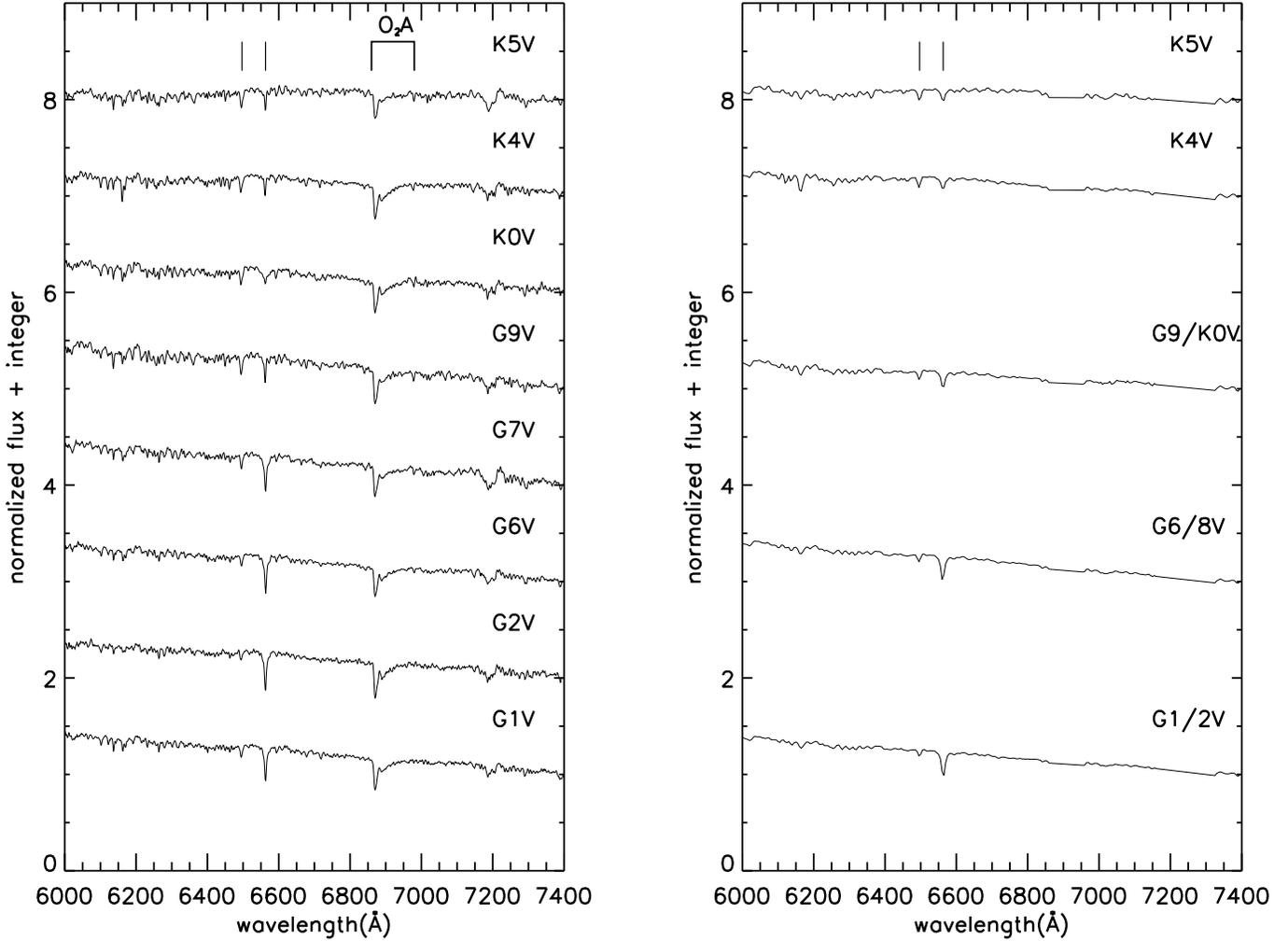}

\caption{Spectra for G and K-type stars from \cite{Jacoby1984} (left
panel, 4\AA~resolution) and \cite{Silva1992} (right panel,
11\AA~resolution) are shown.  Because the red cutoff is 7400\AA~in
\cite{Jacoby1984}, the results from \cite{Silva1992} are also plotted
to 7400\AA, and both sets of spectra are normalized at 7400\AA.  The
two tick marks indicate the Ba I (left) and H$\alpha$ (right)
absorption features.  \cite{Silva1992} have removed telluric
absorption features.}

\label{fig.jacoby.silva}
\end{figure}

Although we focus primarily on the K and M subdwarfs for this study,
we include a few G-type subdwarfs among those selected from
metallicity measurements in the literature.  For reference, we
consider early K-type stars to have types K0 to K2, mid K-types to be
K3 to K5, and late K-types to be K6 and later.  We find that it is
difficult, but possible, to separate late G from early K-type stars
using our spectral coverage and resolution.  Their continuum slopes
have only slight differences across our wavelength window coverage,
and there are no noticeable absorption differences beyond 7500\AA.
Spectra for G1V to K5V types from \cite{Jacoby1984} (resolution
$\sim$4\AA) and \cite{Silva1992} (resolution $\sim$11\AA) are plotted
in Figure~\ref{fig.jacoby.silva}.  The only strong features are the
absorption lines of Ba I (6497\AA)\footnote{\cite{Turnshek1985} noted
that this feature at 6497\AA~is a blend of different atomic lines,
including Fe I, Ba I, Ca I, Mn I, Co I, Ti I and II, and Ni I.  Ba I
is likely the dominant absorber because it has the largest Einstein
coefficient in the NIST atomic spectra database.} and H$\alpha$
(6563\AA), with gradually increasing Ba I absorption and decreasing
H$\alpha$ absorption as the effective temperature drops.  These
effects can be seen in both sequences, regardless of the spectral
resolution.  Our spectral resolution of 6--9\AA~falls between the
resolution of the two sequences shown, so we can use the relative
absorption strengths of Ba I and H$\alpha$ to separate G and K-type
stars.  In total, we have identified 88 K and M subdwarfs and five G
type subdwarfs, using spectra with coverage from 6000\AA~to 9000\AA.

\subsection{Sorting Spectra}
\label{sec.sorting.spectra}

\begin{figure}
\centering
\includegraphics[scale=0.7, angle=0]{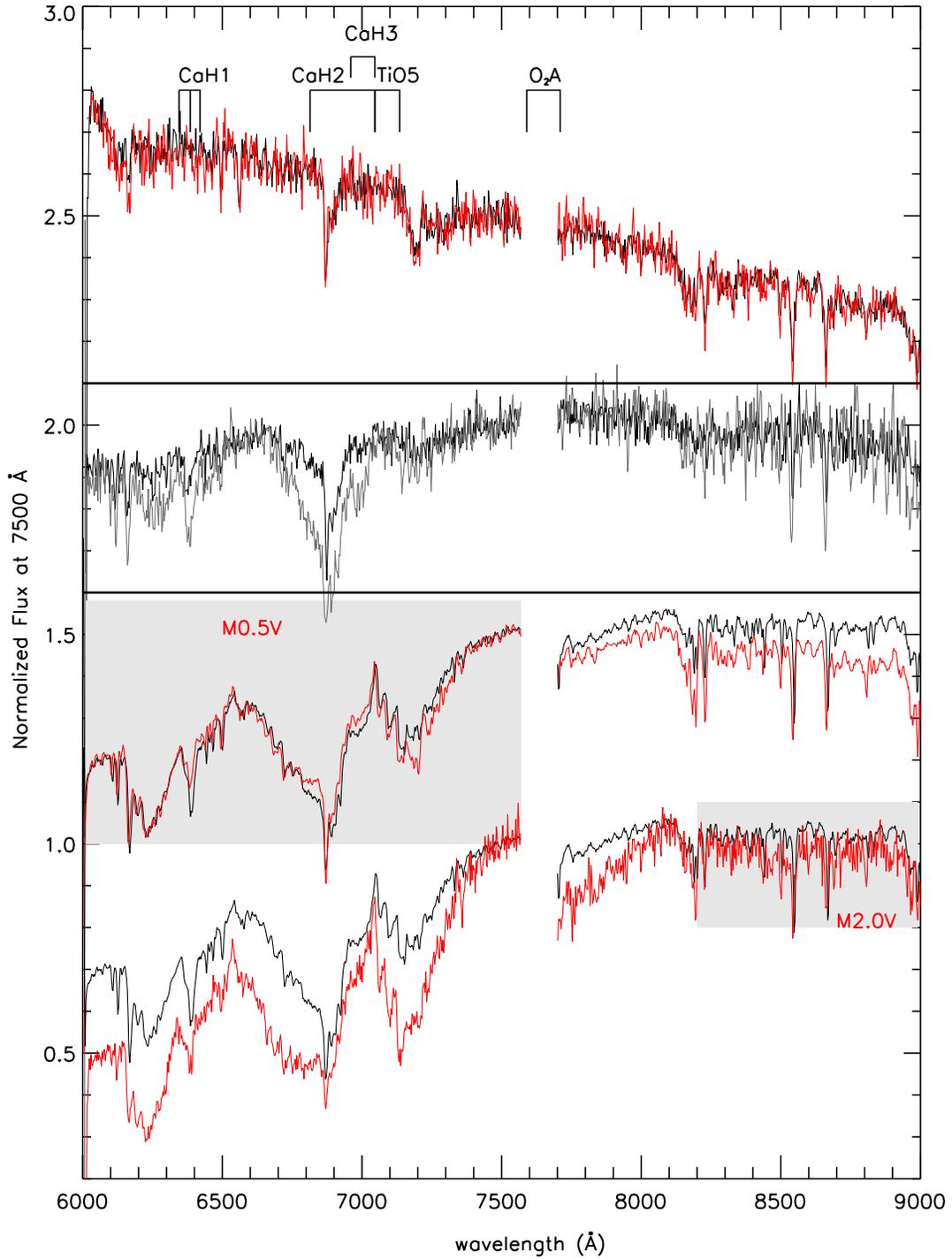}

\caption{Top: An mid K subdwarf spectrum (black) is virtually
identical to an mid K dwarf (red).  Middle: The spectra of two early
M subdwarfs differ only at CaH.  Bottom: One subdwarf's spectrum
(black) matches an M0.5V spectral standard (red) at the blue end, but
matches an M2.0V spectral standard (red) at the red end.  The deep
telluric band, O$_{2}$A (7570\AA--7700\AA) has been removed.}
\label{fig.problem.spect}

\end{figure}

After reduction, the spectra were sorted into different bins based
upon similarity in overall slope and features.  This assured that
stars in each bin had approximately the same temperature.  However,
several impediments to clean sorting were encountered:

\begin{itemize}

\item All mid K-type subdwarfs had spectra virtually indistinguishable
from dwarf standard stars, yet they had low metallicity measurements
and/or were found below the main sequence on the HR diagram (see top
figure in Figure~\ref{fig.problem.spect}).

\item Many subdwarf spectra placed into the same bin showed
differences only in CaH (see middle figure in
Figure~\ref{fig.problem.spect}).

\item Many subdwarf spectra matched different dwarf standards at the
blue ($\lambda$ $<$ 7570\AA) and red ($\lambda$ $>$ 8200\AA) ends (see
bottom figure in Figure~\ref{fig.problem.spect}).

\end{itemize}

\noindent In order to understand what factors caused these anomalies,
we next examine theoretical studies that provide synthetic spectra
that can be compared directly to the observed spectra.

\section{Grids of Synthetic Spectra}

\begin{figure}
\centering
\includegraphics[scale=0.5, angle=90]{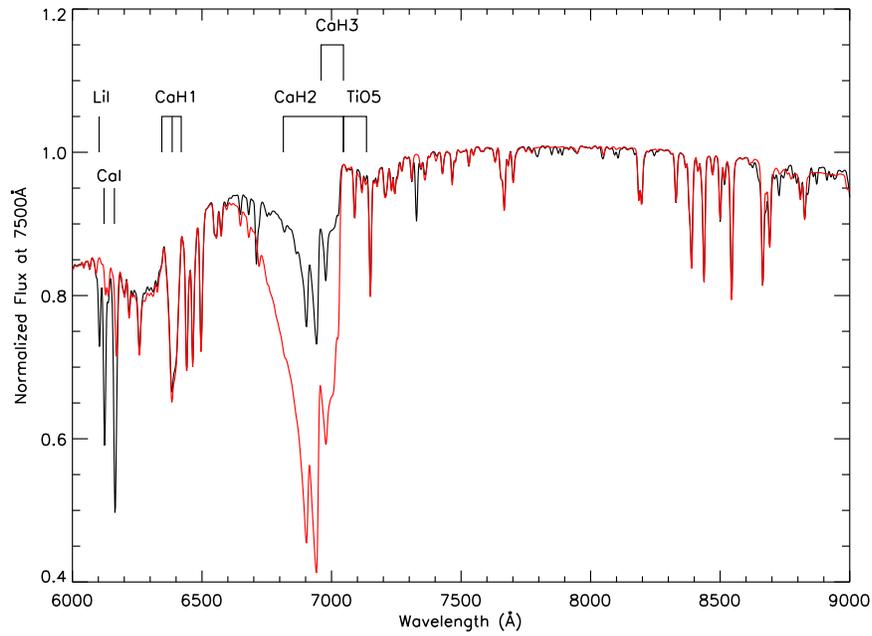}

\caption{Two synthetic spectra from old (red) and new (black) model
grids are shown.  The physical parameters for both models are $T_{eff}
=$ 3600K, {\it log g} = 5.0 and [m/H]$ = -$2.0.  Major differences are
seen between 6500\AA~and 7000\AA, and in the absorption lines of Li I
(6103\AA) and Ca I (6122\AA~and 6162\AA).}

\label{fig.new.old}
\end{figure}

We use grids of synthetic spectra computed with PHOENIX codes
(hereafter, GAIA model grids) to understand how subdwarfs' physical
parameters (temperature, metallicity and gravity) affect their
spectra.  The most recently released GAIA model grids
\citep{Brott2005} are available at an FTP site in
Hamburg\footnote{ftp://ftp.hs.uni-hamburg.de/pub/outgoing/phoenix/GAIA/v2.6.1/}.
\cite{Gizis1997}, \cite{Woolf2005}, and \cite{Burgasser2006} have all
used these synthetic model grids to characterize subdwarfs, but an
older version of the grids was used in all three cases.  The version
we employ here, 2.6.1, was released in late 2004.  A comparison of one
pre-2004 spectral model (provided by V. Woolf, private communication)
and a new spectral model (from the GAIA model grids) for a cool
subdwarf is shown in Figure~\ref{fig.new.old}.  Improvements to the
new models include (1) an enlarged and enhanced version of the
equation of state, (2) more atomic, ionic, and molecular line
opacities, (3) inclusion of the formation of dust particles for cool
stars, and (4) microturbulence calculations.  Additional water and TiO
opacities and the inclusion of dust are enhancements particularly
applicable to the low mass stars discussed here.

The two spectra shown in Figure~\ref{fig.new.old} are virtually
identical redward of 7000\AA, but there are significant differences
between 6500\AA~and 7000\AA, where the CaH2 and CaH3 bands are found,
and these differences will certainly affect evaluations done with the
older models.  The new model in this region has much shallower
absorptions than the old one, which will affect metallicity and
gravity estimates.  A few narrow absorption features (Li I at 6103\AA,
Ca I at 6122\AA~and 6162\AA) are also changed.  As outlined in the
discussion section $\S$\ref{sec:discussion}, even the latest version
of the model grids does not provide ideal matches to real spectra, so
further progress can still be made.

\subsection{Identifying Mid K-type Subdwarfs}
\label{sec.K.subdwarf}

\begin{figure}
\centering
\includegraphics[scale=0.5, angle=90]{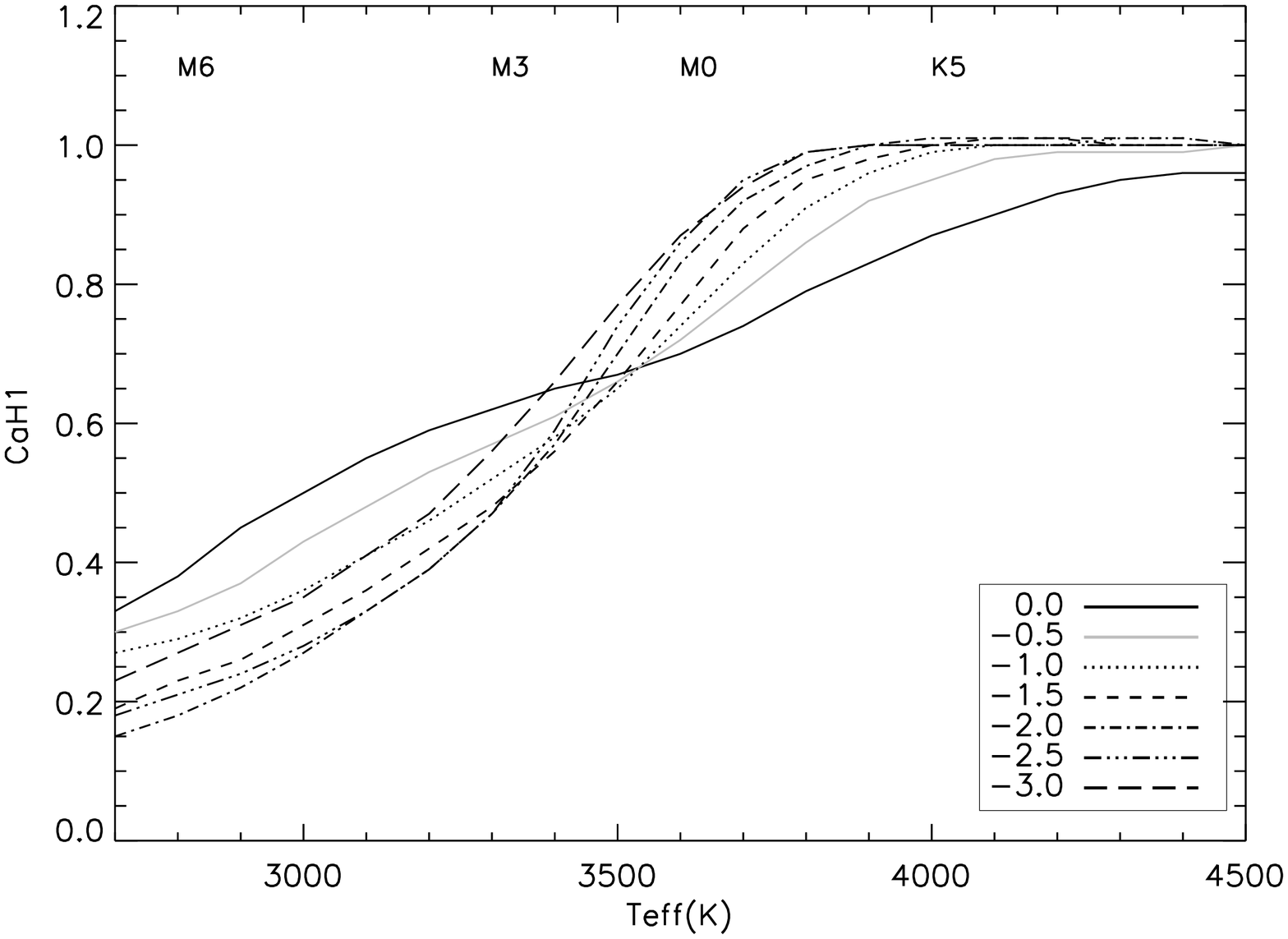}
\includegraphics[scale=0.5, angle=90]{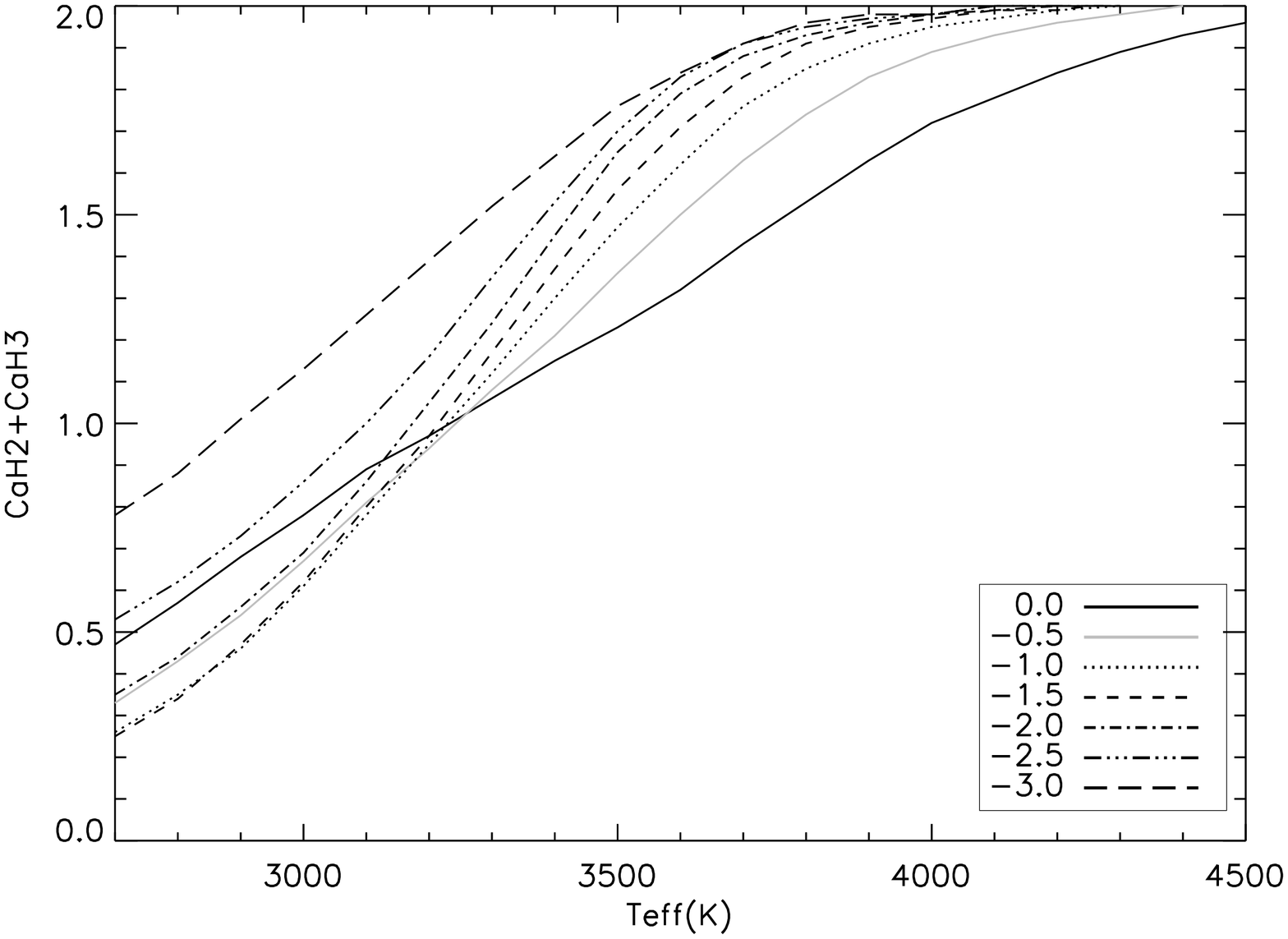}

\caption{CaH1 (top) and CaH2$+$CaH3 (bottom) indices from GAIA model
grids are plotted against $T_{eff}$. The different line styles
represent different [m/H]. The spectral types for dwarfs are given at
the top of the figure as references.}

\label{fig.model.index.plot}
\end{figure}

We use the latest GAIA model grids\footnote{GAIA model grids also
provide various values of $\alpha$-elements (O, Ne, Mg, Si, S, Ar, Ca,
and Ti), which yield abundance ratios such as O/Fe, Ne/Fe, etc.
\cite{Nissen1997} showed that most field halo stars have
[$\alpha$/Fe]$_{*}$$\approx$[$\alpha$/Fe]$_{\sun}$, so we select
[$\alpha$/$\alpha_{\sun}$]=0.0.} to calculate predicted CaH band
strengths and plot the derived indices against effective temperatures
in Figure~\ref{fig.model.index.plot}.  We evaluate stars with
$T_{eff}$ between 2700K and 4500K and [m/H] between 0.0 and $-$3.0.
For the moment, we adopt {\it log g} $=$ 5.0 generically for
subdwarfs.  This gravity value does not apply to all types of
subdwarfs, but we are presently interested in outlining the behavior
of the CaH features with metallicity alone.

When $T_{eff}$ is less than about 3300K,
Figure~\ref{fig.model.index.plot} shows that the CaH1 index decreases
(stronger absorption) when metallicity decreases from 0.0 to $-$2.0 at
fixed $T_{eff}$.  However, the trend reverses for [m/H] $=$ $-$2.5 and
$-$3.0.  When $T_{eff}$ is between 3300K and 3500K, there is a very
weak relation between metallicity and the CaH1 index, but from 0.0 to
$-$1.0, the relation (CaH1 index decreases when metallicity decreases)
still holds. This relationship is degenerate for lower
metallicities. For temperatures hotter than 3500K, the CaH1 index
increases (weaker absorption) as metallicity decreases, in contrast to
the low temperature region.  For the CaH2$+$CaH3 index, the trends are
generally the same, except that (1) at temperatures less than 3200K
the index decreases (stronger absorption) only for 0.0 to $-$1.0, with
a reversal for lower metallicities, and (2) the trend for higher
temperature stars (increasing index, weaker absorption with lower
metallicity) is the same, but the turnover is near 3200K rather than
3500K.

Even more important than these subtleties is that overall, the hotter
the subdwarf, the weaker its CaH bands.  Note the collapse of any
differences between indices for hotter stars of various metallicities
in Figure~\ref{fig.model.index.plot}.  This collapse makes separating
mid K-type subdwarfs from dwarfs based only on spectroscopic indices
difficult using our spectral coverage (6000\AA--9000\AA) and
resolution (6\AA--8.6\AA).  An alternative method, such as the HR
diagram shown in Figure~\ref{fig.hr.plot}, sufficiently solves the
problem for mid K-type subdwarfs and will be discussed in the next
section.  Independent metallicity measurements via high resolution
spectroscopic observations \citep{Bonfils2005, Bean2006} can also be
utilized.  Our own spectra are not of sufficiently high resolution to
measure metallicities, but all selected K-type subdwarfs have measured
[m/H] $\leqslant$ $-$0.5 from other publications.

\subsection{Mid K-type Subdwarf Sample from Our Observations}

There are 31 subdwarfs having CaH1 indices larger than 0.9 enclosed by
the dashed box in Figure~\ref{fig.cahn.plot}.  In this region, there
are no subdwarfs with previously measured CaH/TiO5 indices to compare
to our new sample of mid K-type stars, making it difficult to separate
the dwarfs and subdwarfs based on the CaH1 index alone.  Fortunately,
21 of these 31 stars have trigonometric parallaxes, and are plotted
with concentric circles in Figure~\ref{fig.hr.plot}.  At least 15 of
these stars are subdwarfs based on their locations one or more
magnitudes below the main sequence on the HR diagram.  The star with
the largest offset is DEN0515-7211\footnote{This star was first
reported in \cite{Costa2006} as reference star \#4 in the LHS 1749
parallax field.  We identify it henceforth as DEN0515-7211 (DENIS$-$P
J051545.1$-$721122).}, located at $(V-K)$ = 3.3, $M_{Ks} =$ 9.7.  This
star is a full 4.5 magnitudes less luminous than the main sequence,
but has no CaH or TiO5 features.  Two additional stars (G016-009AB and
G026-009ACD) above the main sequence are known to be double-line
spectroscopic binaries with [m/H] $<-$0.5. These individual targets
are discussed in section 6.2.  The dwarf/subdwarf status of only four
stars of the 21 remain ambiguous --- we suspect that most of them are
also subdwarfs, perhaps with as yet undetected companions elevating
them into main sequence territory.

Generally, the spectroscopic index method fails to distinguish
subdwarfs from dwarfs if the derived CaH1 index is greater than 0.9.
With the benefit of additional trigonometric parallax information
and/or metallicity measurements, however, we can conclude that nearly
all of these mid K-type stars are indeed subdwarfs.

\section{Late K-type and M-type Subdwarfs from GAIA Model Grids}
\label{sec.subdwarf.facts}

\begin{figure}
\centering
\includegraphics[scale=0.7]{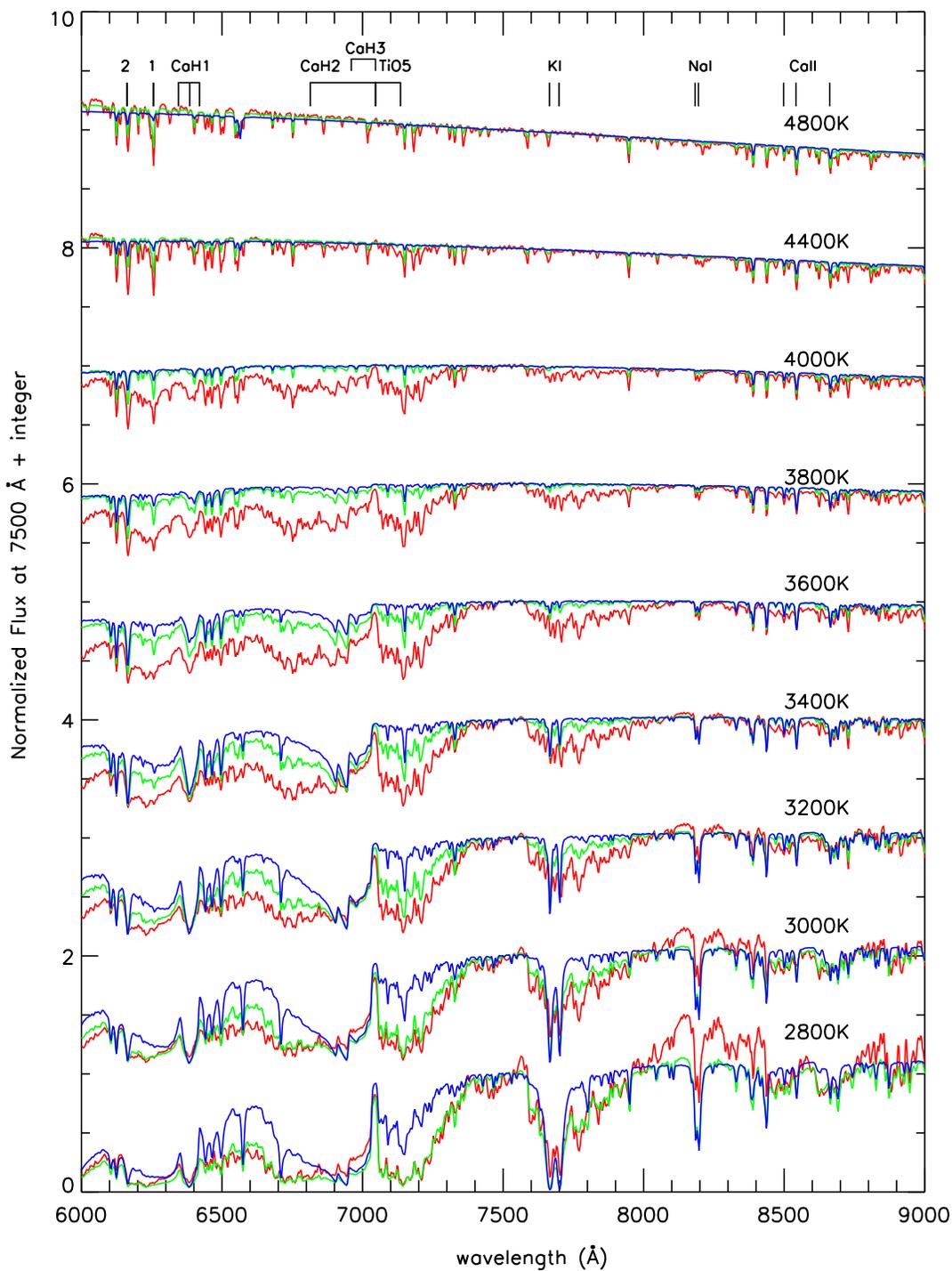}

\caption{GAIA synthetic spectra from 4800K to 2800K are shown. All
spectra are noiseless and have {\it log g} $=$ 5.0. Red, green and
blue lines represent [m/H] $=$ 0.0, $-$1.0 and $-$2.0, respectively.
Effective temperatures for each set of spectra are given above each
group of lines.  The feature marked \#1 is not seen in any of our
spectra.  The Ca I (6162\AA) feature marked \#2 is seen in our
spectra.}

\label{fig.4800-2800.plot}
\end{figure}

\begin{figure}
\centering
\includegraphics[scale=0.5, angle=90]{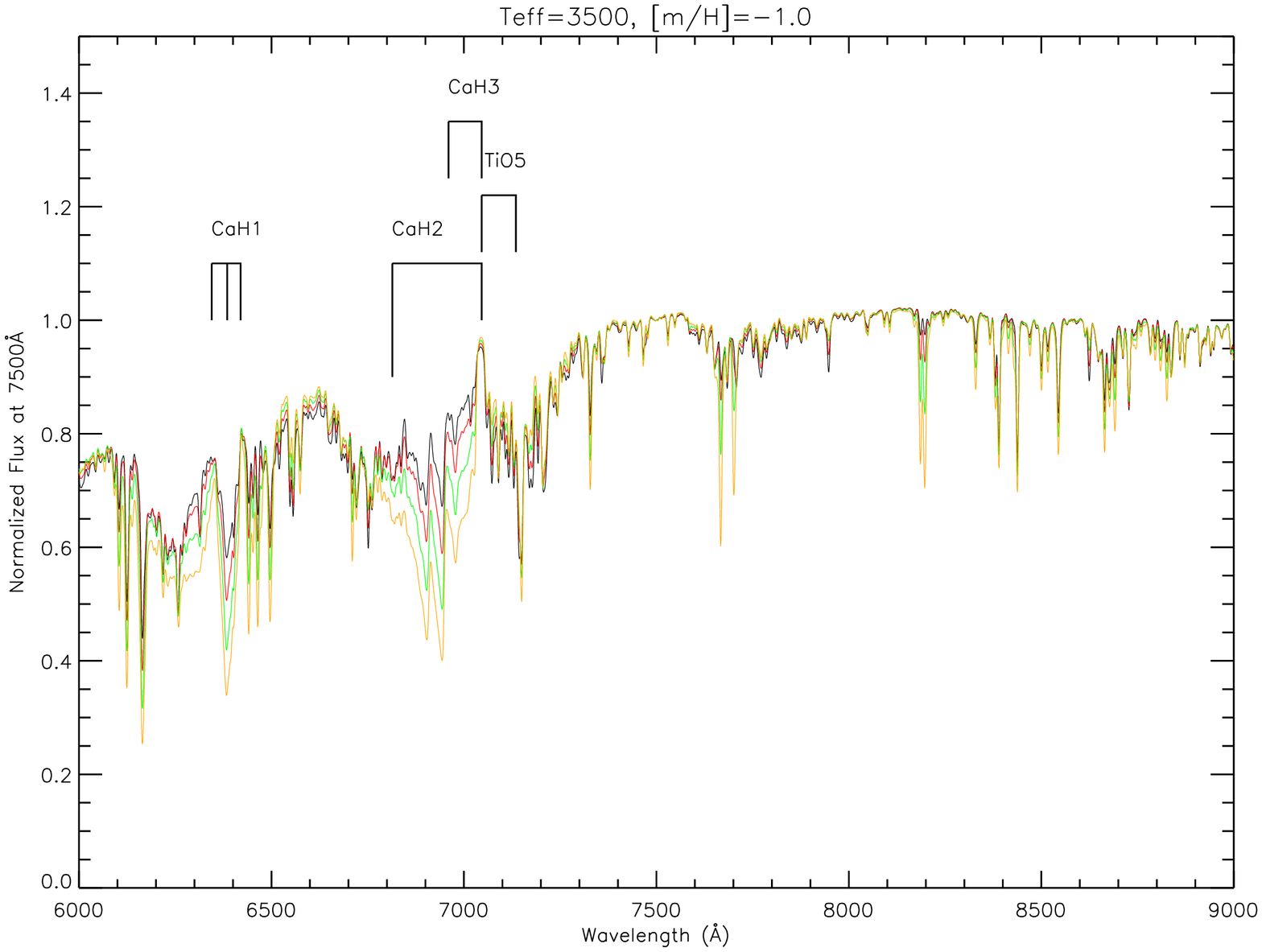}
\includegraphics[scale=0.5, angle=90]{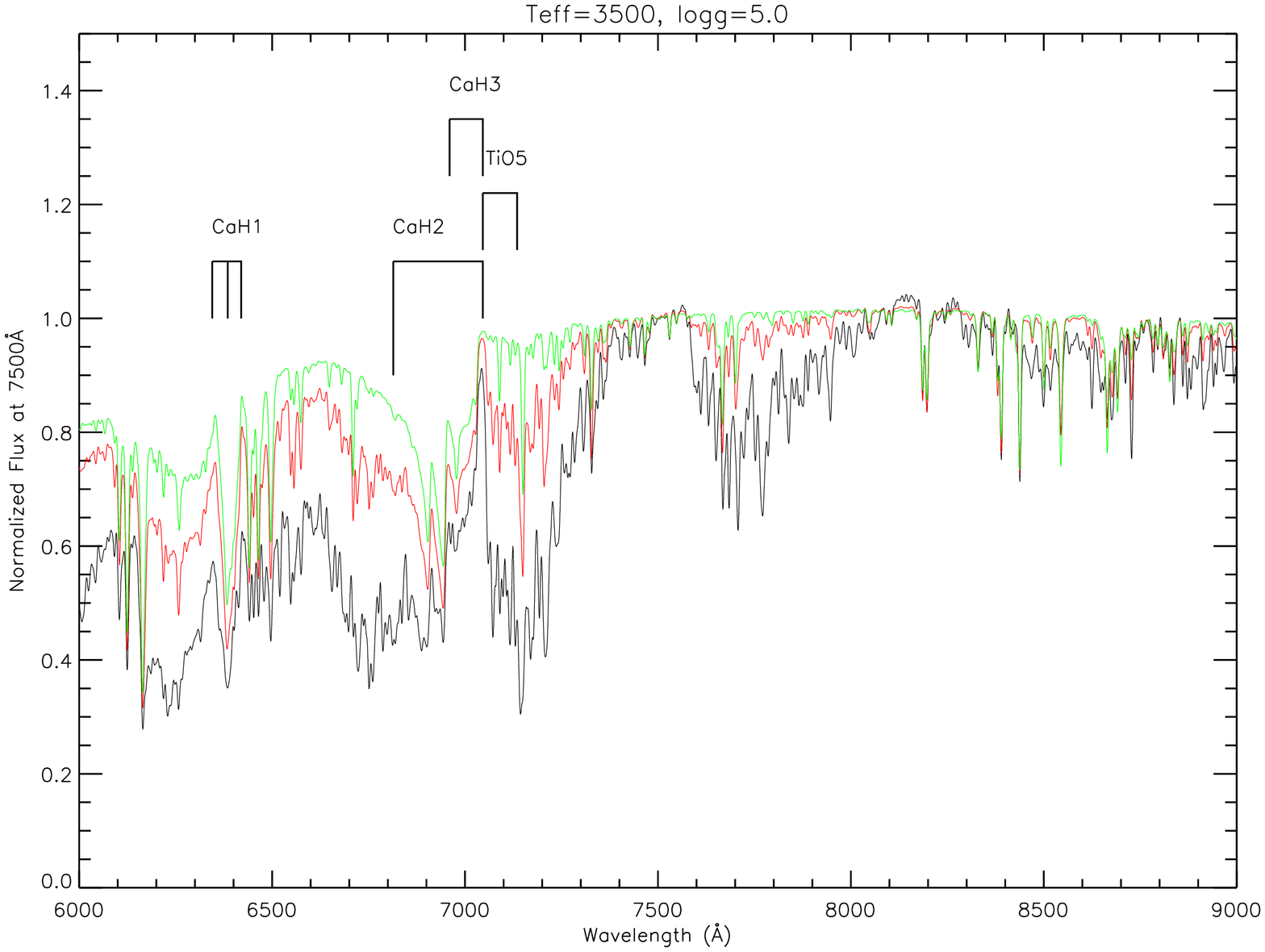}

\caption{The top plot shows GAIA model grids at fixed metallicity
([m/H]$=-$1.0) and effective temperature (3500K).  Black, red, green
and yellow lines represent various {\it log g} = 4.0, 4.5, 5.0 and
5.5, respectively.  It is clear that CaH bands will be affected by
changing gravity but TiO5 is not.  The bottom plot shows model grids
at fixed {\it log g} $=$ 5.0 and $T =$ 3500K.  Black, red and green
lines represent various [m/H] $=$ 0.0, $-$1.0 and $-$2.0,
respectively.  Note that model grids do not have telluric lines. }

\label{fig.3500.plot}
\end{figure}

For late K-type (redder than K5.0) and M-type stars, we use the GAIA
model grids to understand how the effective temperatures,
metallicities, and gravities affect the shapes and features of
subdwarfs' spectra.  This analysis allows us to develop a spectral
sequence for cool subdwarfs of types K6.0 to M6.0, which is
tentatively extended blueward to K3.0 when additional information is
incorporated.  We plot GAIA synthetic noiseless spectra in
Figure~\ref{fig.4800-2800.plot}, at increments of 200K (cooler than
4000K) and 400K (hotter than 4000K) for stars with metallicities of
[m/H]$ =$ 0.0, $-$1.0, and $-$2.0.  Because gravities have very
limited impact on the overall shapes of the spectra (shown in the top
panel of Figure~\ref{fig.3500.plot}), we do not show gravity plots
with fixed temperatures and metallicities (of course, some features do
change markedly with gravity, but not the overall slopes of the
spectra).  Based on the synthetic spectra from GAIA model grids, cool
subdwarf spectra between 6000\AA~and 9000\AA~exhibit the following
trends:

\begin{itemize}

\item The effects of metallicity are minimal in low resolution
subdwarf spectra for stars with temperatures of 4400K and hotter.
This makes it difficult to separate dwarfs and subdwarfs using low
resolution spectra (as discussed in
$\S$\ref{sec.K.subdwarf}). However, from these noiseless spectra, we
can still identify a few metallic lines showing metallicity
trends. The most prominent feature is marked \#1 at 6256\AA~in
Figure~\ref{fig.4800-2800.plot}.  Unfortunately, we do not see this
feature in any of our K-type spectra (nor is it listed in Table 1 of
\citealt{Turnshek1985}), so we consider its validity questionable.
The next prominent feature, marked \#2 in
Figure~\ref{fig.4800-2800.plot}, is Ca I (6162\AA).  This line can
possibly be used to distinguish subdwarfs from dwarfs (see examples
discussed in $\S$\ref{sec.notes}), but in practice it is somewhat
difficult to evaluate in real spectra (with noise) at our resolution.

\item For stars with temperatures of 2800--4000K, metallicity strongly
affects the spectra between 6000\AA~and 8200\AA.  This is the region
that has been historically used to assign spectral types.  In effect,
subdwarfs with decreased metallicities have spectra that are
``brightened'' or ``less blanketed'' at the blue end, relative to
solar metallicity stars.  However, the continuum at wavelengths longer
than 8200\AA~for temperatures 3400--4800K is nearly free of
metallicity effects.  We can therefore use the 8200--9000\AA~region to
establish subtypes in the subdwarf spectral sequence because the slope
is a function of temperature.  This also allows us to mirror the
spectral sequence for dwarfs, providing a useful link between the
dwarf and subdwarf sequences.

\item The TiO5 band strength at 7050--7150\AA~is very sensitive to
metallicity for temperatures cooler than 4000K.  As shown in the top
panel of Figure~\ref{fig.3500.plot}, the TiO5 band strength is
effectively independent of gravity.  We can therefore use the TiO5
feature to separate subdwarfs with different metallicities if their
continua (8200--9000\AA) are the same, regardless of their gravities.

\item For a star of given temperature, stronger CaH bands could be
caused by lower metallicity, as shown in
Figure~\ref{fig.4800-2800.plot}, or higher gravity, as shown in the
top panel of Figure~\ref{fig.3500.plot}.  If two subdwarfs have the
same continua from 8200--9000\AA~and the same TiO5 band strength, but
their CaH bands are different, we can rank them by their relative
gravities.

\end{itemize}

Consequently, the impediments to sorting cool subdwarf spectra at our
resolution discussed in $\S$\ref{sec.sorting.spectra} can be overcome
by understanding the trends revealed in GAIA synthetic spectra: (1)
the mid K-type dwarfs have the same spectra as dwarfs for our spectral
coverage and resolution, (2) CaH features are affected by both
metallicity and gravity, while the TiO5 band is affected by
metallicity but not gravity, and (3) the continuum from 8200\AA~to
9000\AA~is not strongly affected by either metallicity or gravity, so
can therefore be used for spectral sequencing.

\section{Additional Evidence Supporting the Metallicity and Gravity Trends 
Indicated by the GAIA Models}
\label{sec.trend}

\begin{figure}
\centering
\includegraphics[scale=0.65]{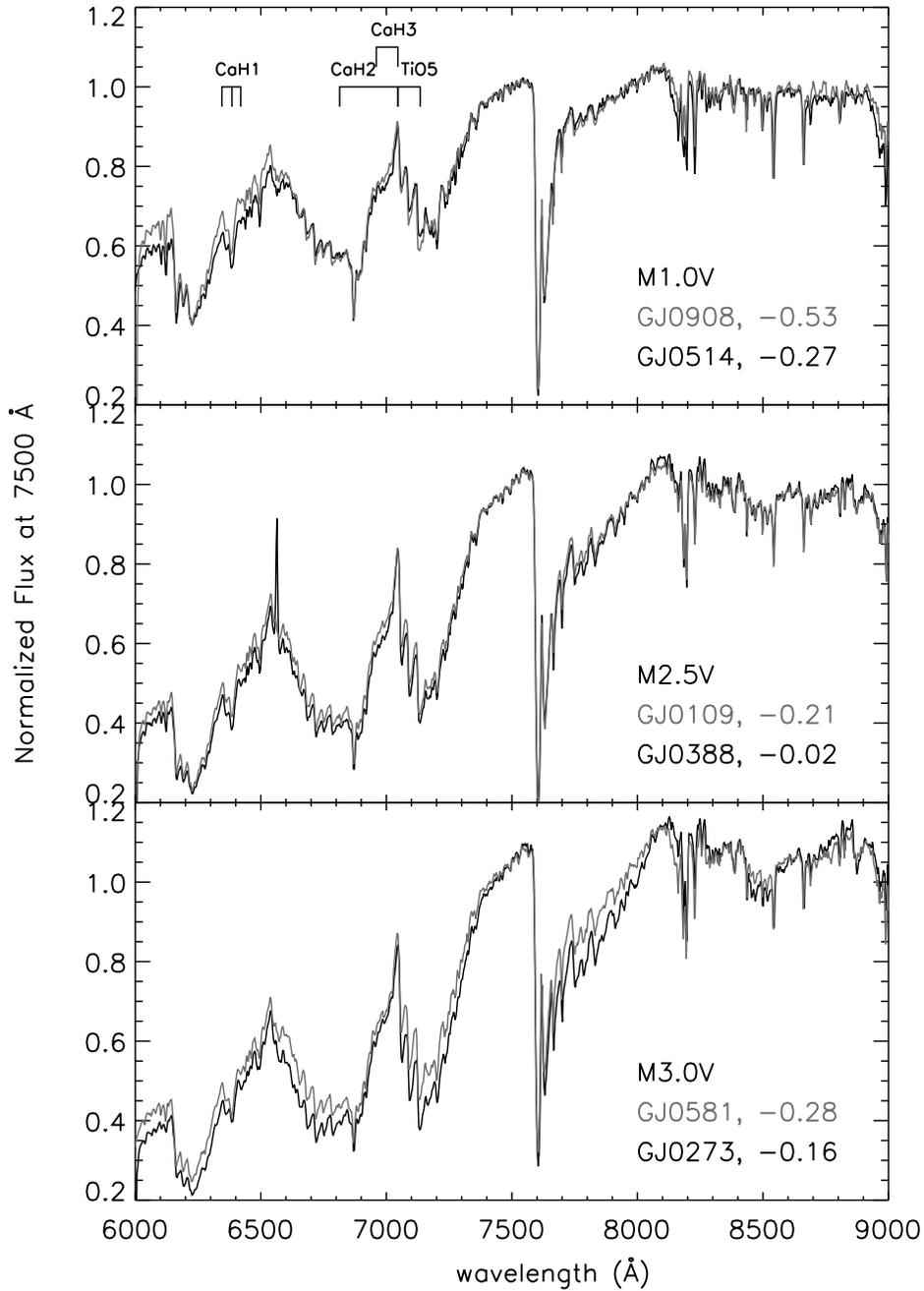}
\caption{Using our spectra and metallicities measured independently by
others, three different types of M dwarfs show the metallicity trend
predicted by GAIA models.  The gray lines represent lower metallicity
stars in each pair.  Metallicities from \cite{Bonfils2005} are given
in each panel for the stars.}

\label{fig:bonfils.recons}
\end{figure}

To investigate the metallicity trends seen in GAIA models, we compare
our available spectra to metallicity measurements provided
independently by others.  Measuring M dwarf metallicities is
difficult, but several recent attempts have made progress
\citep{Valenti1998, Woolf2005, Bonfils2005, Bean2006}.  Six red dwarfs
that we have observed are included in the study by \cite{Bonfils2005},
comprising three pairs of M dwarfs of types M1.0V, M2.5V and M3.0V
shown in Figure~\ref{fig:bonfils.recons}.  Each pair includes a
relatively low metallicity dwarf (gray), and a relatively high
metallicity dwarf (black).  The red ends (from 8200\AA~to 9000\AA) of
each pair match one another, but the blue ends of the lower
metallicity members is brighter in each case, as predicted by the GAIA
models.  The effect is rather subtle, but the derived metallicities
for each pair are not wildly different (none of the six stars are
subdwarfs), and yet the trend is confirmed in all three cases.  One
caveat is that the metallicities from these six stars were determined
from the polynomial relation in \cite{Bonfils2005}, not measured
directly from spectra, but this appears to be the best that can be
done given the available data.  We conclude that the metallicity trend
revealed in the GAIA models (see Figure~\ref{fig.4800-2800.plot}) is
sound because it appears to be confirmed in real M dwarf spectra.

Contrary to the evidence for the metallicity trend, we have found no
direct spectroscopic results to support the gravity trend in dwarfs
and subdwarfs.  Direct gravity measurements are difficult because (1)
stars must have both mass and radius measurements and (2) clean
spectra without contamination from companions must be obtained.  This
limits the available target lists to eclipsing binaries such as those
discussed in \cite{Lopez2007} with cleanly deconvolved spectra or
visual binaries in which individual radii can be measured via long
baseline interferometry.  Neither class of objects yet provides a rich
dataset for cool dwarfs or subdwarfs.

Hence, we rely on other observational or theoretical efforts to
investigate the gravity trend.  The top panel of
Figure~\ref{fig.3500.plot} indicates that other than some sharp
metallic lines, CaH bands show the most prominent changes when gravity
varies (TiO is unaffected by changing gravity).  \cite{Ohman1934}
demonstrated that the CaH2 band is found in the spectra of M-type
dwarfs, but is not observed in the spectra of M-type giants, thus
identifying the CaH2 band as a gravity indicator to separate dwarfs
and giants.  In addition to the GAIA models, \cite{Mould1976} also
showed that for stars with effective temperatures of 3250K, a spectrum
from his atmospheric model with {\it log g} $=$ 5.75 has stronger CaH2
than a spectrum with {\it log g} $=$ 4.75.  If the CaH2 band is a
gravity indicator, we may presume that the same gravity effects for
CaH1 and CaH3 bands will be seen.

As shown in the two panels of Figure~\ref{fig.3500.plot}, the GAIA
models imply that the CaH band strengths are indicators of both
gravity and metallicity differences.  In reality, if two red subdwarfs
have spectra with the same overall continua and slopes, as well as
matching TiO band strengths (e.g.~middle panel in
Figure~\ref{fig.problem.spect}), the only remaining discrepancies will
be at the CaH bands.  We believe that such differences are caused by
different gravities.

Obviously, there is not yet a wealth of accurate direct observational
results of red dwarfs and subdwarfs that can be used to stress test
the metallicity and gravity trends seen in GAIA models.  Nonetheless,
what little we do have supports the trends, so we use these trends to
assist us in establishing the subdwarf spectral sequence discussed
below.

\section{Subdwarf Spectral Sequence}

\subsection{Procedures}

In addition to presenting 88 subdwarfs, a goal of this project is to
establish a subdwarf spectral sequence that mirrors the sequence for
dwarfs and considers the latest available synthetic models.  Although
synthetic models are not yet capable of fully representing the
complicated spectra of these cool stars, the models can be used to
investigate the primary factors that affect subdwarfs' spectra.
Specifically, we examine the spectra in the framework of what appear
to be the three main drivers of the trends observed --- temperature,
metallicity, and gravity.

We first separate our available subdwarf spectra into several groups
that have similar overall slopes.  Known subdwarfs such as LHS 12, GJ
161, and LHS 2734A are used as anchor points.  Within each group,
several stars that have high surface gravities are obvious because
their spectra match except in CaH, implying nearly identical
temperatures and metallicities.

We then compare each subdwarf's spectrum with our sequence of dwarf
spectral standards\footnote{Our dwarf standard sequence is a hybrid of
cool dwarf standards from \cite{Gray2006} for K stars, and
\cite{Boeshaar1976} and \cite{Kirkpatrick1991} for M stars.  The dwarf
standard sequence is the topic of a future publication
\citep{Beaulieu2009}.  Our spectral sequencing efforts began before
the recent spectral sequence from SDSS \citep{Bochanski2007} was
released.  We continue to use our standard sequence because (1) stars
in our sequence have trigonometric parallaxes so we can understand how
metallicities and gravities affect stars' positions on the HR diagram,
(2) the SDSS sequence does not include K-type stars, (3) we have
acquired spectra for many K-type stars from \cite{Gray2003} and
\cite{Gray2006} so we have benchmark K dwarf spectral standards to
link to the M dwarf sequence, (4) SDSS standard stars have telluric
lines removed (It is difficult for us to make comparisons to SDSS
spectra in the 6800--7100\AA~region, which includes O$_{2}$ and
H$_{2}$O absorption.  This region overlaps with the CaH3 and TiO5
bands.  Thus, all subdwarfs would appear to have high gravity when
compared to the SDSS standards.), and (5) our dwarf spectra have been
acquired using the same telescope/instrument/observing protocols as
used for the subdwarfs, and the data have been reduced identically, so
they are systematically consistent.} that span types from K0.0 to M9.0
using an IDL program to find the closest match to the continuum slope
in the region 8200--9000\AA.  Visual checks of the matches between all
subdwarf and standard dwarf spectra are also made to ensure match
quality.

\subsection{Results and Notes on Objects}
\label{sec.notes}

Based on the 88 confirmed cool subdwarf spectra we have, we present a
sequence of subdwarfs with spectral types spanning K3.0 to M6.0,
listed in Table~\ref{tbl.spectral.type}.  We also identify five
additional G type subdwarfs.  After the five G-type subdwarfs, we sort
the cool subdwarfs from K3.0[VI] to M6.0VI, using double lines in
Table~\ref{tbl.spectral.type} to separate each type.

As discussed in $\S$\ref{sec.K.subdwarf}, because mid K subdwarfs are
virtually indistinguishable from K dwarfs at our spectral resolution,
we use [VI] to indicate their questionable luminosity classes, which
are currently based on their metallicities, kinematics, or locations
on the HR diagram.  We anticipate that higher resolution spectra will
reveal these stars to be subdwarfs.  A colon after the type indicates
that we have had difficulty in assigning a subtype, metallicity, or
gravity.

Within each type for which sufficient spectra are available, we sort
targets by their metallicities (lowest metallicity first).  A letter
``m'' indicates a star having the same metallicity as a main sequence
dwarf, while more negative signs indicate lower metallicities,
e.g. m$--$ is more metal poor than m$-$.  We use as many as six
negative signs, because in the case of the M1.0VI type, we have 23
stars that fall into seven different metallicity categories.
Gravities are indicated by ``g'' with additional plus signs for higher
gravities, e.g. g$++$ is higher gravity than g$+$.  A baseline
subdwarf of low metallicity is assigned m$-$ and g.  If a subdwarf has
a similar metallicity but higher gravity, the designations are m$-$
and g$+$.  A few stars appear to have solar metallicity and are
subluminous only because of high gravities; these stars are assigned m
and g$+$.  We have removed any stars from this study that might have
gravities lower than main sequence stars, i.e.~slightly evolved stars
such as subgiants, that would have gravity g$-$.

Note that values for metallicity and gravity are not comparable across
all spectral types, i.e. M1.0VI with metallicity ``m$-$'' is not
equivalent to M2.0VI with metallicity ``m$-$,'' nor is ``g$+$'' for
M1.0VI the same as ``g$+$'' for M2.0VI.  We can hope to formalize a
definitive subdwarf spectral sequence that includes temperature,
metallicity, and gravity trends when hundreds of systematically
consistent subdwarf spectra and improved model grids are available.
Objects with virtually identical spectra are listed in alphabetical
and/or numerical order.

We discuss each spectral subtype from K3.0[VI] to M6.0VI in the
following sections and highlight noteworthy subdwarfs of various
types.  The order of the highlighted targets is based on their
metallicities or gravities.  Colored spectra in the Figures in these
sections are illustrative only --- they do not represent any numeric
metallicities or gravities, so a red-lined spectrum in one type does
not necessarily have similar attributes to a red-lined spectrum in
another type.

\subsubsection{K3.0[VI] to K5.0[VI] types}

\begin{figure}
\centering
\includegraphics[scale=0.6]{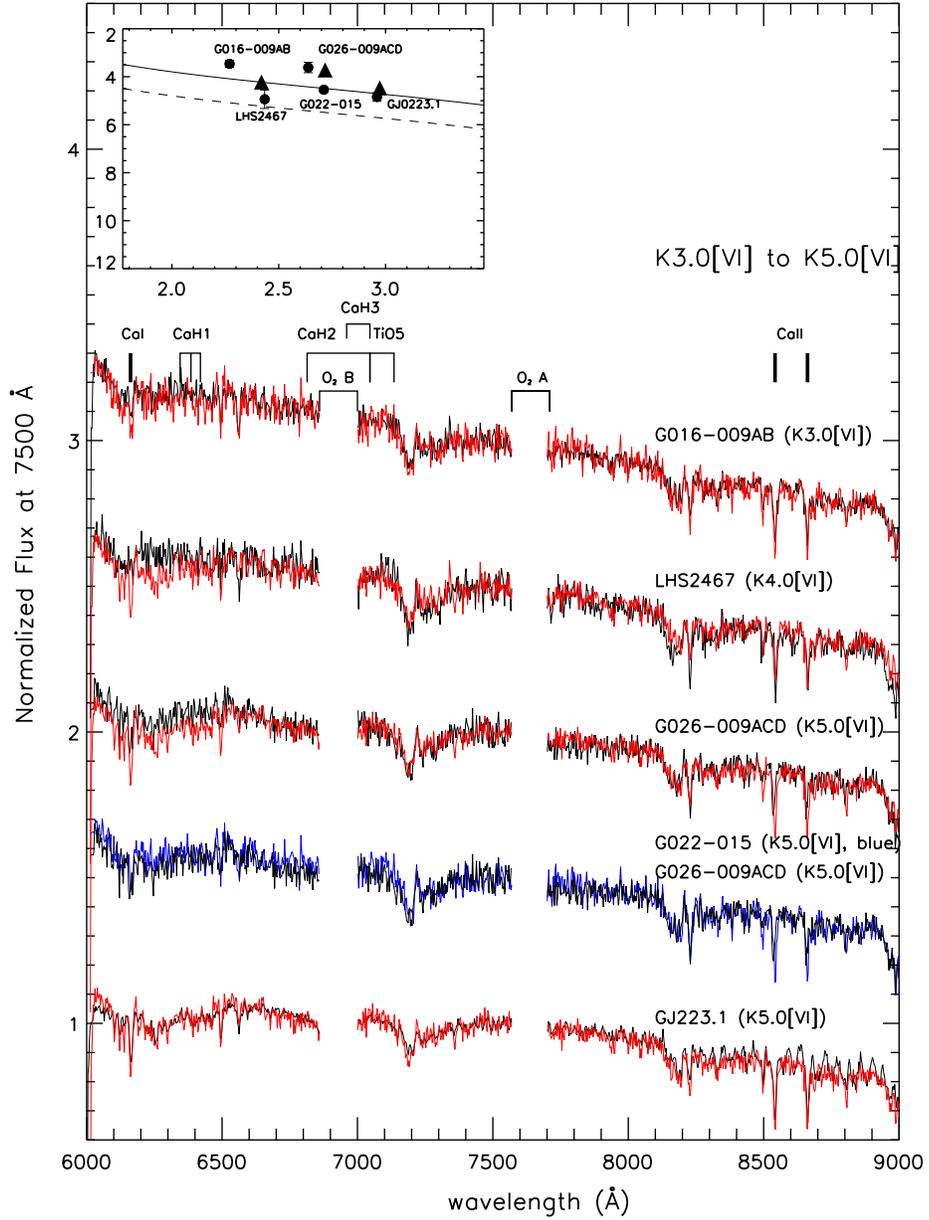}

\caption{K3.0[VI] to K5.0[VI] spectra are shown compared to K dwarf
standard spectra (red).  The two telluric bands (O$_{2}$ A and B) have
been removed for clarity.  The three thick tick-marks represent the
locations of Ca I (6162\AA) and Ca II (8542\AA~and 8662\AA) absorption
features.  The inset plot shows the locations of the stars on the HR
diagram, where filled circles represent subdwarfs and triangles
represent main sequence standards of types K3.0V, K4.0V, and K5.0V
(left to right).  For clarity, the axis labels for the inset plot are
not shown but are always $V-K_{s}$ vs. $M_{Ks}$.  The errors in
absolute magnitudes are shown.  However, because the errors for
$V-K_{s}$ are equal to or smaller than the filled circles, they are
not shown.  The solid line represents a fitted main sequence line and
the dashed line is one magnitude fainter than the solid line.}

\label{fig.K3.K5}
\end{figure}

We begin with K3.0[VI], for which it is still difficult to separate
subdwarfs from dwarfs at our spectral resolution.
Figure~\ref{fig.4800-2800.plot} shows how similar mid K-type subdwarfs
and dwarfs are (temperatures 4400K and 4800K).  For these stars, we
use other independent published measurements, e.g.~[m/H] values, to
confirm their subdwarf natures.  We currently call stars with [m/H]
$\leq$ $-$0.5 subdwarfs.  The K3.0[VI] to K5.0[VI] spectra are shown
in Figure~\ref{fig.K3.K5}.

{\bf G 016$-$009AB (K3.0[VI])} \cite{Goldberg2002} reported this star
to be a double-lined spectroscopic binary with P$=$9.9 days,
$T_{eff}=$ 5903K, and having a mass ratio ($M_{1}$/$M_{2}$) of
$\sim$1.22.  Given that the components are presumably coeval, the
combined spectrum should represent the metallicity for each component.
Although their estimated temperature is that of a G type star, our
spectrum has the same slope as a K3.0V.  \cite{Cayrel2001} and
\cite{Goldberg2002} reported measurements of [Fe/H]$=-$0.7 and $-$1.0
respectively, while \cite{Laird1988} found [m/H]$=-$1.11.  Apparently,
it is a low metallicity star and its location on the HR diagram is
elevated because of multiplicity.  This system is also reported to be
a photometrically variable system in \cite{Kazarovets2006}.  This
system demonstrates that low metallicity subdwarfs do not have
different spectra from their main sequence counterparts at our
wavelength coverage and resolution, so we assign its luminosity class
as [VI] as a result its low metallicity measurements.

{\bf LHS 2467 (K4.0[VI])} Our spectrum shows that it is a K4.0[VI],
rather than a G7V type, as reported in \cite{Bidelman1985}.  Its
continuum is not as blue as K0.0V and its Ba I/H${\alpha}$ lines are
not as weak/strong as shown in Figure~\ref{fig.jacoby.silva} for a G
type star.  Its weighted mean parallax from \cite{Hipparcos} and
\cite{YPC} is 10.63$\pm$1.88 mas and its proper motion is 0\farcs98
yr$^{-1}$ \citep{LHS}, indicating $V_{tan}=$ 437 km sec$^{-1}$.
\cite{Ryan1991} reported the star to have $V_{rad}=$ 44 km sec$^{-1}$.
After removing the solar motion, LHS 2467 has $(U,V,W)=$
(281.4,$-$304.8,$-$41.9) km sec$^{-1}$.  The extremely high tangential
and space velocities are indicative of a subdwarf.

{\bf G 026$-$009ACD (K5.0[VI])} The wide (132\arcsec, corresponding to
6520 AU at the system's distance of 49.4 pc) common proper
motion companion, B, in this quadruple system is a white dwarf known
as G026$-$010 or G026$-$009B, that during its planetary nebula phase
should have had limited impact on the metallicity of the close ACD
triple.  \cite{Peterson1980} reported G026-009AC to be double-lined
spectroscopic binary with P$=$3.75 days, with a mass ratio
($M_{1}$/$M_{2}$)$\sim$1.25.  \cite{Allen2000} reported a third
component, D, 0\farcs7 away from G026-009AC.  \cite{Morrison2003} and
\cite{Allen2000} both report the system (ACD) to have low metallicity,
measuring [Fe/H]$=-$0.91 and $-$1.19, respectively.  Both the $V$ and
$K_{s}$ magnitudes include three stars, causing the point to be
improperly placed on the HR diagram.

{\bf G 022-015 (K5.0[VI])} \cite{Woolf2005} reported this system to
have [Fe/H]$=-$0.61.  G022-015's spectrum is virtually identical to
G026-009ACD, as shown by plotting both together in
Figure~\ref{fig.K3.K5}.

{\bf GJ 223.1 (K5.0[VI])} This star has the same slope as the K5.0V
standard and its location on the HR diagram is on the main sequence
line.  Nonetheless, \cite{Woolf2005} reported it to have
[m/H]$=-$0.62, so we tentatively consider it to be a subdwarf until
further information indicates otherwise.

Although not definitive, the Ca I line at 6162\AA~in these spectra
seems to show the metallicity trend seen in the noiseless model
spectra plotted in Figure~\ref{fig.4800-2800.plot} and discussed in
section~\ref{sec.subdwarf.facts}.  LHS 2467 has a much weaker Ca I
line, and G016$-$009AB and G026$-$009AB have slightly weaker Ca I
lines than seen in the dwarfs' spectra.  However, G022$-$015 and
GJ223.1 both have approximately the same Ca I line strengths as
dwarfs.  Contrary to the Ca I line, the Ca II lines at (8542\AA~and
8662\AA) do not show any metallicity trend from our spectra.  Because
these prominent lines fail to separate subdwarfs from dwarfs at our
resolution and S/N, other evidence, i.e.~kinematics, parallaxes or
independent metallicity measurements, are required to identify the
subdwarfs.  An alternative method to separate the subdwarfs from the
dwarfs is to obtain spectra that include the MgH bands at 4845\AA~,
5211\AA, and 5621\AA~that \cite{Bessell1982} pointed are sensitive to
metal abundance for temperatures hotter than 4500K.

Several possibilities may explain the locations of the three
subdwarfs, LHS 2467, G 022-015, and GJ 223.1 on the HR diagram just
below or on the main sequence line: (1) they are subdwarfs with unseen
companions that brighten their $M_{Ks}$ magnitudes, (2) they are
slightly evolved and have moved significantly from the subdwarf zero
age main sequence line, or (3) their metallicity measurements are not
accurate and they are, in fact, main sequence dwarfs.

\begin{figure}
\centering
\includegraphics[scale=0.7]{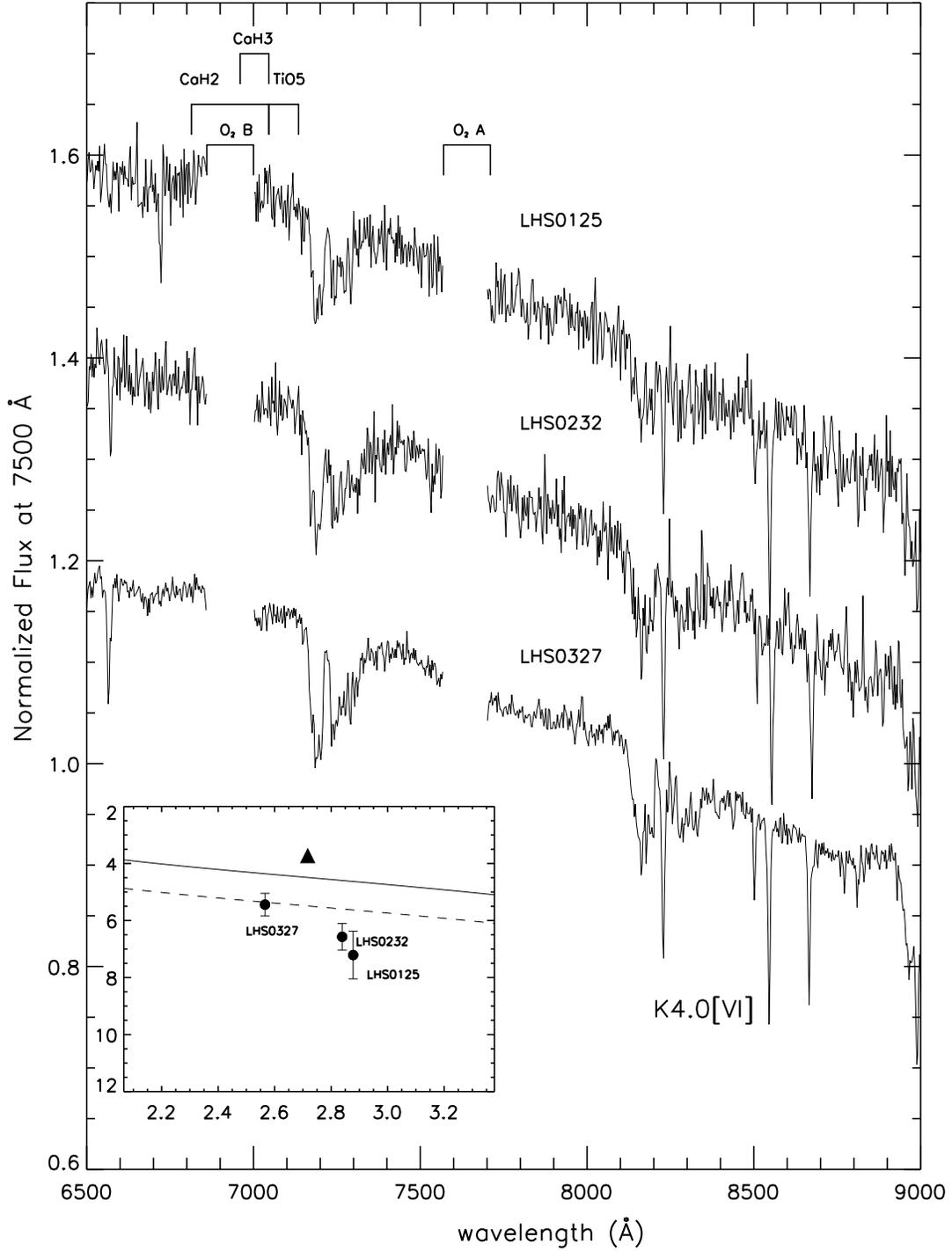}

\caption{Spectra of LHS 125, LHS 232 and LHS 327 (all K4.0[VI]), are
shown.  Note that the blue end of these spectra are at 6500\AA~because
of problems with these particular spectra between 6000\AA~and 6500\AA.
The two telluric O$_{2}$ A and B bands have been removed.  The
triangle in the inset plot represents a K4.0V standard star.  Symbols
and lines have the same meanings as in Figure~\ref{fig.K3.K5}.}

\label{fig.K4}
\end{figure}

{\bf LHS 125, LHS 232 and LHS 327} The spectra for these three stars,
shown in Figure~\ref{fig.K4}, match neither the LHS 2467 K4.0[VI]
spectrum nor a K4.0V spectrum, although the spectra have the overall
slopes and the stars have $V-K_s$ colors indicative of K4.0 stars.
They are tentatively assigned types of K4.0[VI] based on their
locations on the HR diagram. 

\subsubsection{K6.0VI}

\begin{figure}
\centering
\includegraphics[scale=0.7]{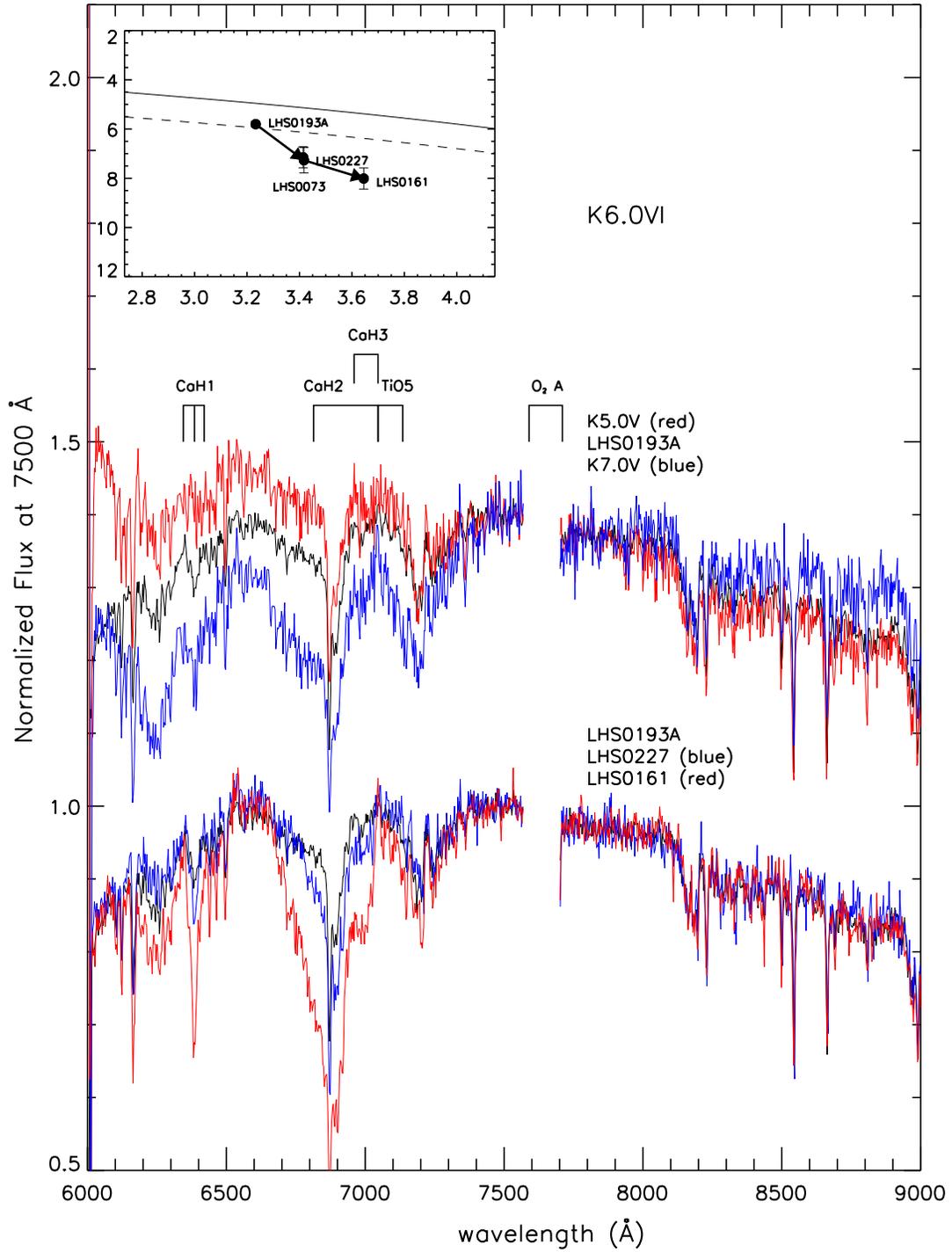}

\caption{Spectra of LHS 193A, LHS 227 and LHS 161 (all K6.0VI), are
shown with our K5.0V and K7.0V standard spectra.  The telluric
O$_{2}$ A band has been removed.  The thick arrows in the HR
diagram point toward sequentially higher gravity subdwarfs.
Symbols and lines have the same meanings as in
Figure~\ref{fig.K3.K5}.}

\label{fig.K6}
\end{figure}

We identify four stars to have spectral type K6.0VI, shown in
Figure~\ref{fig.K6}. Their spectral slopes fall between K5.0V and
K7.0V.

{\bf LHS 193A} This is a binary with a separation of 12\farcs59
comprised of a cool subdwarf (A) and a DC-type white dwarf (B) with a
featureless spectrum \citep{Monteiro2006}.  At the system's distance
of 31.2 pc, the large separation of the pair corresponds to 393 AU,
indicating that significant pollution of the subdwarf by the white
dwarf during its planetary nebula phase seems unlikely.  From
Figure~\ref{fig.K6}, it is clear that LHS 193A's spectral slope is
between K5.0V and K7.0V.  Although the spectrum is a near match to
K5.0V between 8200\AA~and 9000\AA, the blue end (6000\AA--7500\AA) is
too low to be a K5.0V (as a subdwarf, the blue end of the spectrum
would have to be above the dwarf standard).  As shown in
Figure~\ref{fig.4800-2800.plot}, for a star with 4400K (a K5.0V from
\citealt*{Cox2000}), there is effectively no difference in the
continua of spectra for dwarfs and subdwarfs.  Hence, we assign LHS
193A a type of K6.0VI because its TiO5 band is ``brightened'' (low
metallicity), as shown in Figure~\ref{fig.4800-2800.plot}.  Given its
low metallicity, location on the HR diagram (shown in the inset of
Figure~\ref{fig.K6}), its $V_{tan}=$147.3 km sec$^{-1}$
\citep{Jao2005}, and its age of 6--9 Gyr based on its white dwarf
companion's cooling age \citep{Monteiro2006}, it is most likely a
thick disk subdwarf.

{\bf LHS 73 and LHS 227} These two objects have virtually identical
spectra, so we only show LHS 227 in Figure~\ref{fig.K6} for clarity.
The only differences between these spectra and that of LHS 193A are in
CaH.  Using results from the top panel of Figure~\ref{fig.3500.plot},
we find that LHS 73 and LHS 227 have higher gravities than LHS 193A.

We have acquired spectra for both LHS 72 and LHS 73, a wide
($\sim$96\arcsec~separation) common proper motion pair.
Unfortunately, the spectrum of LHS 72 ($V-K_{s} =$ 3.29, $M_{Ks} =$
6.69) is poor and requires re-observation, although it would fall in
the subdwarf region of the HR diagram.  LHS 73 has $V-K_{s}=$ 3.42 and
$M_{Ks}=$ 7.27 and is plotted in Figure~\ref{fig.K6}, also clearly in
the subdwarf region.  \cite{Rodgers1974} previously identified both
objects as subdwarfs, but no spectral types were given.
\cite{Bidelman1985} identified LHS 72 and LHS 73 as K4 and K5 dwarfs,
respectively.  \cite{Reyle2006} reported LHS 72/73 (they identified
LHS 73 as sdK7) to be the nearest subdwarf binary based on their
spectroscopic parallax (18.7 pc), although the trigonometric parallax
from YPC is 37.6$\pm$8.9 mas (21.5 pc $<$ $d$ $<$ 34.8 pc).  The
nearest known subdwarf binary system is actually $\mu$ Cas AB, at a
distance of 7.5 pc ($\pi$$_{trig}$ $=$ 132.4$\pm$0.6 mas,
\citealt{Hipparcos}).

{\bf LHS 161} This star was previously reported to be an esdM2.0 extreme
subdwarf \citep{Gizis1997}.  However, its slope and spectral features,
like LHS 227, are too hot for type M2.0VI.  As shown in
Figure~\ref{fig.K6}, LHS 161 has stronger CaH than LHS 193A and LHS
227, but the rest of the spectrum is the same.  We conclude that LHS
161 has higher gravity than the other two subdwarfs.  LHS 161's very
strong CaH lines and consequent CaH and TiO5 indices place it in the
previously called ``extreme'' subdwarf region, which typically implies
that the star has very low metallicity.  It appears, from our current
understanding, that the strong CaH in LHS 161 may not be linked to
metallicity, because the TiO5 band is not significantly different from
LHS 193A, LHS 73 or LHS 227.

In the inset of Figure~\ref{fig.K6} we connect LHS 193A, LHS 73, LHS
227, and LHS 161 on the HR diagram with thick arrows to outline a
sequence of increasing gravity effects that shift their locations to
the lower right (less luminous and redder).  We will soon see more
such gravity effects, which are always towards lower luminosities.

\subsubsection{K7.0VI}
\label{sub.sec.K7.0VI}

\begin{figure}
\centering
\includegraphics[scale=0.7]{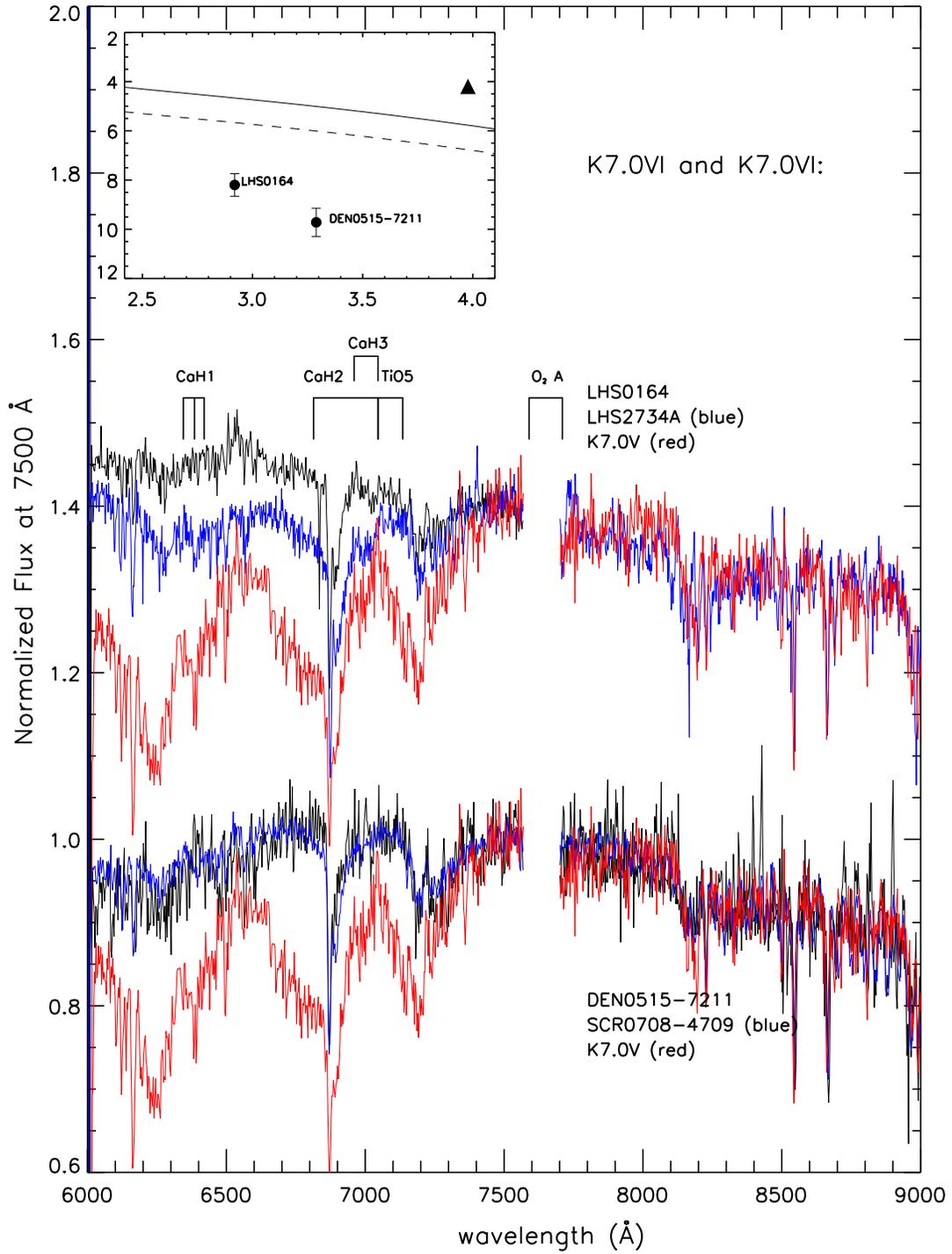}

\caption{Spectra of LHS 164, LHS 2734A (both K7.0VI), DEN 0515$-$7211,
and SCR0708-4709 (both K7.0VI:) are shown, with our K7.0V standard
spectrum (represented by a triangle in the inset plot).  The telluric
O$_{2}$ A band has been removed.  Symbols and lines have the same
meanings as in Figure~\ref{fig.K3.K5}.}

\label{fig.K7}
\end{figure}

We identify four stars to have spectral type K7.0VI, shown in
Figure~\ref{fig.K7}.  This is the earliest spectral type for which we
see clear metallicity effects in the blue regions of our spectra.

{\bf LHS 2734A} This is the primary in a new common proper motion
system discovered during CTIOPI, for which we measure the components
to have $\mu$ = 0$\farcs$59 and 0$\farcs$60 yr$^{-1}$ at position
angles 228.2$^{\circ}$ and 228.4$^{\circ}$ for the A and B components,
respectively.  The B component is located 68\farcs8 away at position
angle 162.4$^{\circ}$.  Unfortunately, both components are beyond our
current trigonometric parallax limit (100 pc), but the A component
provides a useful subdwarf anchor point even though its location on
the HR diagram is unknown.

There are three reasons to adopt LHS 2734A as a reliable subdwarf
anchor.  First, we measure a zero parallax, implying that its
$V_{tan}$ is at least 280 km/sec (assuming a distance of 100 pc).  A
tangential velocity of this size is indicative of a subdwarf.  Second,
LHS 2734B $(V-K_{s}=$ 3.90), is redder than LHS 2734A ($V-K_{s}=$
3.23), and later type subdwarfs are more easily separated from dwarfs,
as shown in Figure~\ref{fig.4800-2800.plot}.  As discussed in
$\S$\ref{sub.sec.M1}, LHS 2734B is clearly a subdwarf.  Assuming this
common proper motion pair formed at the same time with similar
``genetics'', both components are subdwarfs.  Third,
Figure~\ref{fig.K7} shows the comparison between K7.0V and LHS 2734A,
in which the blue end of LHS 2734A's spectrum is clearly
``brightened''.  This indicates that LHS 2734A has lower metallicity
than K7.0V, as expected for a subdwarf.

{\bf LHS 164} This star is four full magnitudes below the main
sequence line in the inset in Figure~\ref{fig.K7}.  We find a good
match at the red end between our K7.0V standard and LHS 164's
spectrum, while the blue end is elevated.  We conclude that LHS 164's
metallicity is even lower than LHS 2734A's.

Thus, our K7.0V standard, LHS 2734A, and LHS 164 form a sequence of
decreasing metallicity.  The CaH and TiO5 band depths significantly
decrease as the metallicity drops, matching the effects discussed in
section~\ref{sec.K.subdwarf} and Figure~\ref{fig.model.index.plot}.

{\bf DEN0515$-$7211 and SCR 0708$-$4709} These two spectra are
virtually identical, and are assigned type K7.0VI:
(Figure~\ref{fig.K7}) because the spectra do not match either the
K7.0V standard, LHS 164, or LHS 2734A.  DEN0515$-$7211 is nearly five
magnitudes less luminous in $M_{Ks}$ than the main sequence line, so
it is certainly a subdwarf.  SCR 0708$-$4709 ($\mu=$0\farcs402
yr$^{-1}$, \citealt{Subasavage2005a}) has almost the same spectrum as
DEN0515$-$7211 and falls in the subdwarf region on the reduced proper
motion diagram (see \citealt{Subasavage2005a}).

\subsubsection{M0.0VI}

\begin{figure}
\centering
\includegraphics[scale=0.7]{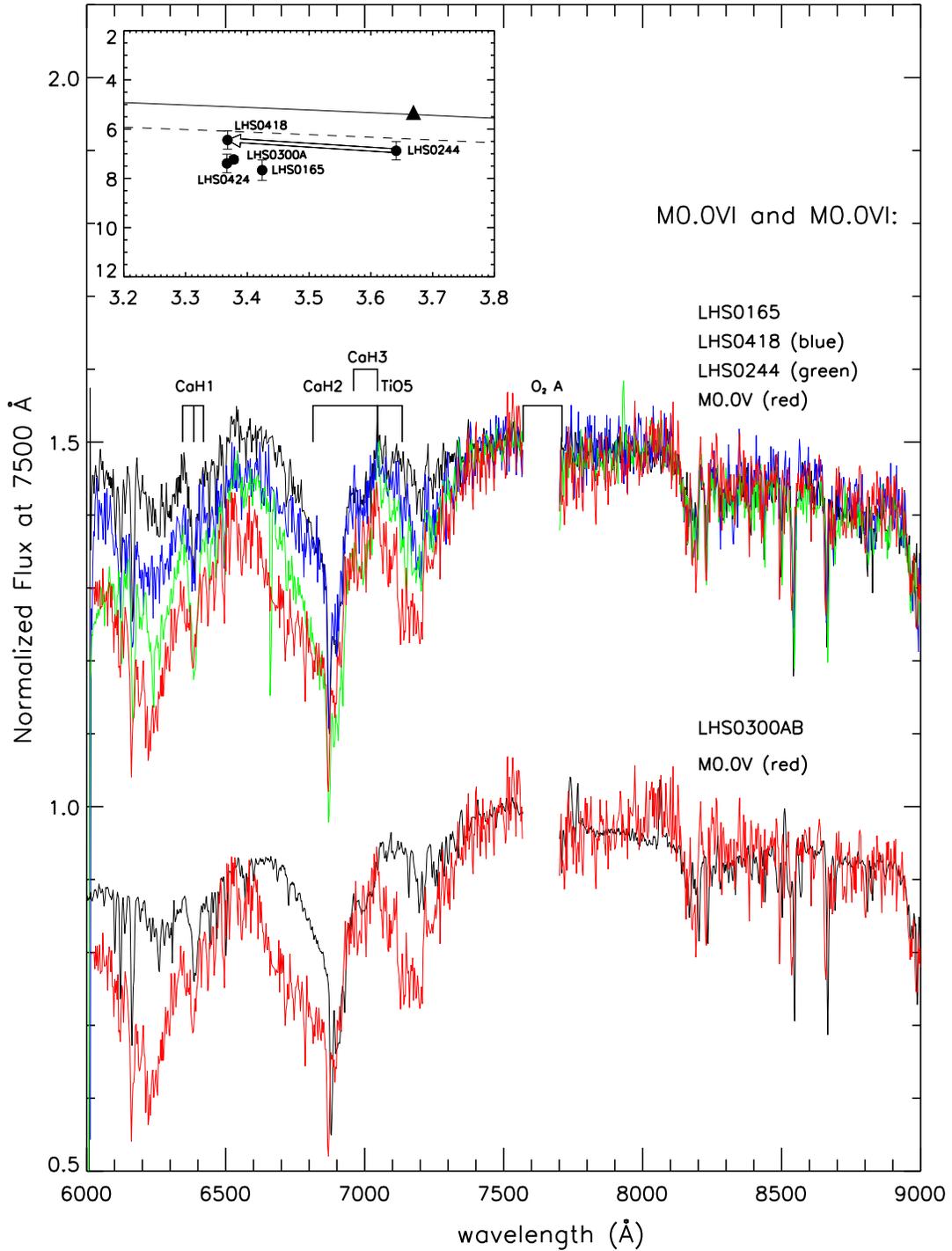}

\caption{Spectra of LHS 165, LHS 418, LHS 244 (all M0.0VI), and LHS
300AB (M0.0VI:) are shown with our M0.0V standard spectrum
(represented by a triangle in the inset plot).  The telluric O$_{2}$ A
band has been removed.  The hollow arrow indicates the shift on the HR
diagram caused by decreasing metallicity.  Symbols and lines have the
same meanings as in Figure~\ref{fig.K3.K5}.}

\label{fig.M0}
\end{figure}

We show four stars with spectral type M0.0VI in Figure~\ref{fig.M0}.
This is the first group in this manuscript for which we see clear
effects of both metallicity and gravity.

{\bf LHS 244, LHS 418, and LHS 165} These three spectra have the same
continua at the red end but are very different at the blue end.  The
differences are caused by metallicity, as seen in
Figure~\ref{fig.4800-2800.plot}, with the main sequence standard--LHS
244--LHS 418--LHS 165 trending to lower metallicities.  The blue ends
of the subdwarf spectra have clearly ``brightened'' along this
sequence, which is consistent with the theoretical models.  In
addition, as discussed in $\S$\ref{sec.K.subdwarf}, CaH1 absorption
decreases (increasing index) as metallicity decreases if $T_{eff}$ is
greater than 3500K.  LHS 165 is redder in $V-K_{s}$ and has stronger
CaH1 absorption than LHS 418, so we suspect LHS 165 has a higher
gravity (similar to gravity effects shown in Figure~\ref{fig.K6}).  We
also find that LHS 418 has weaker CaH1 absorption than LHS 244,
indicating that LHS 244 could also possibly have higher gravity than
LHS 418.  LHS 244 is much redder than LHS 418 and LHS 165 in the HR
diagram is because of a combination of gravity and metallicity
effects.

{\bf LHS 424} This star has the same spectrum and color as LHS 418, so
we assign the same spectral type.  We note that LHS 424 is a full
magnitude less luminous in $M_{Ks}$, perhaps hinting that LHS 418 is a
multiple system.

{\bf LHS 300AB} This is a close binary with a separation of
$\sim$4\farcs3 \citep{Jao2003} comprised of a cool subdwarf (A) and a
DC-type white dwarf (B) with a featureless spectrum
\citep{Monteiro2006}.  The spectrum obtained includes both components,
but with $\Delta$VRI $=$ 4.61, 4.85, 4.96 mag \citep{Monteiro2006},
the contamination of the cool subdwarf spectrum from the white dwarf
is negligible.  At a distance of 31.0 pc, the projected
separation implies a distance of 133 AU between the two
components, indicating that the subdwarf's composition was unlikely to
be significantly contaminated by the evolved star during its planetary
nebula phase.  The spectrum does not match other M0.0VI stars at the
blue end, but it matches the red end of M0.0V and M0.0VI well.  We
therefore assign it a type of M0.0VI:.

\subsubsection{M0.5VI}

\begin{figure}
\centering
\includegraphics[scale=0.7]{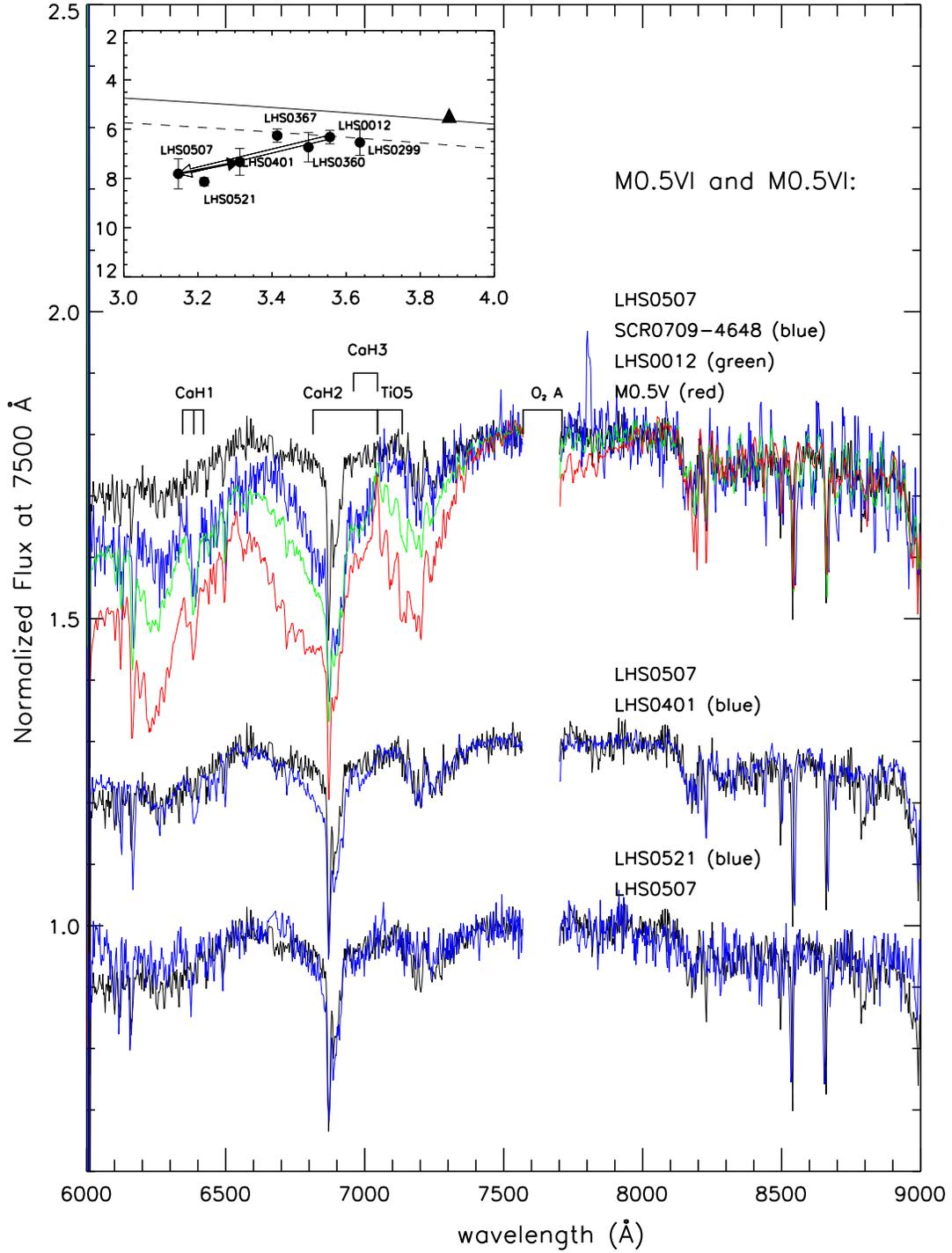}

\caption{Spectra of LHS 507, SCR 0709-4648, LHS 12, LHS 401 (all
M0.5VI), and LHS 521 (M0.5VI:) are shown with our M0.5V standard
spectrum (represented by a triangle in the inset plot).  The telluric
O$_{2}$ A band has been removed.  The hollow arrow indicates the shift
on the HR diagram caused by decreasing metallicity.  The solid arrow
indicates the shift caused by higher gravity.  Symbols and lines have
the same meanings as in Figure~\ref{fig.K3.K5}.}

\label{fig.M0.5.1}
\end{figure}

\begin{figure}
\centering
\includegraphics[scale=0.7]{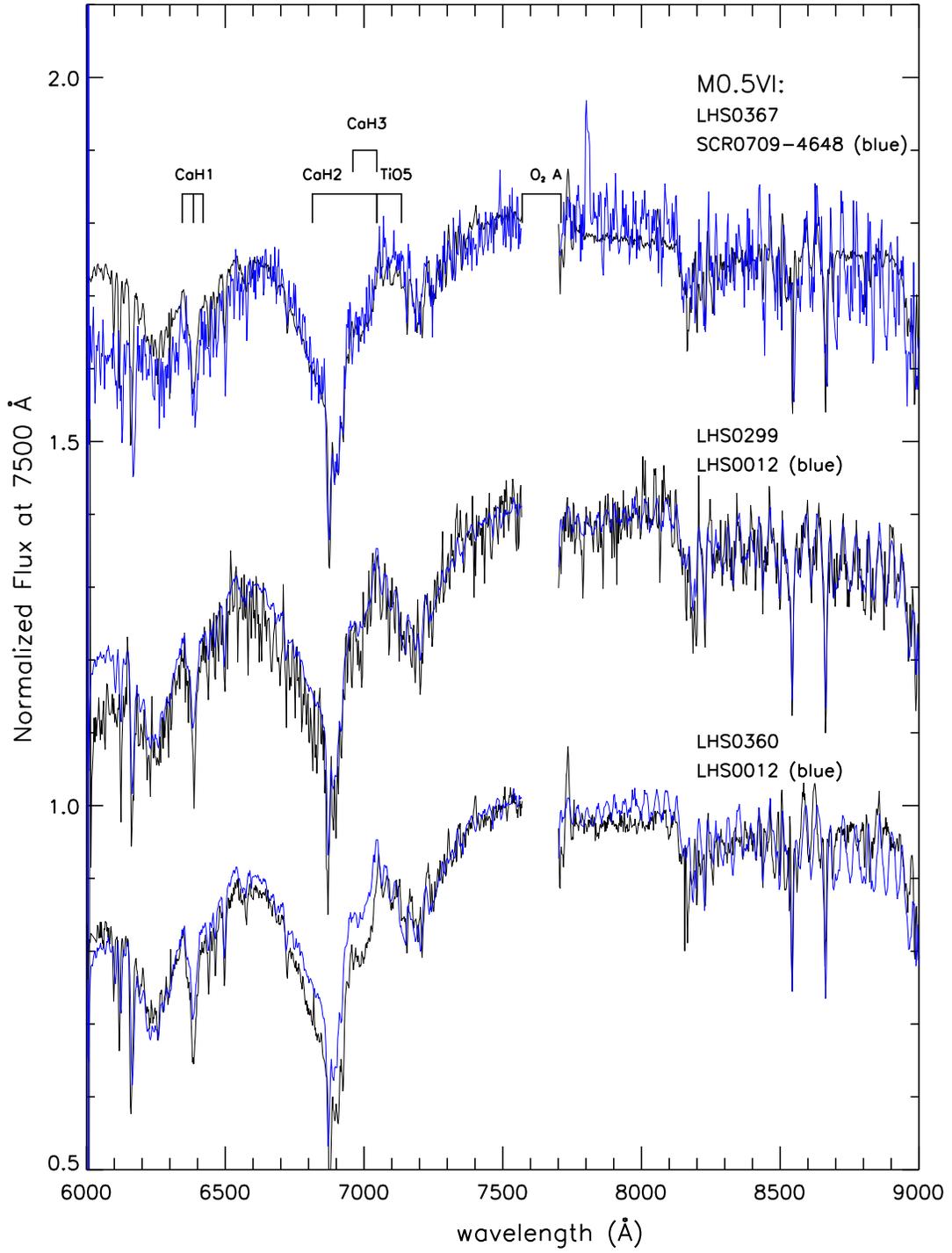}

\caption{Spectra of LHS 367, LHS 299, and LHS 360, are shown (all
M0.5VI:).  The telluric O$_{2}$ A band has been removed.}

\label{fig.M0.5.2}
\end{figure}

This is the first half spectral type we currently assign for
subdwarfs.  We see both metallicity and gravity effects for M0.5VI, as
shown in Figures~\ref{fig.M0.5.1} and \ref{fig.M0.5.2}.

{\bf LHS 507, SCR 0709$-$4648, and LHS 12} These three stars, along
with our M0.5V standard, illustrate a nice four-step trend in
metallicity.  With the lowest metallicity, LHS 507 is significantly
bluer than LHS 12, which is, in turn, bluer than our main sequence
standard.  We do not yet have a trigonometric parallax for SCR
0709$-$4648, so we cannot confirm its location on the HR diagram.  LHS
12 was previously reported to be a type sdM0.0 in \cite{Gizis1997}.

{\bf LHS 401} This star has stronger CaH bands and therefore higher
gravity than LHS 507, which has extremely weak CaH1.  LHS 401 is
redder than LHS 507, as expected, but is slightly more luminous, which
is not consistent with the gravity effects seen for types K6.0VI and
M0.0VI.  We believe this inconsistency is a result of the large
absolute magnitude errors for these two objects.

{\bf LHS 521} This star has a spectrum nearly identical to LHS 507,
except at 6000-6200\AA.  Its CaH2 feature is slightly different from
LHS 507, so we assign it a type of M0.5VI:.

{\bf LHS 367 and LHS 299} Both objects have trigonometric parallaxes.
Their spectra are similar to SCR 0709$-$4648 and LHS 12, respectively
(shown in Figure~\ref{fig.M0.5.2}), so they also have similar offsets
in metallicities from one another.  We assign them as M0.5VI: because
of (1) discrepancies at 6000\AA--6200\AA, (2) LHS 299 has deeper CaH1
absorption than LHS 12 and different CaH2 absorption, and (3) their
locations on the HR diagram are redder than their respective
comparison stars.  LHS 299 may have a higher gravity than LHS 12, but
a smaller absolute magnitude error is needed to confirm this
speculation.

{\bf LHS 360} This star appears to have higher gravity than LHS 12
because the CaH features are deeper, but its location on the HR
diagram does not match its stronger gravity because its $V-K_{s}$
color is slightly bluer than LHS 12.  \cite{Costa2006} reported that
LHS 360 has $V_{tan}\sim$ 524 km s$^{-1}$.  Based on its very high
$V_{tan}$ and location on the HR diagram, we assign it a type of
M0.5VI:.

\subsubsection{M1.0VI}
\label{sub.sec.M1}

\begin{figure}
\centering
\includegraphics[scale=0.7]{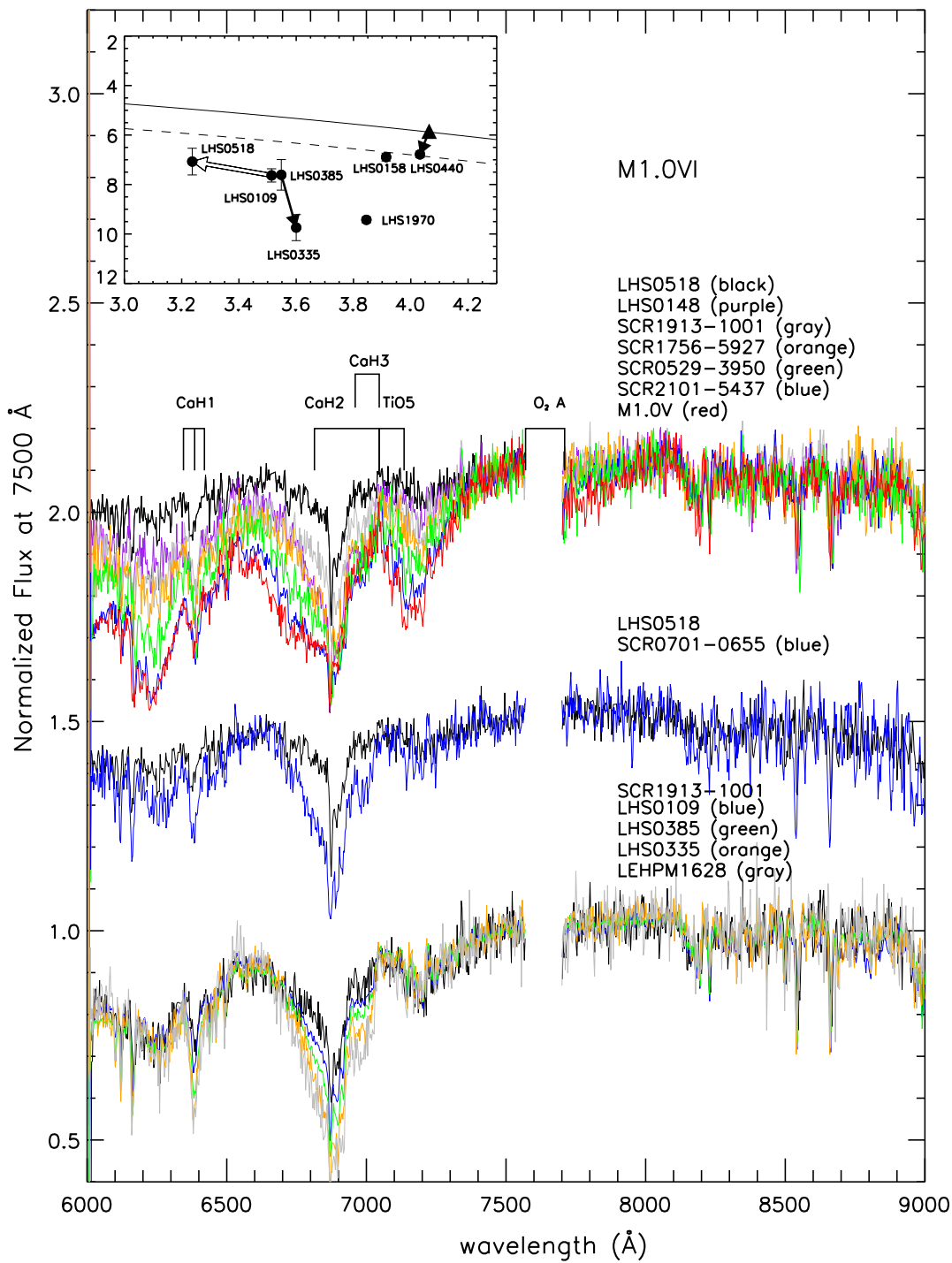}

\caption{Progressions in metallicity and gravity effects in M1.0VI
subdwarf spectra are shown.  The top set of spectra illustrates
spectra with metallicities increasing from black to red (top to
bottom, metallicity scale from m$------$ to m).  The red line is our
M0.5V standard spectrum (represented by a triangle in the inset
plot). The bottom two sets of spectra illustrate gravity effects for
metallicity scales m$------$ and m$----$.  The telluric O$_{2}$ A band
has been removed.  Symbols and lines have the same meanings as in
Figure~\ref{fig.M0.5.1}.}

\label{fig.M1.1}
\end{figure}

\begin{figure}
\centering
\includegraphics[scale=0.7]{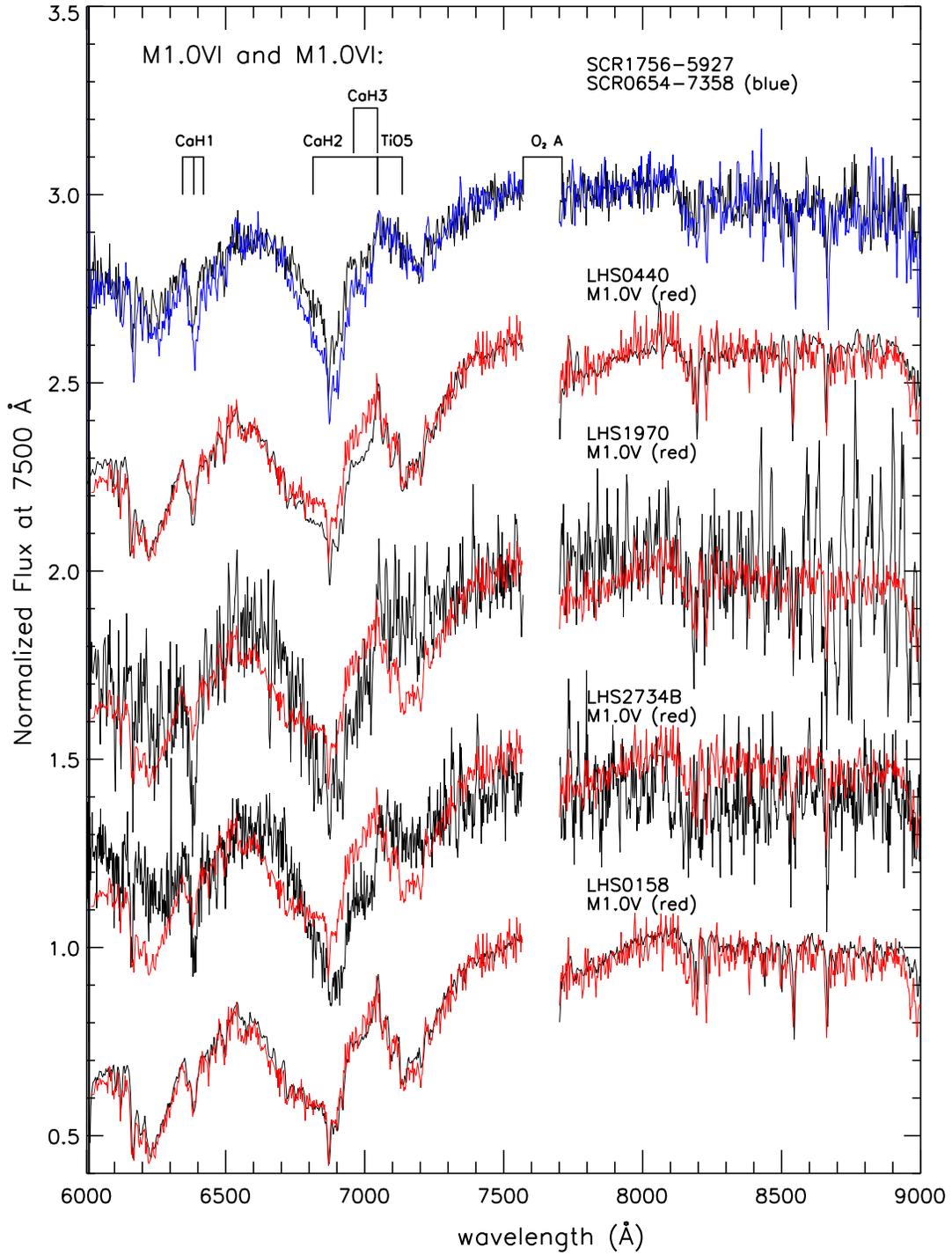}

\caption{The top set of spectra illustrates gravity effects for
metallicity scale m$---$.  Spectra of LHS 440, LHS 1970, LHS LHS 2734B
(all M1.0VI), and LHS 158 (M1.0[VI]), are also shown compared
individually to our M1.0V standard. The telluric O$_{2}$ A band has
been removed.}

\label{fig.M1.2}
\end{figure}

The M1.0VI type has the largest number (23) of confirmed subdwarfs in
our current sample.  The spectra of 16 different M1.0VI stars are
shown in Figures~\ref{fig.M1.1} and~\ref{fig.M1.2}.  Many of these
stars have similar colors, as identified in a reduced proper motion
diagram derived using results from our SuperCOSMOS-RECONS (SCR) survey
\citep{Subasavage2005a, Subasavage2005b}. Given the rich dataset for
this type, we first discuss the sample in terms of metallicity and
gravity effects, then discuss individual targets.

{\it Metallicity effects} Perhaps better than for any other spectral
type available, the M1.0VI stars show a beautiful trend in metallicity
in their spectra, as shown in the top of Figure~\ref{fig.M1.1}.  The
red ends of the spectra match the M1.0V standard spectrum, but the
blue ends of the spectra are very different because of metallicity
effects.  We assign their metallicities on a scale of m to m$------$,
where m is indistinguishable from the main sequence standard and 6
``-'' indicate the most severely metal poor subdwarf.

As metallicity drops, the TiO5 band gradually weakens, as predicted in
the model spectra of Figure~\ref{fig.4800-2800.plot} for stars cooler
than $\sim$4000K.  Our M1.0V standard (metallicity m), LHS 109
(metallicity scale m$----$), and LHS 518 (metallicity scale m$------$)
have trigonometric parallaxes, and as predicted from the models, their
positions on the HR diagram shift bluer with decreasing metallicity.
At metallicity m$-$, SCR 2101$-$5437 is slightly metal poor compared
to M1.0V (TiO5), and has slightly higher gravity than M1.0V (CaH).
Unfortunately, none of the SCR objects in Figure~\ref{fig.M1.1} have
trigonometric parallaxes, yet.  Thus, they cannot be plotted on the HR
diagram, and the progressive trend in metallicity effects for these
stars cannot yet be shown.

{\it Gravity effects} The effects of gravity can be seen at three
different metallicities for type M1.0VI.  Spectral for metallicity
scale m$------$ subdwarfs are shown in the middle of
Figure~\ref{fig.M1.1}.  SCR 0701$-$0655 has higher gravity than LHS
518.  Both spectra are identical, except at CaH.  Spectra for
metallicity scale m$----$ subdwarfs are shown in the bottom of
Figure~\ref{fig.M1.1}.  A clear trend can be seen.  Three of these
five objects have trigonometric parallaxes --- LHS 109, LHS 385 and
LHS 335 --- and their gravity differences shift their locations on the
HR diagram toward redder and less luminous territory, as with
previously discussed types.  (LHS 385's high parallax error is the
likely cause of the slight offset between its HR diagram position and
LHS 109's.)  Finally, as shown at the top of Figure~\ref{fig.M1.2} for
metallicity scale m$---$, SCR 0654$-$7358 has higher gravity than SCR
1756$-$5927.

{\bf LHS 440} was reported by \cite{Bidelman1985} to be type M1.0V.
The second set of spectra in Figure~\ref{fig.M1.2} shows that LHS 440
has deeper absorptions in all three CaH bands than the M1.0V standard,
but is virtually identical at all other wavelengths.  This indicates
that LHS 440 probably has higher gravity than M1.0V, making it less
luminous than main sequence stars on the HR diagram.  Because it is
just barely one magnitude below the main sequence fit, we currently
assign it a type of M1.0VI:.

{\bf LHS 1970 and LHS 2734B} both have noisy spectra because they are
faint.  As shown in Figure~\ref{fig.M1.2}, their strong CaH bands
indicate that they are high gravity (strong CaH), low metallicity
(weak TiO5) subdwarfs.  LHS 1970's location on the HR diagram confirms
that it is a subdwarf and \cite{Gizis1997} reported its spectral type
is esdM2.5. LHS 2734B's spectral similarity to LHS 1970 indicates that
it also has low metallicity and high gravity.  Unfortunately, we
measure zero parallax for the LHS 2734AB pair, within the errors, so
we can not plot them on the HR diagram (see
section~\ref{sub.sec.K7.0VI} for discussion of LHS 2734A).  We assign
both LHS 1970 and LHS 2734B types of M1.0VI: because of their noisy
spectra.

{\bf LHS 158} has a spectrum nearly identical to an M1.0V star, but
falls one magnitude below the main sequence line on the HR diagram.
It also has $V_{tan}=$ 191 km sec$^{-1}$ \citep{Jao2005}.  The low
luminosity and tangential velocity together imply that LHS 158 may be
a subdwarf, so we assign it a type of M1.0[VI], representing its
uncertain assignment as a subdwarf, as is the case for the mid K-type
subdwarfs.

\subsubsection{M2.0VI}

\begin{figure}
\centering
\includegraphics[scale=0.7]{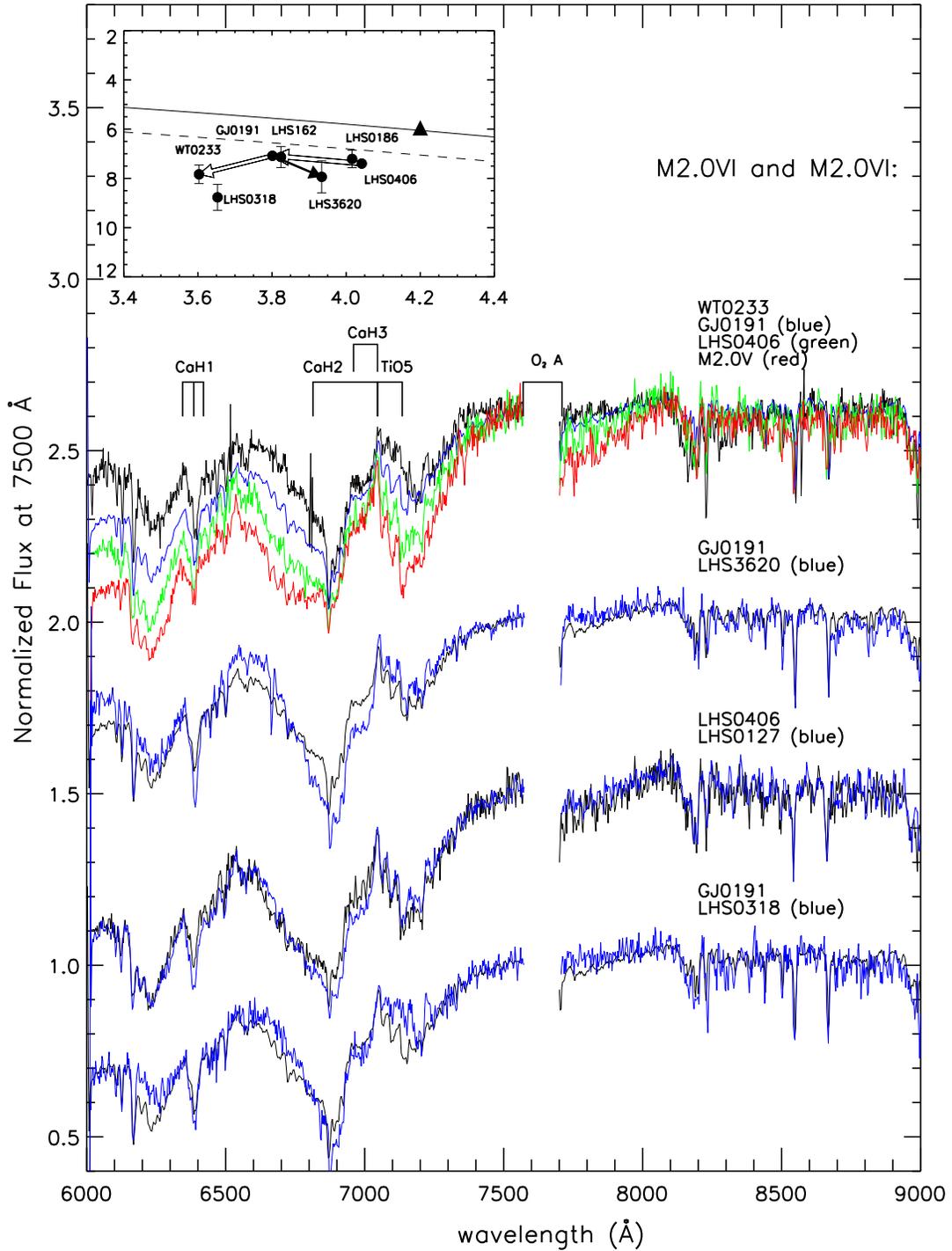}

\caption{Spectra for WT 233, GJ 191, LHS 406, LHS 3620, LHS 127 (all
M2.0VI) and LHS 318 (M2.0VI:) are shown with our M2.0V standard
spectrum (represented by a triangle in the inset plot).  The telluric
O$_{2}$ A band has been removed.  Symbols and lines have the same
meanings as in Figure~\ref{fig.M0.5.1}.}

\label{fig.M2}
\end{figure}

This is the earliest type for which we see that CaH1 absorption
deepens as the metallicity decreases, as shown by the curves in
Figure~\ref{fig.model.index.plot} for $T_{eff}$ less than 3500K.
Figure~\ref{fig.M2} shows the spectra for stars discussed in this
section.

{\bf LHS 406, GJ 191 (Kapteyn's Star) and WT0233} form a sequence of
decreasing metallicity.  These three stars have parallaxes, and their
positions on the HR diagram show a clear trend toward the blue for
stars with decreasing metallicity (shown with open arrows).  One of
the prototypes of the subdwarf class, Kapteyn's Star, which has
[Fe/H]$=$$-$0.99 from \cite{Woolf2005}, has previously been reported
to be a type sdM1.0 in \cite{Gizis1997}.  We assign a somewhat later
type for this famous star, which at 3.9 pc is the nearest known
subdwarf of any type.

{\bf GJ 191/LHS 3620 and LHS 406/LHS 127} The relative effects of
gravity can be seen in the spectra of these two pairs.  LHS 3620 has
much stronger gravity than GJ 191, while LHS 127 has slightly stronger
gravity than LHS 406.  On the HR diagram, LHS 3620 is redder and less
luminous than GJ 191, as expected.  LHS 162 has a spectrum (not
plotted) virtually identical to LHS 3620, but is only slightly redder
and less luminous than GJ 191.

{\bf LHS 318} is almost identical to GJ 191 except for weaker TiO5
absorption.  This weaker TiO5 indicates LHS 318 is slightly more metal
poor than GJ 191, so the blue end of the spectrum should be
``brightened''; however, it is not.  LHS 318's location on the HR
diagram in relation to GJ 191 suggests a higher gravity that is not
apparent in its spectrum.  We would expect LHS 318 to appear brighter
and bluer than GJ 191 if metallicity is the only difference in these
two stars, so we assign it as M2.0VI:.

\subsubsection{M3.0VI}

\begin{figure}
\centering
\includegraphics[scale=0.7]{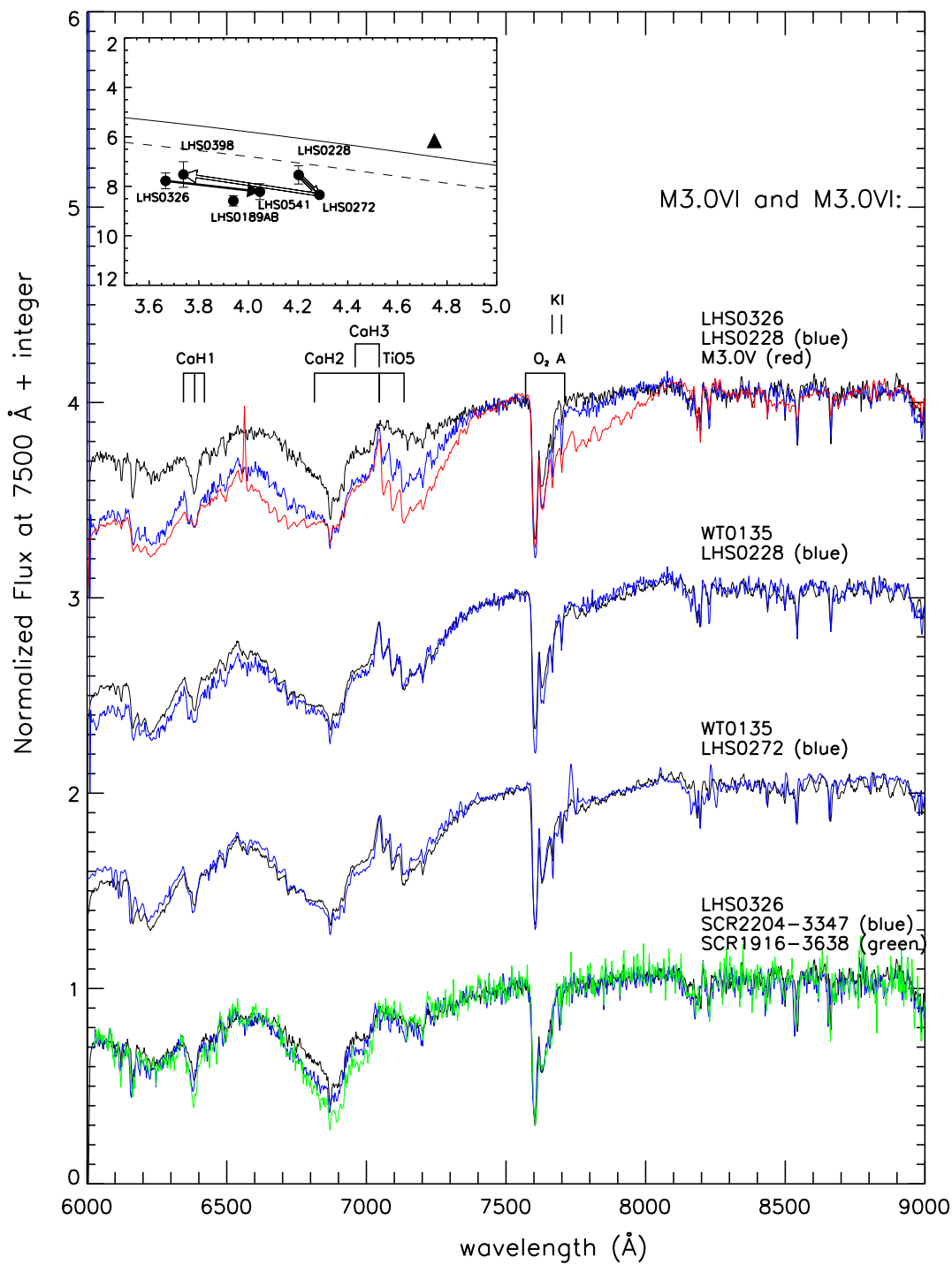}

\caption{Spectra for six M3.0VI are shown, with our M3.0V standard
spectrum in the top set of spectra (represented by a triangle in the
inset plot).  LHS 326 and LHS 228 show metallicity changes, while LHS
326, SCR2204$-$3347 and SCR1916$-$3638 show effects of gravity.  WT
135 and LHS 272 are M3.0VI:.  Note that the K I lines at 7665\AA~and
7699\AA~have appeared and blended in the O$_{2}$ A band.  Symbols and
lines have the same meanings as in Figure~\ref{fig.M0.5.1}.}

\label{fig.M3}
\end{figure}

{\bf LHS 228/LHS 189AB and LHS 326/LHS 398} are pairs of stars with
nearly identical spectra.  As shown in Figure~\ref{fig.M3}, LHS 326,
LHS 228 and our M3.0V standard form a metallicity sequence.  As in the
case of M2.0VI, CaH1 absorption deepens for M3.0VI stars with
decreasing metallicity.  In the case of LHS189AB, \cite{Costa2006}
reported their separation is about 3\arcsec~ and their
$\Delta$$R\sim$0.5mag. At a distance of 22.1 pc, the projected
separation implies a distance of $\sim$66 AU between the two
components. This spectrum has combined both components, so we assign
it type M3.0VI: until completely ``clean'' data can be acquired for
the components, and their individual metallicity scales can be
determined.

{\bf WT 135} was previously identified by \cite{Henry2002} as type
M2.5V.  We assign its metallicity m$-$: because its TiO5 band is the
same as seen in LHS 228, but the blue end of its spectrum is
``brightened'' (making it somewhat metal poor compared to LHS 228).
It lacks a trigonometric parallax, so we cannot plot it on the HR
diagram.

{\bf LHS 272} with $V-I=$ 4.29 is similar in color to WT 135 ($V-I=$
4.32) but appears to have slightly higher gravity (deeper CaH bands).
Overall, the blue end of its spectrum (6000\AA--6300\AA) is comparable
to WT 135, although with portions slightly brighter and portions
slightly fainter.  We therefore assign its metallicity to be the same
as WT 135, at m$-$:.  It is slightly redder and less luminous than LHS
228.  We use both hollow (metallicity) and solid (gravity) arrows to
indicate its location relative to LHS 228.  The competition between
metallicity and gravity effects in this case seems to indicate that
gravity is the dominant factor because LHS 272's position moves to the
red, rather than the blue.  LHS 272 was reported as sdM3.0 in
\cite{Gizis1997}.

{\bf LHS 326, SCR 2204$-$3347/LHS 541 and SCR 1916$-$3638} form a
clear sequence of gravity effects.  Other than having a noisier
spectrum (not shown), LHS 541 appears to be identical to SCR
2204$-$3347.  LHS 541's location on the HR diagram relative to LHS 326
reflects the effects of gravity.

\subsubsection{M3.5VI}

\begin{figure}
\centering
\includegraphics[scale=0.7]{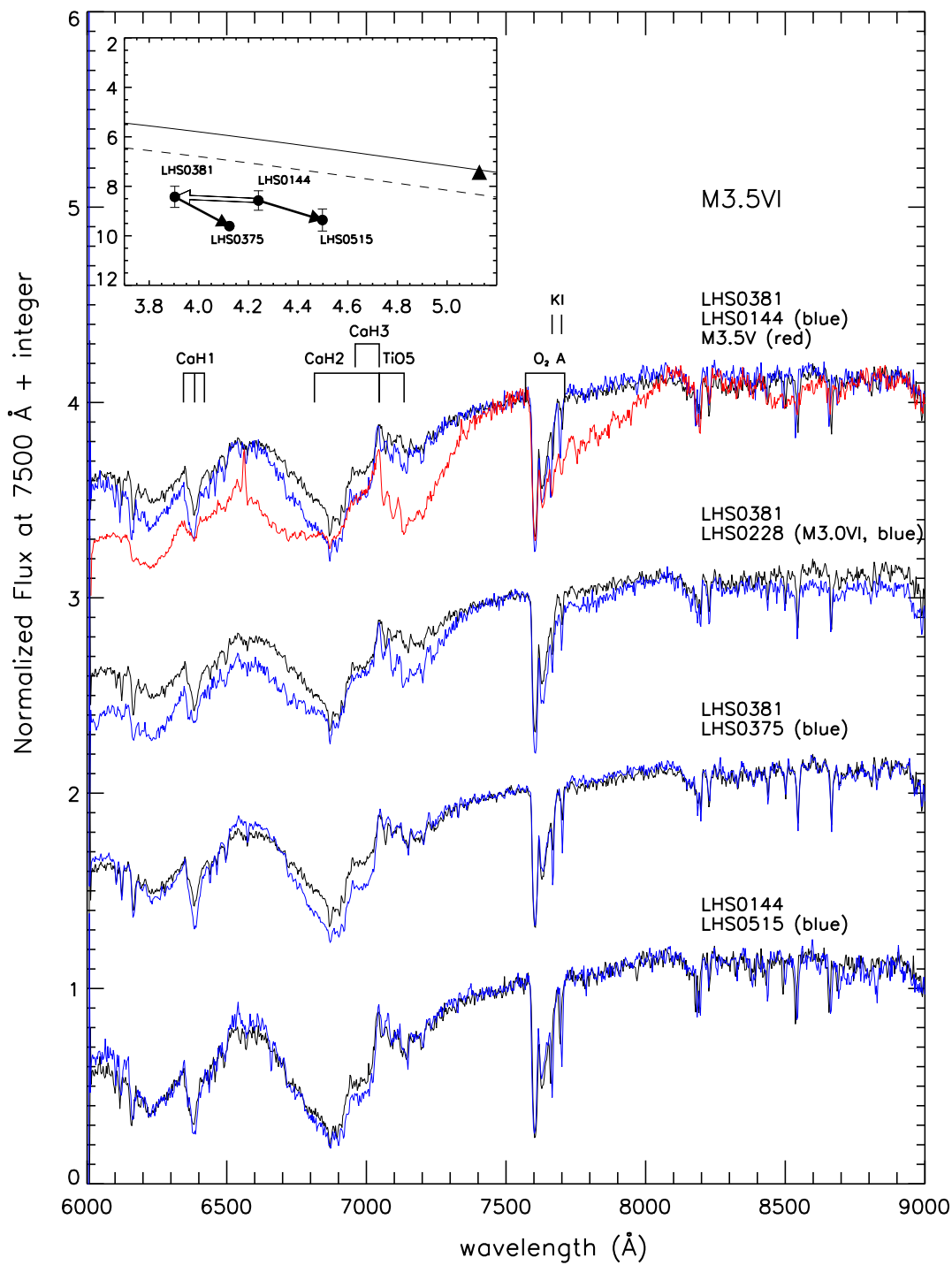}

\caption{Spectra for LHS 381 and LHS 144 (both M3.5VI) are shown with
our M3.5V standard spectrum (represented by a triangle in the inset
plot). These three stars show a dramatic metallicity sequence for
M3.5.  The spectrum of LHS 228 (M3.0VI) is compared to that of LHS 381
to illustrate the slope difference between M3.0VI and M3.5VI
(8200\AA--9000\AA).  LHS 375 and LHS 515 have higher gravities than
LHS 381 and LHS 144, respectively.  Note that K I lines (7665\AA~and
7699\AA) have appeared and blended in the O$_{2}$ A band.  Symbols and
lines have the same meanings as in Figure~\ref{fig.M0.5.1}.}

\label{fig.M3.5}
\end{figure}

This half type is assigned because four stars have spectra shown in
Figure~\ref{fig.M3.5} with redder slopes than M3.0VI (compare to LHS
228, a M3.0VI subdwarf in the second set of spectra), but not as steep
as M4.0VI (see next section).  Conveniently, all four stars of this
type have trigonometric parallaxes.

{\bf LHS 381 and LHS 144} have lower metallicities than observed for
our M3.5V standard.  Their spectra redward of 8200\AA~match M3.5V
relatively well and show the same trends predicted by the GAIA models
(Figure~\ref{fig.4800-2800.plot}) for stars with temperatures of about
3200K.  Their CaH1 indices also match the trend discussed in
section~\ref{sec.K.subdwarf}.

{\bf LHS 375 and LHS 515} have higher gravities than LHS 381 and LHS
144, respectively.  Both objects are redder and less luminous than
their low gravity counterparts on the HR diagram.  \cite{Gizis1997}
reported LHS 375 to be an esdM4.0 subdwarf and \cite{Reid2005}
reported LHS 515 to be an esdM5.0 subdwarf.  We classify both as
M3.5VI with high gravities.

\subsubsection{M4.0VI to M6.0VI}

We identify only four subdwarfs with types M4.0VI to M6.0VI, only one
of which, LHS 2067A, currently has a trigonometric parallax.  Because
of the paucity of such objects, we do not yet have a sample
sufficiently large to map out the effects of metallicity and gravity.

\begin{figure}
\centering
\includegraphics[scale=0.7]{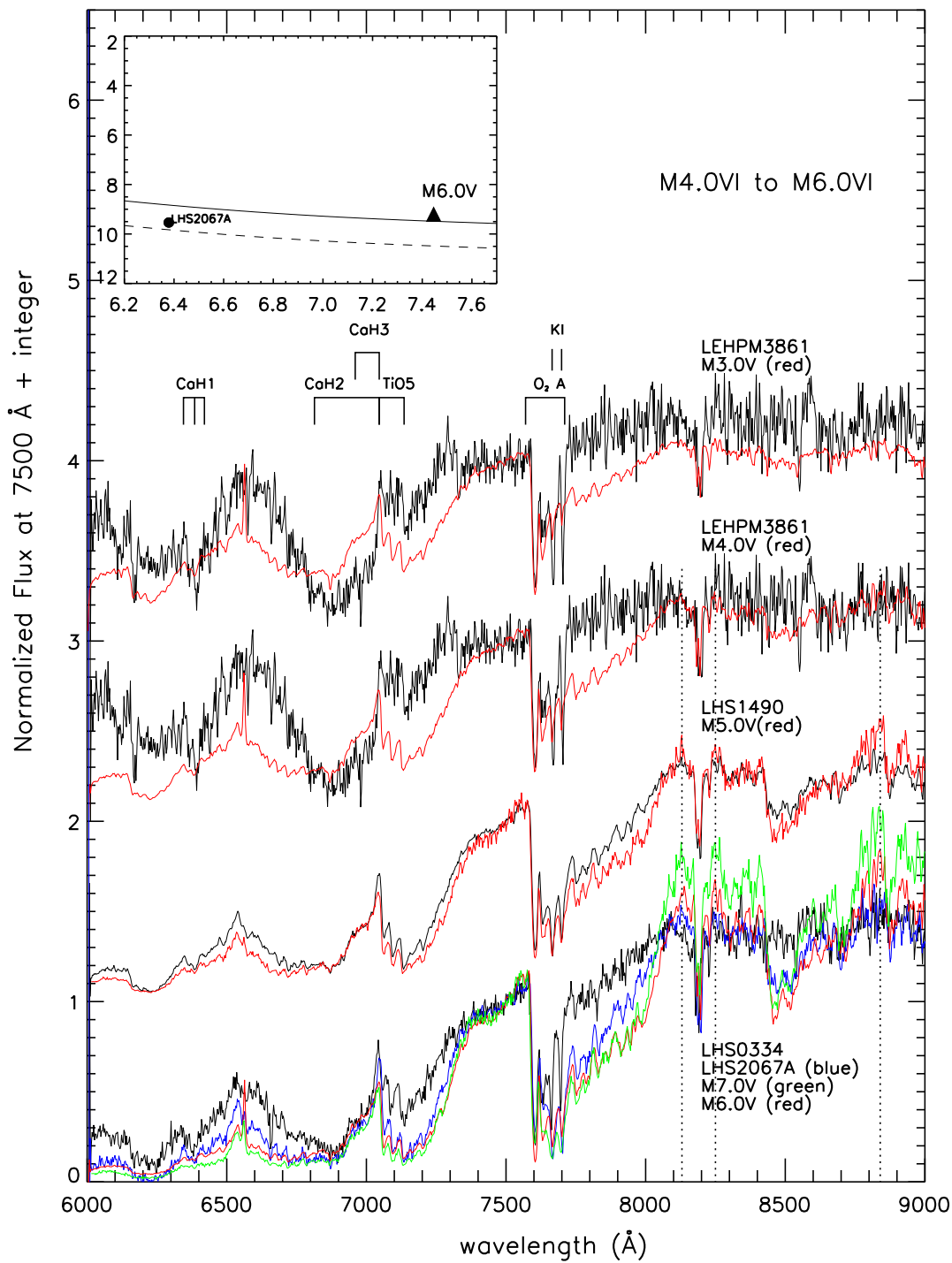}

\caption{LEHPM 3861 (M4.0VI), LHS 1490 (M5.0VI), LHS 334 (M6.0VI), and
LHS2067A (M6.0VI) subdwarf spectra are shown compared to our standard
M3.0-7.0V standards.  Only LHS2067A and the M6.0V standard (filled
triangle) are shown in the HR diagram.  Vertical dotted lines
represent the three peaks we use to assist us to assign spectral
types.  The K I lines at 7665\AA~and 7699\AA~are clearly seen blended
in the O$_{2}$ A band.  Symbols and lines have the same meanings as in
Figure~\ref{fig.M0.5.1}.}

\label{fig.M4.6}
\end{figure}

When stars are cooler than $\sim$3200K, their spectra redward of
7500\AA~change radically between [m/H]=0.0 and $-$1.0 (see
Figure~\ref{fig.4800-2800.plot}), but are rather more stable between
[m/H] = $-$1.0 and $-$2.0.  For example, Figure~\ref{fig.M4.6} shows
comparisons between LEHPM 3861 (M4.0VI) and our M3.0V and M4.0V
standards.  Clearly, if its spectral type is incorrectly assigned to
be M3.0VI, there is excess red flux.  \cite{Lodieu2005} reported LEHPM
3861 to be a sdM6.0 subdwarf.

{\bf LEHPM 3861} (M4.0VI), {\bf LHS 1490} (M5.0VI), {\bf LHS 334}
(M6.0VI), and {\bf LHS 2067A} (M6.0VI), whose spectra are shown in
Figure~\ref{fig.M4.6} are assigned subdwarf spectral types later than
M3.5VI.  The types are based on fluxes redward of 7500\AA, with
emphasis on three pseudo-continuum peak points at 8130, 8250 and
8840\AA~(marked in Figure~\ref{fig.M4.6} with dotted lines).  As shown
in Figure~\ref{fig.4800-2800.plot} for stars cooler than 3200K, the
flux decreases at these three points as the metallicity drops from
[m/H]=0.0 to $-$1.0 and this trend becomes even more prominent as
$T_{eff}$ decreases.  Thus, these three peaks for a low metallicity
subdwarf will not be brighter than a dwarf with a comparable spectral
type.

{\bf LHS 2067A} is bluer than M6.0V on the HR diagram and is slightly
below the main sequence line.  LHS 2067A was previously identified to
be a subdwarf in \cite{Kirkpatrick1995}, but no spectral type was
given.  Although its CaH1 and CaH2+CaH3 indices are not in the
subdwarf region (see Figure~\ref{fig.cahn.plot} and
Table~\ref{tbl.cahn.tio5.index}), its spectrum is clearly different
from M6.0V and M7.0V, with stronger CaH1 absorption.  It forms a wide
($\sim$55\arcsec~NE) common proper motion pair with a white dwarf, LHS
2067B, so it, like LHS 193AB and LHS 300AB, forms an intriguing pair
that can be used for comparing metallicity and white dwarf ages.  At
the system's distance of 25.5 pc, the large separation of the
pair corresponds to 472 AU, indicating that significant pollution of
the subdwarf by the white dwarf during its planetary nebula phase
seems unlikely.

{\bf LHS 334} was reported to be a sdM4.5 subdwarf in \cite{Reid2005}.
We assign it a type of M6.0VI because of the good match of the three
pseudo-continuum peak points to LHS 2067A, which appears to be
slightly more metal rich than LHS 334.

\section{Application to SDSS Subdwarfs}
\label{sec:SDSS}

\begin{figure}[h]
\hspace{-2cm}
\includegraphics[scale=0.8, angle=90]{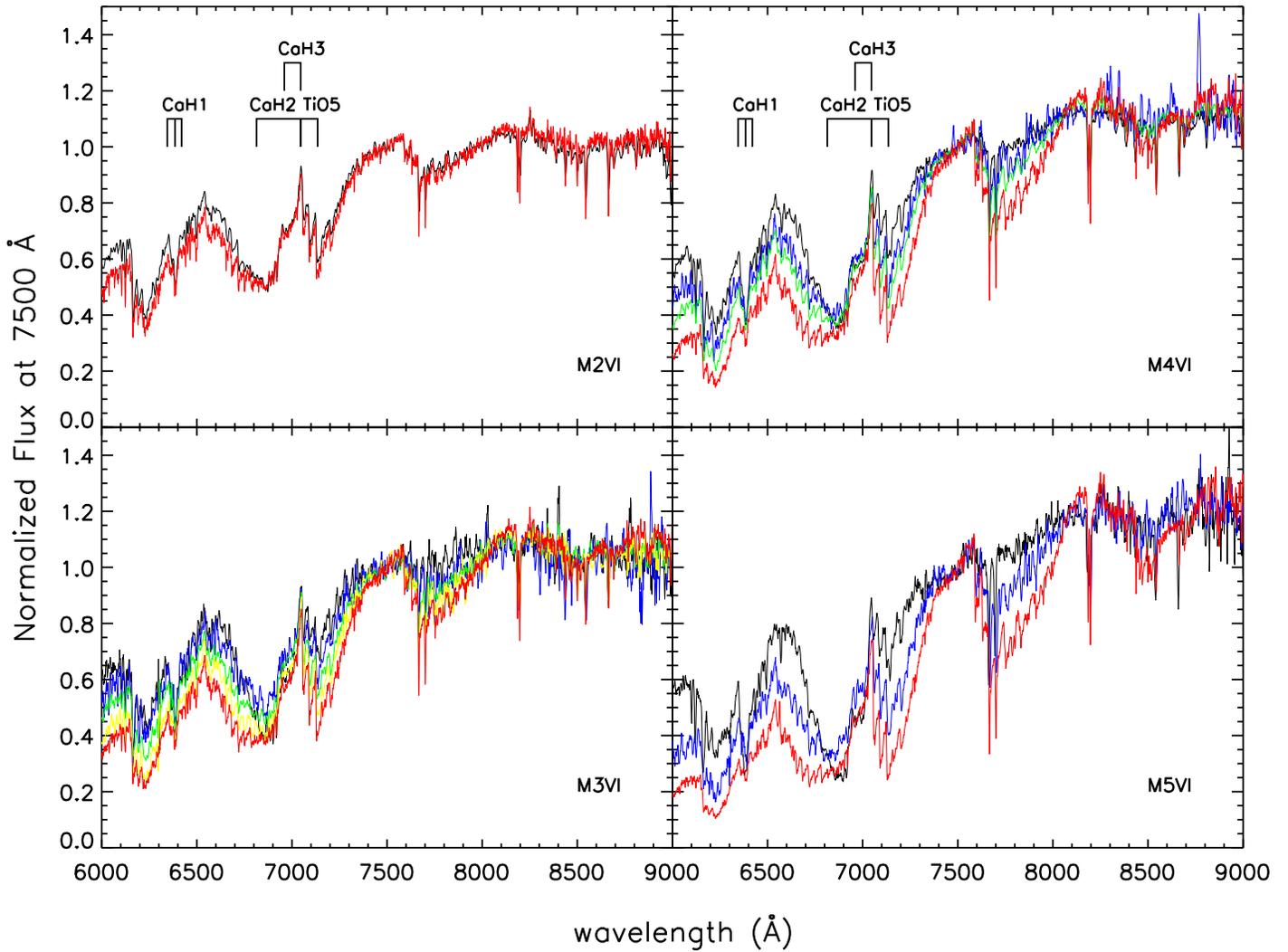}

\caption{Spectra are shown for subdwarfs from \cite{West2004}
separated into types assigned by us by matching standard spectra (red)
from \cite{Bochanski2007} in the region 8200\AA--9000\AA.  Other
colors (black, blue, green, and yellow) represent different
metallicities at each type.  Metallicity increases from black to red.}

\label{fig.sdss.subdwarf.1}
\end{figure}

\begin{figure}[h]
\centering
\includegraphics[scale=0.7]{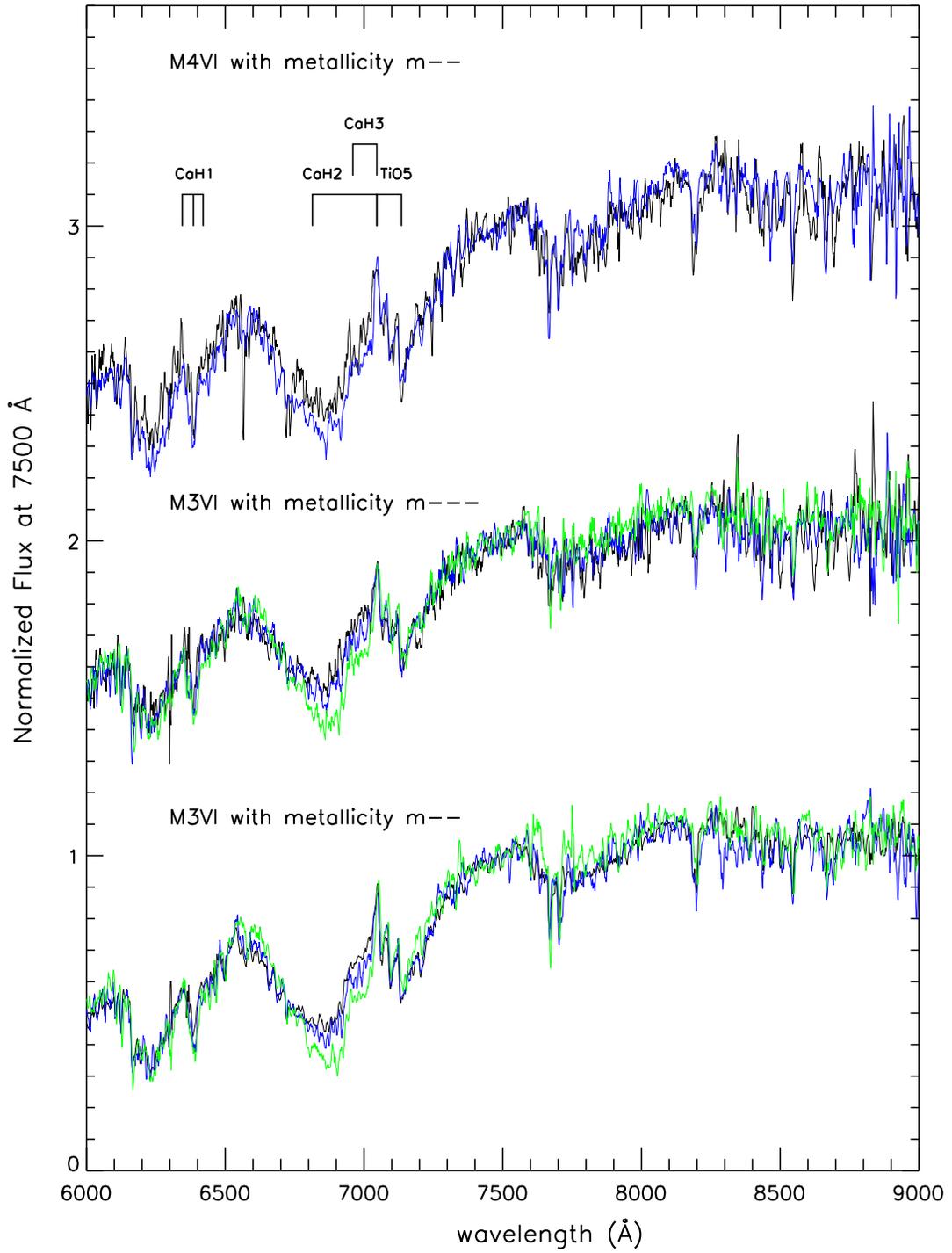}

\caption{These plots indicate the gravity effects for SDSS subdwarfs
of type M3VI and M4VI.  The black, blue and green lines represent
increasing gravities: g, g$+$ and g$++$.}

\label{fig.sdss.subdwarf.2}
\end{figure}

It is useful to apply our spectral typing methodology to the recent
work of \cite{West2004}, who have provided a significant sample of
sixty new subdwarf spectra, all acquired and reduced in a homogeneous
way. The subdwarfs were selected and spectral types were assigned via
their CaH and TiO indices.  For comparison, we retrieved
fifty\footnote{The remaining 10 subdwarfs could not be retrieved using
the SDSS DR4 website, even though we used a 1\arcmin~search radius and
coordinates from their table.} of the publicly available spectra and
use our new method to assign types.  The brightest star among these
subdwarfs has $r$=17.0, resulting in somewhat noisy spectra for the
sample, so we smoothed the spectra by averaging the flux over five
pixels and normalizing at 7500\AA.  Representative spectra are shown
in Figures~\ref{fig.sdss.subdwarf.1} and~\ref{fig.sdss.subdwarf.2}.

The SDSS spectra have telluric lines removed, so we use M dwarf
standard spectra from \cite{Bochanski2007} that also omit the telluric
lines, rather than our own standards.  In addition,
\cite{Bochanski2007} do not present half-type standard spectra, so we
do not present any half-type spectra for the SDSS subdwarfs.  Previous
and current types are listed in Table~\ref{tbl.sdss.subdwarf}.  The
subdwarfs have types between M1.0 and M3.0 in \cite{Bochanski2007},
while our efforts yield types between M2.0 and M5.0.

Our point here is not to assign definitive spectral types, but to
illustrate the gravity and metallicity effects in the SDSS dataset,
and provide examples of the application of our proposed spectral
typing method.  Stars of a given spectral type, of course, may have a
range of metallicities, as discussed at length in previous sections,
and as is evident in Figure~\ref{fig.sdss.subdwarf.1}.  Gravity
effects are also seen, as shown in Figure~\ref{fig.sdss.subdwarf.2}.
Three stars (SDSSJ085843.89$+$511210.1, SDSSJ093141.85$+$453914.5 and
\\SDSSJ145447.32$+$011006.8) previously identified as subdwarfs have
spectra identical to SDSS M dwarfs.  Two other stars,
SDSSJ083217.77$+$522408.2 and SDSSJ113501.76$+$033720.3, appear to
have slightly higher gravities (stronger CaH1 line) than dwarfs, but
the rest of these two spectra are almost identical to SDSS M dwarfs.
Therefore, to these two stars we assign types of M2.0VI: and M3.0VI:
with g+:, respectively.

Unfortunately, the SDSS dwarfs do not have parallaxes that can be used
to confirm their locations on the HR diagram.  In addition, we find
many SDSS subdwarfs with identical spectral types that have $g-z$
colors as different as 0.53.  It is therefore difficult to verify how
metallicities and gravities affect the SDSS absolute magnitudes and
colors.  Nonetheless, the robust sample of \cite{West2004} provides
many additional subdwarfs that can be targeted for further work.

\section{Discussion}
\label{sec:discussion}

\subsection{Why Subdwarfs' Physical Parameters Are Not Listed Explicitly}

Stellar spectra follow trends primarily defined by temperature, yet as
we have seen, metallicities and gravities also have significant
effects on the spectra of cool subdwarfs.  Here we have provided a
consistent spectral sequence for subdwarfs that is based on linking
observed subdwarf spectra to spectra acquired for main sequence dwarfs
using the same telescope/instrument/observing protocols.  Here we
compare synthetic to observed spectra to evaluate how well we can
assign values for the temperatures, metallicities, and gravities of
cool dwarfs and subdwarfs.

\subsubsection{Matching Synthetic and Observed Spectra}

To test the reliability of the model grids and our fitting procedures,
we use our six standard spectral sequence dwarfs with types M0.0V to
M5.0V as test spectra (see Figure~\ref{fig.observed.model}).  All are
currently believed to be uncorrupted single red dwarfs because they
show no evidence of multiplicity from combinations of (1) HST/NICMOS
observations, (2) optical speckle observations, (3) optical CCD
imaging, and/or (4) three or more years of astrometric observations
that would reveal perturbations from unseen companions that
contributed significant light to the system.

We compared the spectral region from 6000\AA~to 9000\AA~after
normalizing both the GAIA grid spectra and ours at 7500\AA, and
applying a Gaussian function to the much higher resolution GAIA model
spectra to match our resolution.  If a star had a significant spectral
shift because of radial velocity, we manually offset the spectrum to
match features obvious in the synthetic spectra.  We then calculated
the reduced $\chi^{2}$ differences (hereafter, simply $\chi^{2}$)
between the model grid spectra and ours.  Because the synthetic
spectra do not have the telluric lines of O$_{2}\alpha$
(6270\AA--6330\AA), O$_{2}$B (6860\AA--6980\AA), O$_{2}$A
(7590\AA--7710\AA) and water (7150\AA--7330\AA~and 8952\AA--9000\AA),
these absorption regions were excluded in the $\chi^{2}$ calculations.

The selected model grids have effective temperatures of 2400K to
4500K, [m/H] from $-$2.0 to $+$0.5, and {\it log g} from 4.0 to 5.5 in
steps of 100K, 0.5 dex, and 0.5 dex, respectively.  The upper panel of
Figure~\ref{fig.3500.plot} shows that for a metallicity of $-$1.0 and
temperature 3500K, varying the gravity causes changes in only certain
wavelength regions (and some specific lines).  This means $\chi^{2}$
is only sensitive to {\it log g} in relatively small spectral regions,
but not for the overall spectrum.  On the other hand, as shown in the
lower panel of Figure~\ref{fig.3500.plot}, for the same temperature
star with {\it log g} fixed at 5.0, there are large differences
between spectra when the metallicity is varied.

Thus, for a star of a given temperature, the gravity changes the
overall spectral shape minimally while the metallicity changes it a
great deal, so we first secure a star's metallicity and then its
gravity.  Figure~\ref{fig.GJ701.chi2} is an example for M1.0V standard
star, showing $\chi^{2}$ curves at various metallicities and
gravities.  Each curve represents a specific {\it log g} and grid
spectra have temperatures incremented by 100K.  The smallest scatter
in the plots is found when [m/H] is $-$0.5, which is adopted as the
star's metallicity.  We then examine each point in the [m/H]$=$$-$0.5
panel to find the best fit with the smallest $\chi^{2}$ at this fixed
metallicity.  In this case, the best fit has $T_{eff}=$ 3600K, {\it
log g} $=$ 4.5 and $[m/H] =$ $-$0.5.

\subsubsection{Discrepancies Between the Best Fitting Synthetic and 
Observed Spectra}

\begin{figure}
\centering
\includegraphics[scale=0.7]{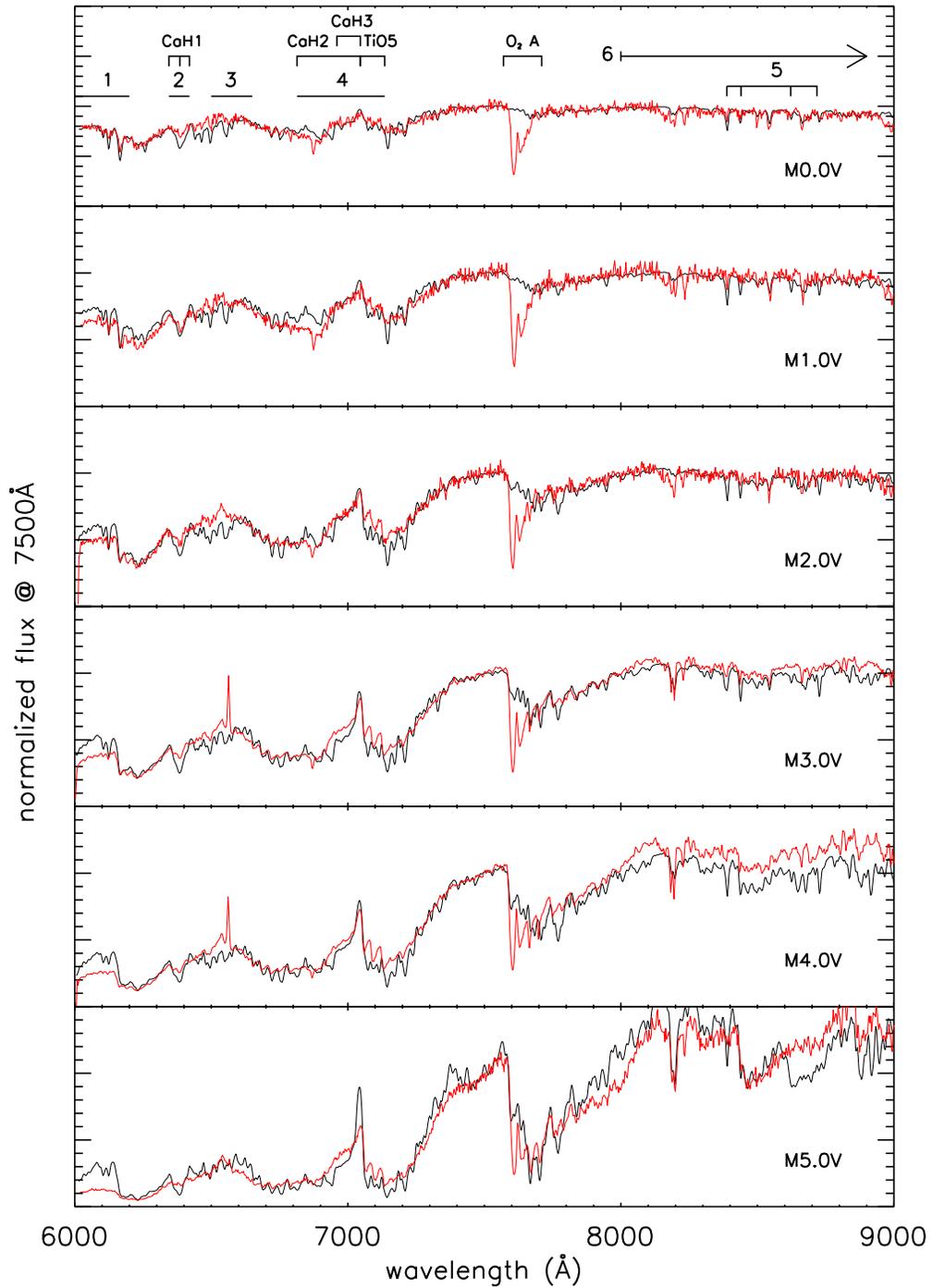}

\caption{Observed spectra of dwarf standard stars (red lines)
compared with the best fitting synthetic spectra (black lines).
Telluric lines are not present in the synthetic spectra.  Six regions
labeled from 1 to 6 are discussed in the text.}

\label{fig.observed.model}
\end{figure}

\begin{figure}
\centering
\includegraphics[scale=0.7, angle=90]{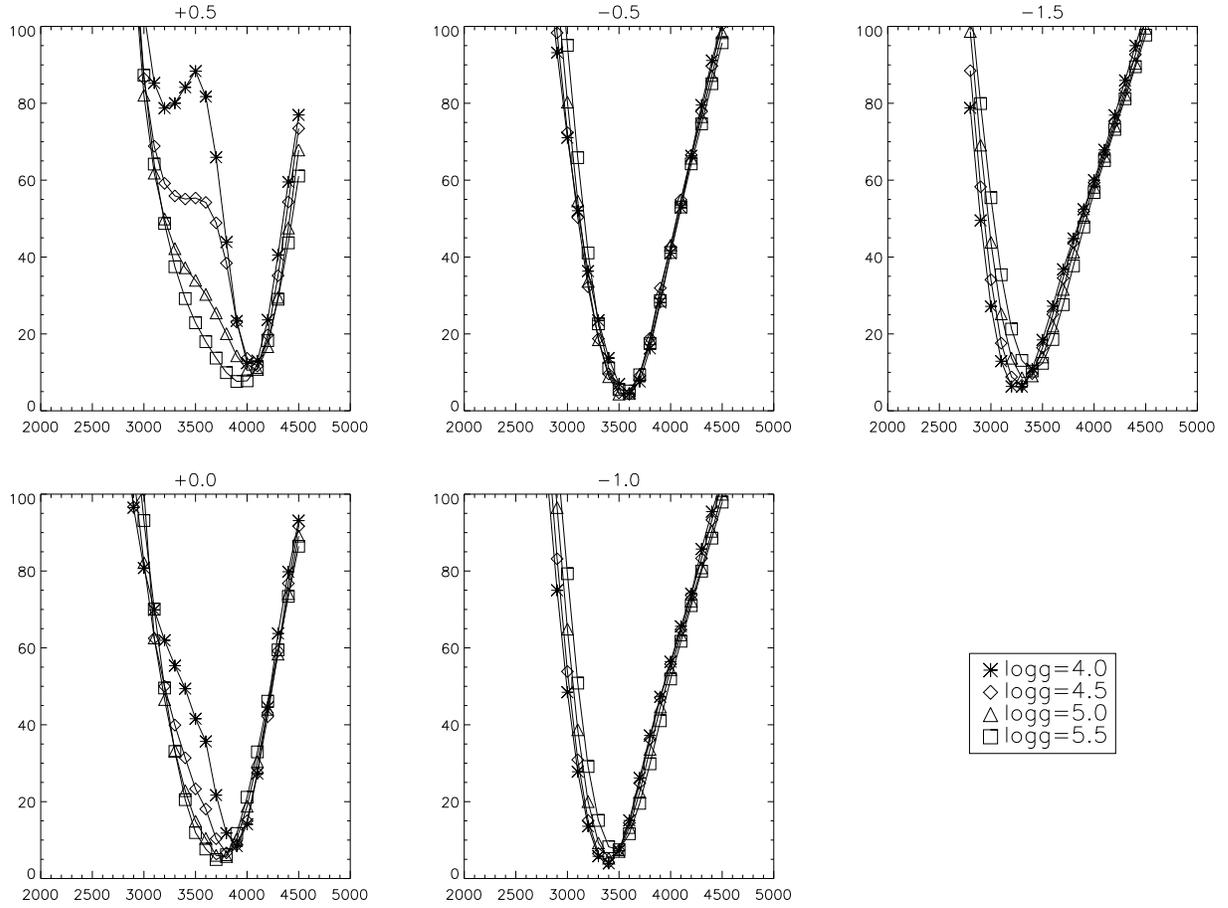}

\caption{Synthetic models were compared to the observed spectrum of
our M1.0V standard.  The resulting $\chi^{2}$ values are plotted
against temperature for models with five different metallicities, at
four different gravities each.  The [m/H]$=-$0.5 plot has the tightest
curves at different {\it log g}.}

\label{fig.GJ701.chi2}
\end{figure}

In this paper we do not explicitly list temperatures, metallicities,
or {\it log g} values for subdwarfs.  A few examples support our
reasoning for not listing these physical parameters.
Figure~\ref{fig.observed.model} shows observed spectra and the best
fit synthetic spectra for six main sequence spectral standard stars
with types M0.0V to M5.0V (in steps of 1.0 subtypes).  In general, the
overall slopes of the model spectra fit fairly well, especially for
the earlier types.  However, we discuss here six regions labeled at
the top of Figure~\ref{fig.observed.model} that do not match, which
are particularly relevant for the subdwarfs that are the focus of this
paper.

\begin{enumerate}

\item Region 1: The observed spectrum is always less luminous than the
model, except for M0.0V.

\item Region 2: This is the CaH1 absorption region.  Observed spectra
have shallower CaH1 features than the models.  The depth of CaH1 is
determined by a combination of gravity and metallicity.  Decreasing
CaH1 absorption can be caused by either decreasing gravity or
decreasing metallicity.  However, fine tuning the gravity or
metallicity affects not only the CaH1 feature, but the TiO5 feature in
region 4.

\item Region 3: The pseudo-continuum always peaks near 6530\AA~in the
observed spectra.  However, the synthetic spectra have this peak
``red-shifted'' to $\sim$6650\AA.

\item Region 4: Containing the CaH2, CaH3 and TiO5 features, this is
the most important region for examining the interplay of metallicity
and gravity in cool dwarfs.  Unfortunately, the CaH2 feature is
blended with O$_{2}$B so it is not entirely reliable for analysis of
spectra taken through the Earth's atmosphere.  The CaH3 and TiO5
features are usually weaker in observed spectra than in the models.
The strength of TiO5 is primarily driven by metallicity, not gravity
(see the bottom panel of Figure~\ref{fig.3500.plot}).  Therefore, this
region reveals valuable information about a star's metallicity, in
particular at types later than M2.0.

\item Region 5: Several Fe I absorption lines (8388\AA, 8440\AA~and
8718\AA) and a Mg I line (8718\AA) in the model spectra redward of
8300\AA~are deeper than observed.  This indicates poor metallicity
matches and/or poor modeling of those particular lines.

\item Region 6: Overall, redward of 8000\AA~the M4.0V and M5.0V
matches are not as good as other regions.  This likely indicates
fundamental problems with the strengths of some opacity sources
(e.g.~H$_{2}$O and TiO) in the models.

\end{enumerate}

\begin{figure}
\centering
\includegraphics[scale=0.5, angle=90]{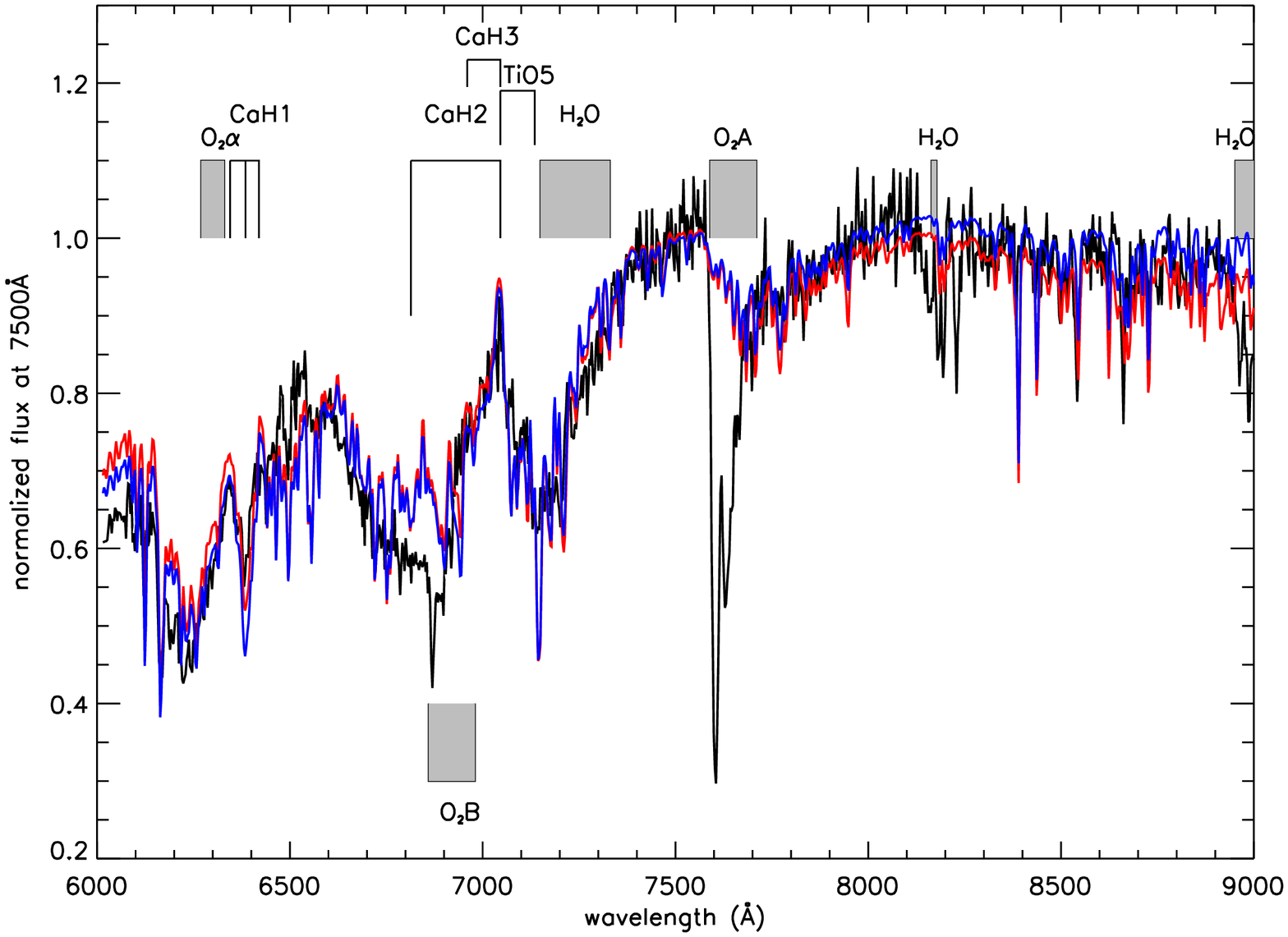}
\includegraphics[scale=0.5, angle=90]{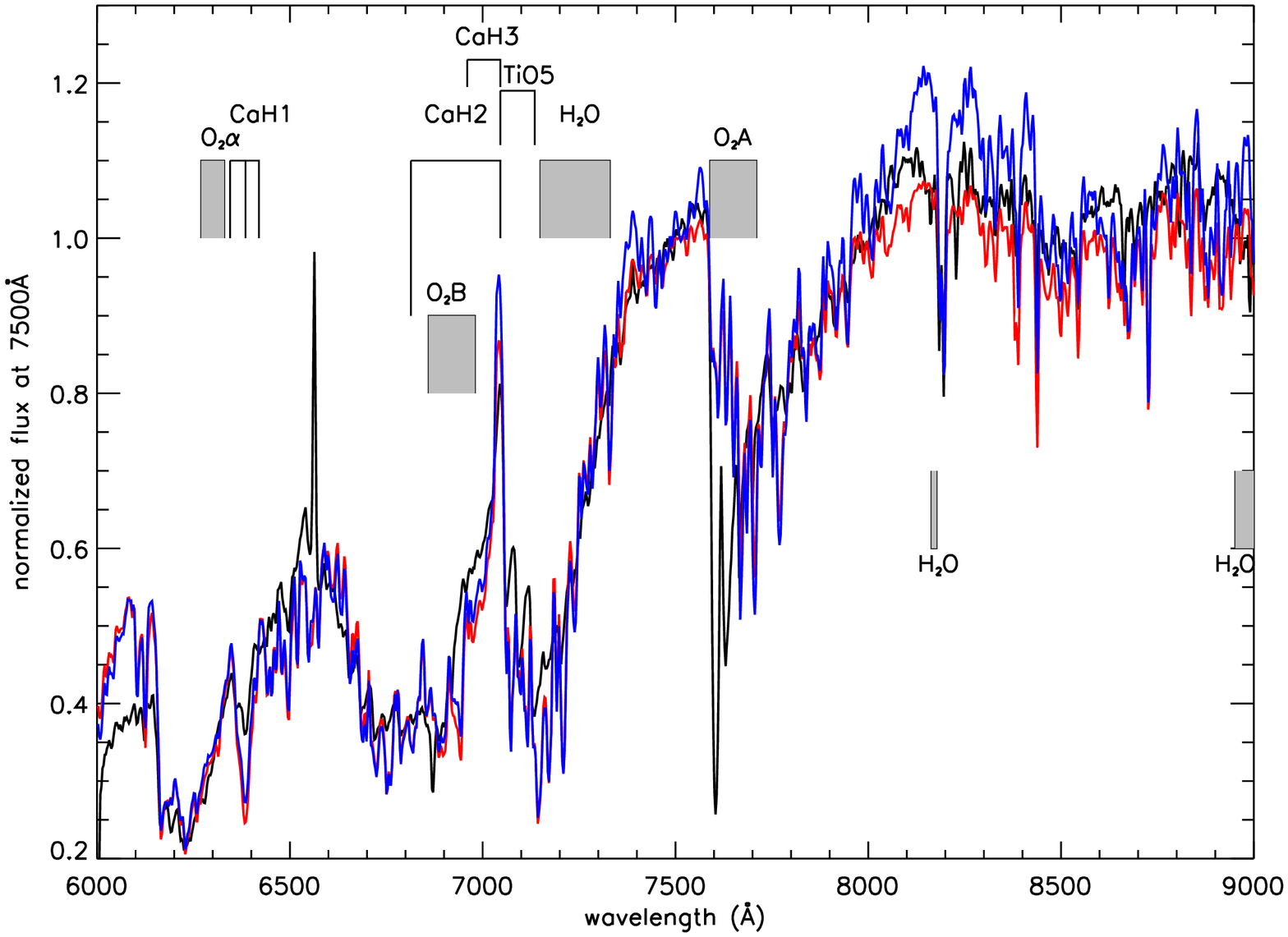}

\caption{Comparisons of observed spectra for M1.0V (top) and M3.0V
(bottom) are shown against the two best fitting synthetic spectra
for each.  The red line is the best fit, where $T_{eff}$/[m/H]/{\it
log g} $=$ 3600/$-$0.5/4.5 for M1.0, and 3200/$-$0.5/4.5 for M3.0.
The blue line is the second best fit, where the values are
3400/$-$1.0/4.0 for M1.0 and 3300/0.5/5.5 for M3.0.  Telluric bands
are marked as gray boxes.}

\label{fig.GJ701.LP776}
\end{figure}

The top panel of Figure~\ref{fig.GJ701.LP776} shows the two best
fitting synthetic spectra for our M1.0V standard.  The red spectrum
(3600K, [m/H]$ = -$0.5, {\it log g}$ = $4.5) provides the best fit and
the blue spectrum (3400K, [m/H]$ = -$1.0, {\it log g}$ = $4.0) is
second best.  There are only slight differences between the two
synthetic spectra --- the CaH1 strengths and continuum fluxes redward
of 8000\AA~are the only notable differences.  However, the effective
temperatures for the two models differ by 200K, while the
metallicities and {\it log g} each differ by 0.5 dex.  The lower panel
of Figure~\ref{fig.GJ701.LP776} shows the two best fit spectra for our
M3.0V standard.  The best fit (red line) yields 3200K, [m/H]$=-$0.5,
{\it log g}$=$4.5 and the second best fit (blue line) yields 3300K,
[m/H]$=$0.5, {\it log g}$=$5.5.  However, neither model fit is an
ideal match to the observed spectrum (which is why the $[m/H]$ value
for the two best matches differs by an order of magnitude), and
matches become even poorer for cooler stars.

As a whole, many of the best fits for the main sequence dwarfs in
Figure~\ref{fig.observed.model} are for metallicities of $-$0.5, which
is somewhat lower than studies that have specifically attempted to
assign metallicities to cool dwarfs in the solar neighborhood.  For
example, the mean [Fe/H] from 21 M dwarf secondaries in
\cite{Bonfils2005} is $-$0.09, and the mean [Fe/H] from five M dwarf
secondaries in \cite{Bean2006} is $-$0.17.  Also worthy of note is
that the fits were made for metallicities incremented by 0.5 dex, and
stars with metallicities between $-$0.5 and 0.0 may have slightly
better fits for $-$0.5.  One might think that interpolation from
existing model grids would allow better fits.  There are, however,
degeneracy problems in matching model spectra to observations --- when
fitting the three-dimensional space of temperature, metallicity, and
gravity, two or more synthetic spectra yield low points with similar
$\chi^{2}$ , as is the case for the M1.0V example discussed above.
More worrisome is that the overall discrepancies discussed above
(items 1 to 6) cannot be removed through interpolation in the existing
model grids.

\begin{figure}
\centering
\includegraphics[scale=0.5, angle=90]{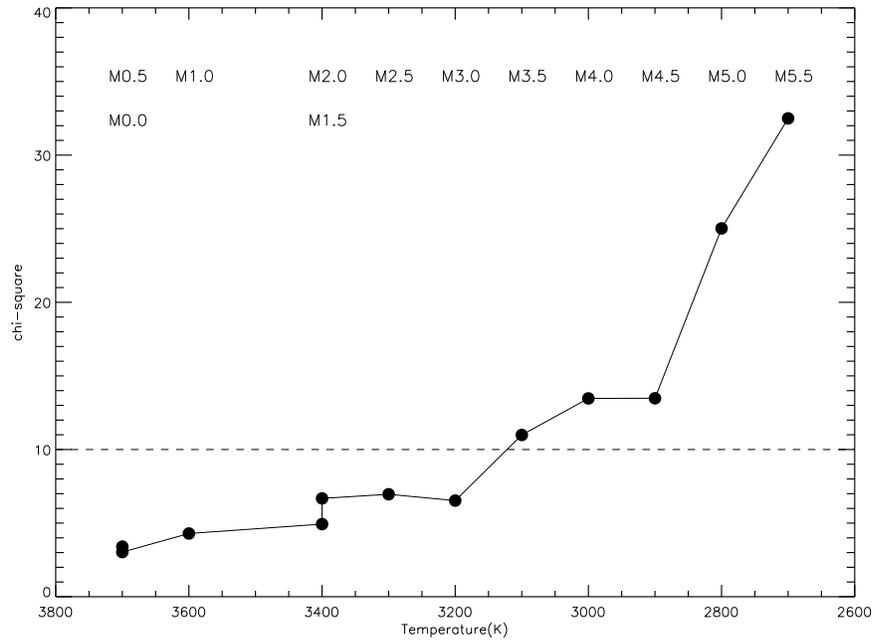}

\caption{The best $\chi^{2}$ values are shown for synthetic spectra
fits to observed spectra for stars of types M0.0V to M5.5V (labeled
relative to each type).  The dashed line indicates the selected limit
for reliability of fitting model spectra to observed spectra --- above
this line fits are deemed unreliable.}

\label{fig.chi2.curve}
\end{figure}

Figure~\ref{fig.chi2.curve} illustrates how the $\chi^{2}$ values
change for best fit matches of the GAIA models to our spectral
standards with types M0.0V through M5.5V.  Spectral type M3.0V is the
latest type for which a reasonably tight plot like the one shown in
Figure~\ref{fig.GJ701.chi2} can be identified.  The dotted line in
Figure~\ref{fig.chi2.curve} provides a dividing point between
reasonable matches and poor matches, indicating that once $\chi^{2}$
exceeds 10, the $\chi^{2}$ plots are too scattered to choose a unique
set of model spectra parameters to match observed spectra.

\begin{figure}
\centering
\includegraphics[scale=0.5, angle=90]{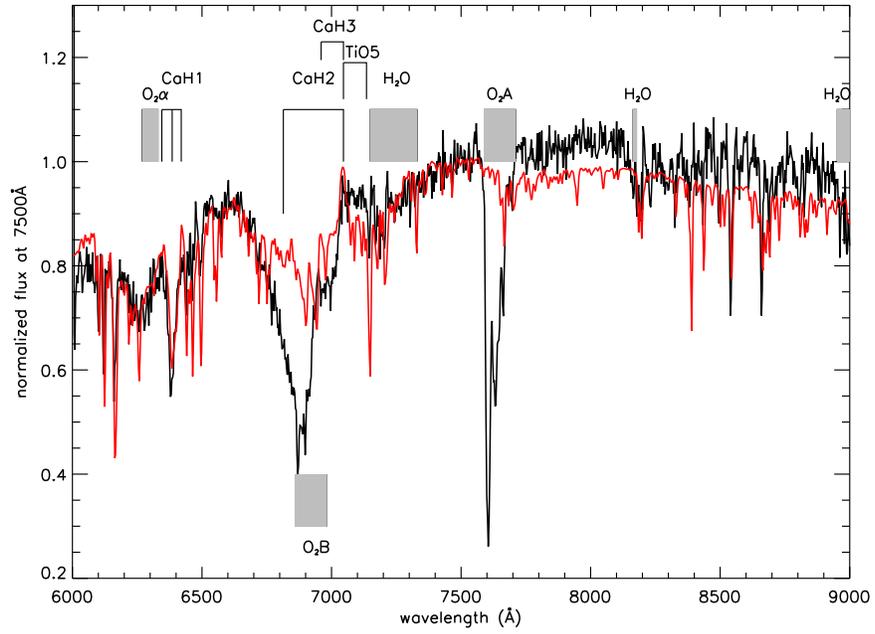}

\caption{The spectrum of the subdwarf LHS 335 (black) is shown with
the corresponding best fitting synthetic spectrum (red).}

\label{fig.lhs0335}
\end{figure}

A specific example of applying the models to one of our subdwarfs
illustrates the large discrepancies that must be overcome to derive
reliable parameter values for metallicity and gravity.  When we
applied our fitting algorithm to our observed spectrum for LHS 335
(M1.0VI, shown in Figure~\ref{fig.lhs0335}), we discovered that the
``best fit'' was quite poor.  Although the formal $\chi^{2}$ value was
5.9 (less than our cutoff of 10, indicating a reliable fit), the
model's CaH3 band is not deep enough, the TiO5 band is too deep, and
the continuum flux redward of 7700\AA~is less than observed.

For myriad reasons we conclude that we cannot strictly determine
reliable metallicities and {\it log g} values for cool dwarfs using
the fitting method discussed here.  Thus, until improved model grids
are available, we defer assignment of numerical values for
temperatures, metallicities and {\it log g} values.  However, we
certainly {\it can} use the GAIA model grids to mimic {\it trends}
(discussed in $\S$\ref{sec.subdwarf.facts} and~\ref{sec.trend}) in the
spectra of subdwarfs to compare the stars within a framework of
changing physical parameters.

\subsection{The Confusion Between {\it sd}, {\it esd} and {\it usd} Prefixes}

\cite{Gizis1997} proposed that subdwarfs have two subclasses,
``subdwarfs'' ({\it sd}) and ``extreme subdwarfs'' ({\it esd}) based
on their CaH and TiO5 band strengths.  He separated the two classes
using the line shown in the top panel of Figure~\ref{fig.cahn.plot}.
Another term, ``ultra subdwarf'', was proposed by \cite{Caldwell1984}
to describe GJ 59B ($V-K_{s}=$ 3.42, $M_{Ks}=$ 7.89), which is
$\sim$2.5 magnitudes underluminous compared to main sequence stars of
similar color.  This star falls in the region including ``extreme
subdwarfs'' in \cite{Gizis1997}.  More recently, \cite{Lepine2007}
adopted the ``ultra subdwarf'' ({\it usd}) term for subdwarfs found to
have stronger CaH features than ``extreme subdwarfs''.  We believe
these terms confuse the situation and do not address the underlying
astrophysics.  We recommend that they not be used for the following
reasons.

\begin{figure}[h]
\centering
\includegraphics[scale=0.7, angle=90]{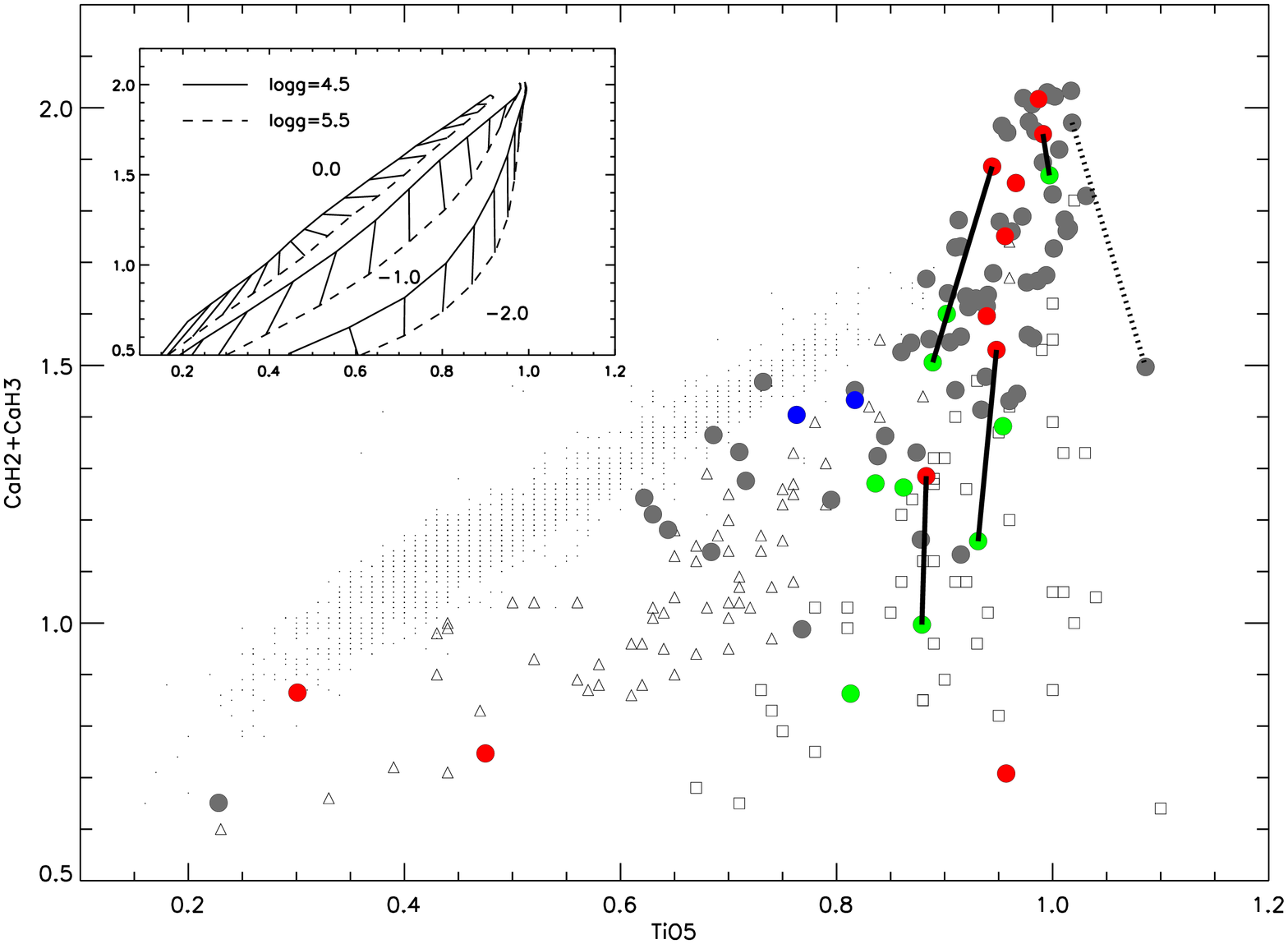}
\vspace{0.5cm}

\caption{The CaH2+CaH3 versus TiO5 indices are plotted for subdwarfs
discussed in this paper (filled circles).  For comparison, known cool
dwarfs (dots), subdwarfs (open triangles), and ``extreme'' subdwarfs
(open boxes) from \cite{Hawley1996}, \cite{Gizis1997}, and
\cite{Reid2005} are also shown.  Green circles are subdwarfs with the
highest gravities at a given type (g with most $+$) in
Table~\ref{tbl.spectral.type}.  The two blue circles represent LHS 400
and SCR 1822$-$4542, which have higher gravities than M dwarfs.
Red circles represent the lowest metallicity stars in
Table~\ref{tbl.spectral.type} at a given type (m with most $-$).
Solids lines indicate pairs of stars having the same number of $-$ in
metallicity, but different gravities.  A dotted line connects LHS
2734A and LHS 2734B, which is the only wide CPM binary in our sample.
For comparison, the inset plot illustrates the indices (TiO5
vs. CaH2$+$CaH3) calculated from GAIA models for temperatures
2800--4400K.  Axis labels are omitted, but both axes have the same
ranges as the main figure.  Three different metallicities, 0.0, $-1.0$
and $-$2.0, and two gravities, $\log g$=4.5 (solid line) and 5.5 (dash
line), are shown.  Connections are drawn for stars with the same
temperatures but different gravities.}

\label{fig.subdwarf.sdm.esdm.index}
\end{figure}

\begin{figure}
\centering
\includegraphics[scale=0.6, angle=90]{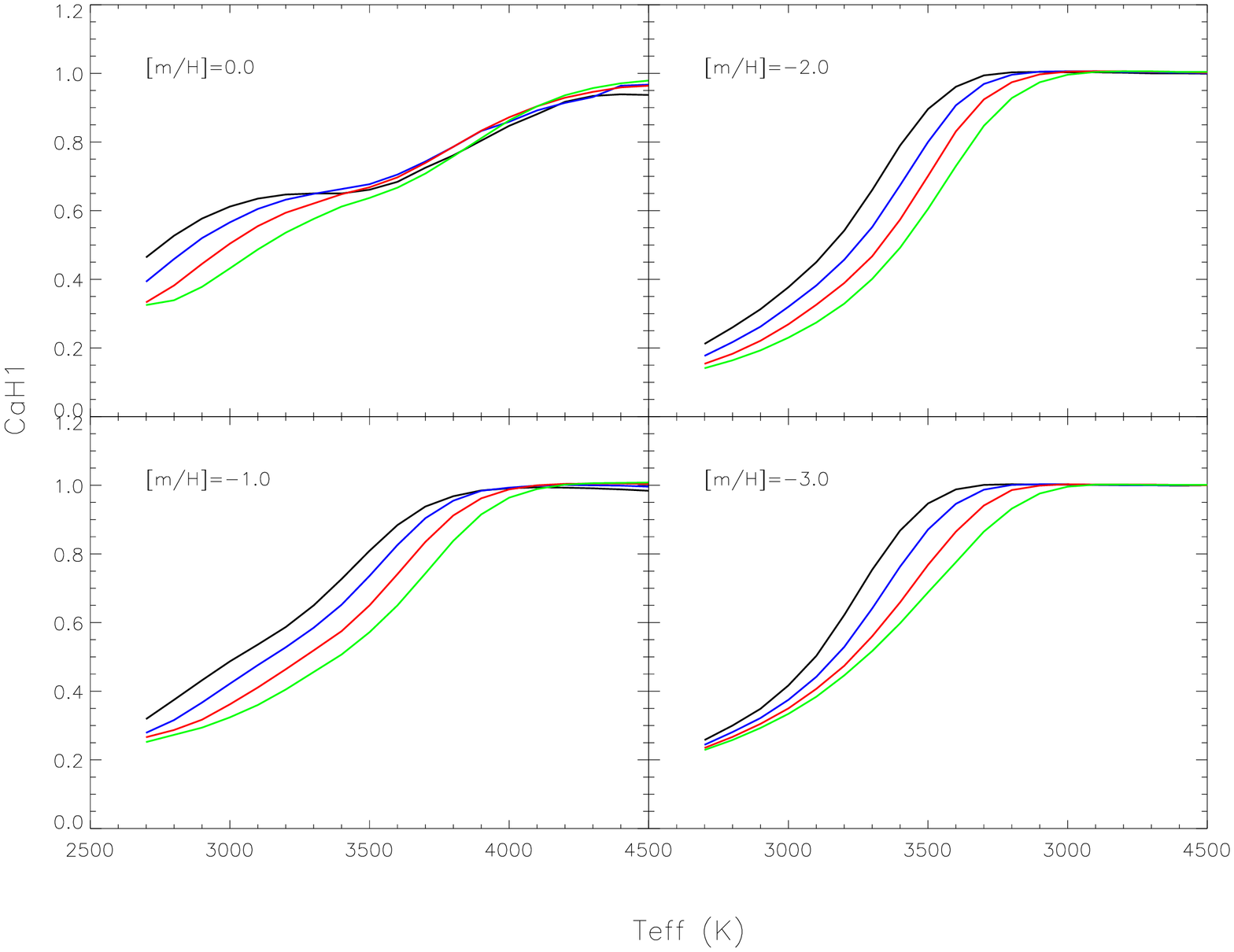}
\\
\includegraphics[scale=0.6, angle=90]{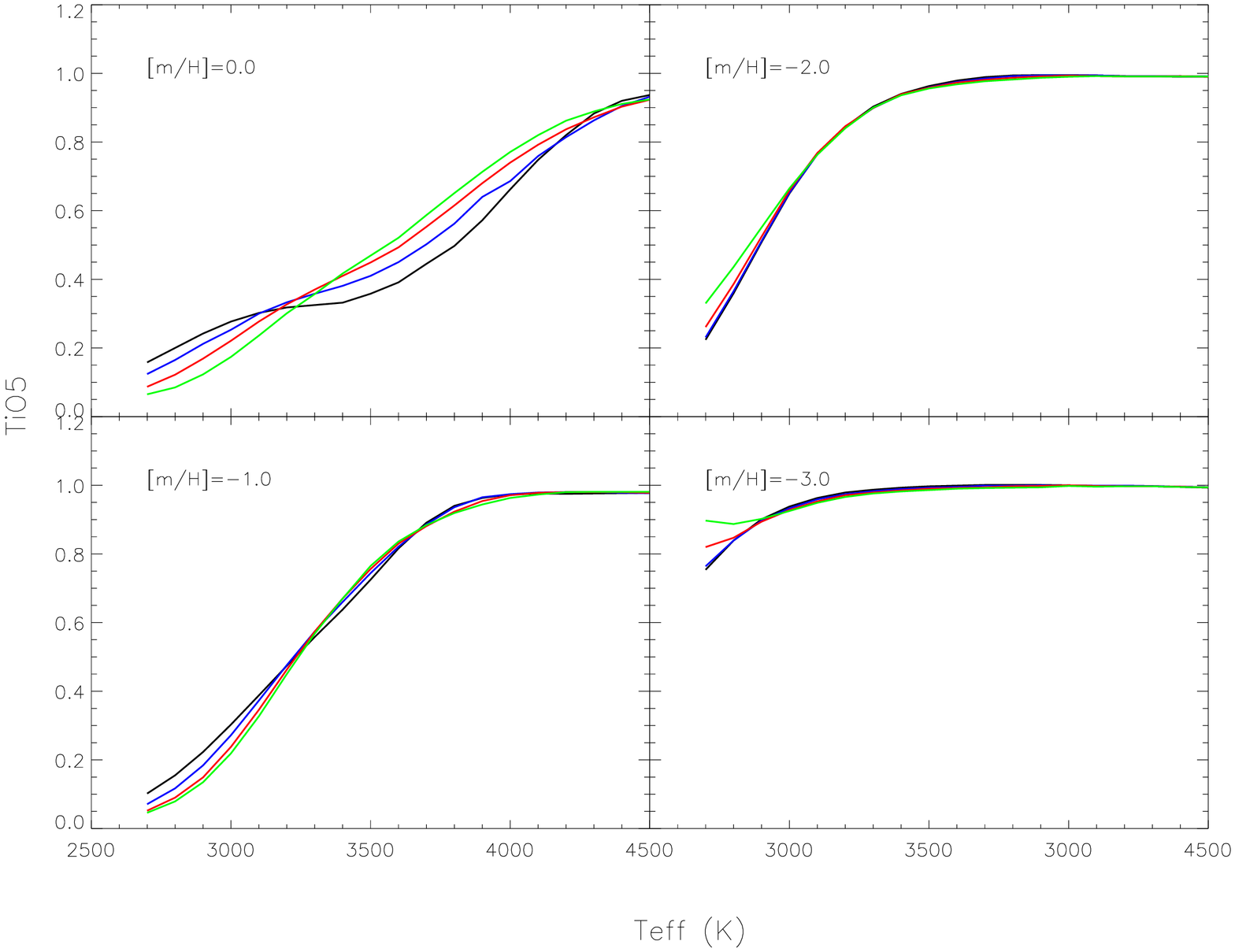}

\caption{CaH1 and TiO5 indices from GAIA model grids are plotted
against effective temperature.  The black, blue, red and green lines
indicate {\it log g}$=$4.0, 4.5, 5.0, 5.5, respectively.  Note that a
given CaH1 and TiO5 value may correspond to many combinations of
metallicities and gravities.}

\label{fig.cah1.tio5.vs.teff}
\end{figure}

\begin{itemize}

\item These terms can only be applied to M-type subdwarfs, not K-type
subdwarfs.  Both observed spectra and models show that K subdwarfs do
not follow the trends in CaH absorption with metallicity that M
subdwarfs follow.  Thus, there is no clear delineation for K-type
subdwarfs.

\item Empirically, the values of the indices are affected by a
complicated interplay of temperature, metallicity, and gravity
effects.  One cannot separate these three factors simply based on the
indices.  Typically, when the term ``extreme'' subdwarf is used, it
refers to ``very low metallicity'' alone.  But that is only one part
of the portrait that needs to be painted for a given subdwarf.  For
example, filled boxes shown in Figure~\ref{fig.hr.plot} indicate stars
previously identified as extreme subdwarfs from their indices.  Note
that the spread for ``extreme'' subdwarfs at a given color
($V-K_{s}\sim$ 3.5) is 3.5 magnitudes, or a factor of $\sim$25 in
luminosity.  There is no clear separation in the fundamental HR
diagram between stars termed ``subdwarfs'' and ``extreme subdwarfs''
classified using spectral indices.

In Figure~\ref{fig.subdwarf.sdm.esdm.index}, we take a detailed look
at astrophysical causes that shift points in the CaH2+CaH3/TiO5
indices plot.  Red circles in Figure~\ref{fig.subdwarf.sdm.esdm.index}
represent the lowest metallicity stars of each spectral type we have
presented here.  Some of these very low metallicity stars are not
located in the ``extreme'' subdwarf region. Green circles represent
subdwarfs in our sample that have highest gravity at a given type.
Solid lines connect stars having the same metallicity rankings in
Table~\ref{tbl.spectral.type} (components of the common proper motion
binary LHS 2734 AB are connected by a dotted line because they
presumably have the same metallicity).  It is clear from these three
pairs that high gravity can push a subdwarf toward or into the extreme
subdwarf region, even if the metallicities of the two objects are
similar.  In addition, many other high gravity subdwarfs (green
circles) that do not have particularly low metallicities are also in
the subdwarf region.  Finally, the two blue circles represent LHS 440
and SCR 1822-45542, which subdwarfs having the same metallicities but
higher gravities than dwarfs.  Their spectroscopic features indicate
that they are subdwarfs, but they do not have low metallicity at all.
Yet, they fall in the subdwarf region of the indices plot, which has
traditionally indicated low metallicity.

The inset in Figure~\ref{fig.subdwarf.sdm.esdm.index} helps explain
this phenomenon.  We have calculated the spectral indices from the
GAIA synthetic spectra with [m/H]$=$0.0, $-$1.0 and $-$2.0 and
temperatures of 2800--4400K.  Two different gravities were selected
for each metallicity.  The models indicate that both low metallicity
and high gravity push stars toward the extreme subdwarf region.  Thus,
if two stars have the same metallicity, the one with higher gravity
will be pushed more toward the extreme subdwarf region, indicating
that this spectral indices plot is not a clear indicator of
metallicity alone.

\item Theoretically, in order to show how complicated the TiO5 vs CaH1
plot is astrophysically, we use GAIA model grids to do a
demonstration.  We have chosen models with 2700K $< T_{eff} <$ 4500K,
$-$3.0 $< [m/H] <$ 0.0 and 4.0 $< {\it log g} <$ 5.5, and have
calculated the output CaH and TiO5 indices, as shown in
Figure~\ref{fig.cah1.tio5.vs.teff}.  As is apparent in the plots, for
a given (TiO5, CaH1) indices pair, there are many possible parameter
combinations.  Both indices are a function of temperature, gravity and
metallicity, which would require a 6-D plot to describe TiO5 vs CaH.
Unfortunately, one cannot classify stars as subdwarfs, extreme
subdwarfs, or ultra-subdwarfs using 2-D index plots that infer trends
in metallicity alone.

\end{itemize}

For these reasons, the indices only indicate a star's location on
these particular spectroscopic indices plots.  They do not address low
or very low metallicity only, as implied by past usage, but instead
incorporate effects of both metallicity and gravity for stars of a
given temperature.  They do not provide obvious or direct links to
positions on HR diagrams, which provide the astrophysical meaning
underlying stellar classification.  Finally, the modifiers ``extreme''
and ``ultra'' themselves have effectively the same meaning and do not
provide meaningful information about their differences.

\subsection{Why Previous Methods Work for Dwarfs but Not Subdwarfs}

In the Palomar-MSU spectroscopy survey, \cite{Reid1995} used the TiO5
index and polynomial equations to assign subtypes for M dwarfs.
\cite{Gizis1997} then applied the same methodology to M subdwarfs.  As
can be seen in Figure~\ref{fig.cah1.tio5.vs.teff}, TiO5 versus
temperature is effectively linear, especially for {\it log g} $=$ 5.0
and [m/H] $=$ 0.0, so this typing method works successfully for a
fairly homogeneous set of dwarfs (possibly with somewhat different
gravities) --- temperature is the main factor affecting the overall
slope of red dwarf spectra.  However, it is not as simple for
subdwarfs because both low metallicity and high gravity stars are
sorted into the same TiO5 values for various combinations.  Thus, the
same methodology that works for dwarfs cannot straightforwardly be
applied to subdwarfs.

\subsection{The Subdwarf Spectral Standards from \cite{Lepine2007}}

\begin{figure}
\centering
\includegraphics[scale=0.7, angle=90]{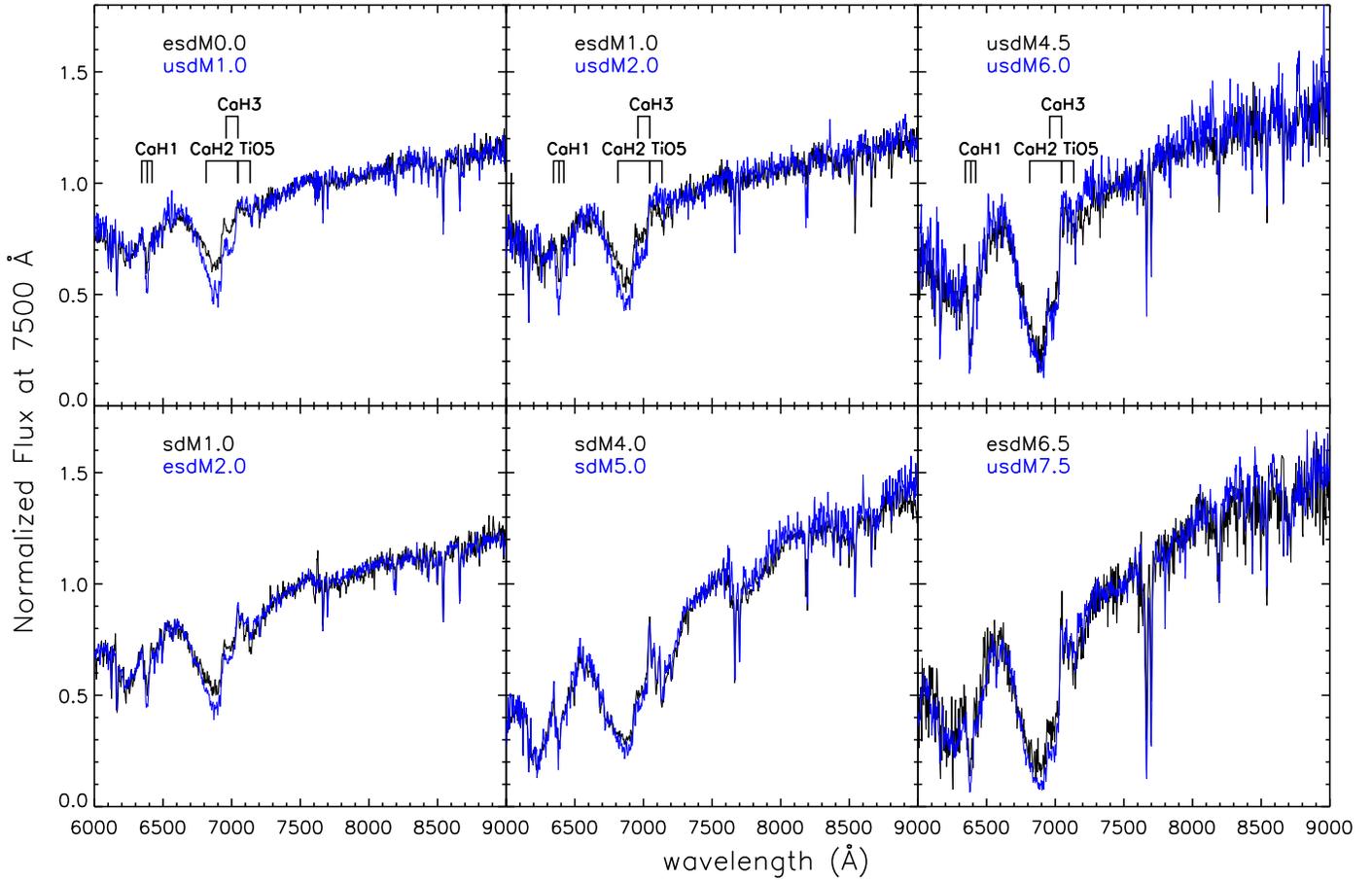}

\caption{Spectra from \cite{Lepine2007} are shown, with their
published types.  The colors (black and blue) represent different
spectra, and their colors match with labels.  Each pair has the almost
the same overall spectral shape, but each star is assigned a different
sub-type and identification as extreme or ultra.  Spectra are
normalized at 7500\AA.}

\label{fig:lepine.1}
\end{figure}

\begin{figure}
\centering
\includegraphics[scale=0.8]{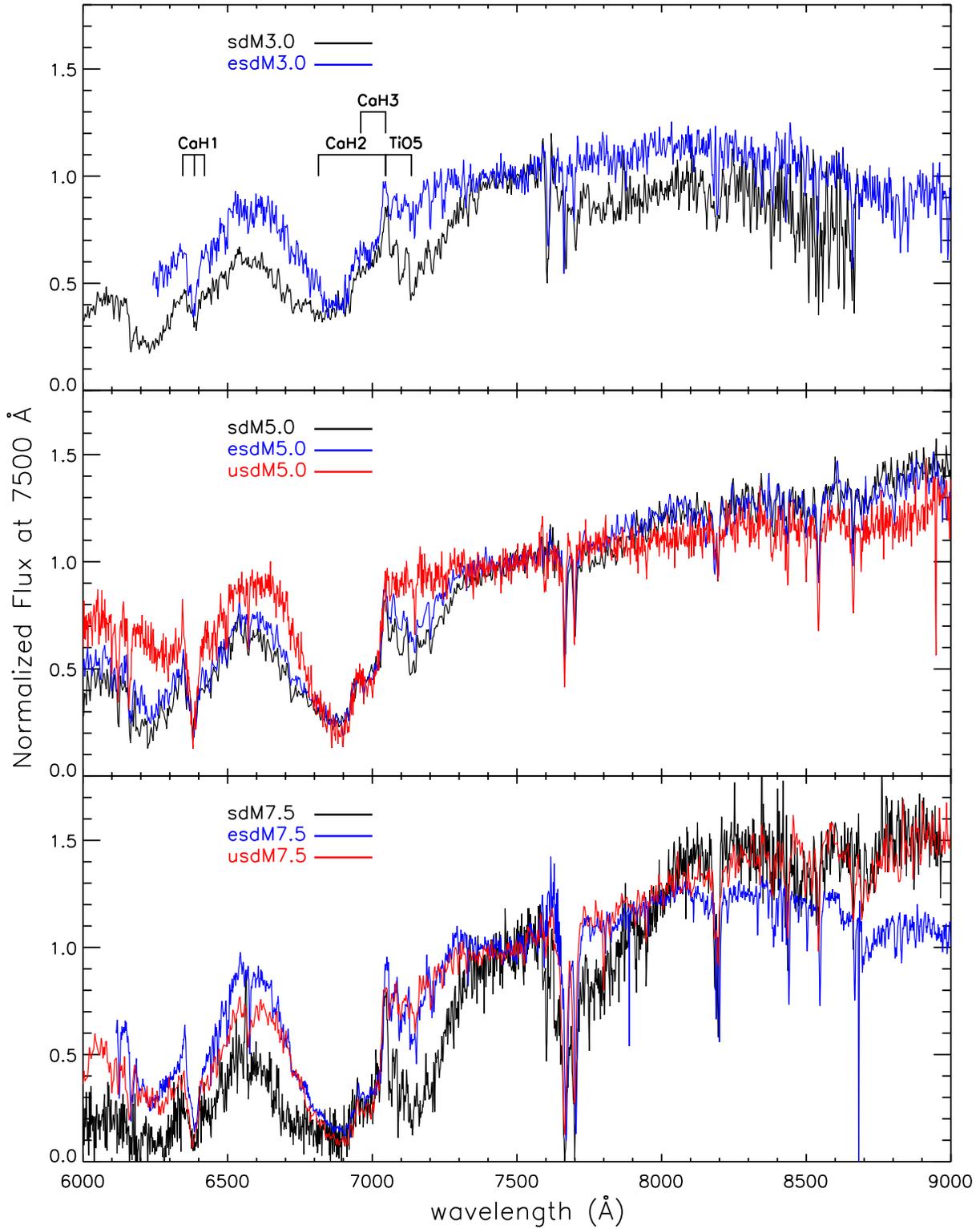}
\vspace{1cm}

\caption{Spectra from \cite{Lepine2007} are shown for spectral types
for M3.0, M5.0 and M7.5.  The black, blue and red spectra represent
sdM, esdM and usdM, respectively. Spectra are normalized at 7500\AA.}

\label{fig:lepine.2}
\end{figure}

\cite{Lepine2007} recently released a set of spectral standards for
the sdM, esdM and usdM subclasses using wider spectral coverage than
used by \cite{Gizis1997} and revised polynomial equations.  Their
project is contemporaneous with ours, but unfortunately there is only
one star in both samples, LHS 228.  We assign it a type of M3.0VI
while their type is sdM2.0.  As shown in Figure~\ref{fig:lepine.1},
gravity effects come into play in their spectra\footnote{Their
telluric lines have been removed.}.  Each pair of spectra shown have
nearly identical continua, with CaH bands being the only significant
spectral difference.  Using their method, each of the pairs of
subdwarfs shown is assigned a different sub-type.  We believe that by
using the overall shape of the spectrum, with knowledge that both
metallicity and gravity affect certain regions, each pair should have
the same spectral sub-types.  Based on the trends from GAIA models,
the pairs of spectra shown have different gravities, as we have also
seen in our spectra and those from SDSS (shown in $\S$\ref{sec:SDSS}).
Figure~\ref{fig:lepine.2} illustrates spectral differences at types
M3.0, M5.0 and M7.5.  Although the spectra in each panel have the same
sub-types for sdM, esdM and usdM subdwarfs, they show very different
continua between 6000\AA~and 9000\AA.  When examining the overall
spectra, there is no clear morphology trend for each sub-type.  We
conclude that using spectral indices alone omits important information
evident in the overall morphology of the spectra that is useful to
spectral classification.

\section{Conclusions}

We have discussed 88 cool subdwarfs using spectra covering
6000--9000\AA.  Based on these spectra and the trends from GAIA model
grids, we have redefined the subdwarf spectral sequence, spanning
types K3.0 to M6.0.  We find that wide spectral coverage is the key to
defining a subdwarf's spectral type.  We consider this to be an
important, but not final, step in defining the subdwarf spectral
sequence.

Through the understanding of GAIA model grids, we find that the key to
assigning a subdwarf spectral type is to compare the spectrum to dwarf
spectral standards in regions affected minimally by metallicity and
gravity, thereby making a direct link between the dwarf and subdwarf
sequences.  Even so, it remains difficult to establish a definitive
sequence for subdwarfs because of the multi-faceted nature of their
spectra.  Until we have surveyed a large number of subdwarfs and
covered a multitude of possible temperatures, metallicities, and
gravities, a definitive sequence will remain elusive.

From an analysis of the history of the term ``subdwarf,'' and the
layout of the fundamental HR diagram, we propose that the suffix
``VI'' be used, rather than the ``sd'' prefix, as the preferred
spectral classification notation.  This reduces the confusion between
cool subdwarfs and hot OB subdwarfs.  This is also prudent because
subdwarfs really do form an independent class of stars on the HR
diagram, for which five Roman numerals are in common use, with cool
subdwarfs naturally falling beneath the main sequence V types.

Overall, we find that trigonometric parallaxes are crucial for
identifying mid K-type subdwarfs, and allow us to understand how the
complex interplay of temperatures, metallicities, and gravities
affects the positions of cool dwarfs on the HR diagram.  We found that
mid K-type subdwarfs can not be identified spectroscopically using our
data.  There are many G and M-type subdwarfs, so K-type subdwarfs
presumably exist, but additional information is needed to identify
them.  Consequently, we use metallicities, kinematics, and parallaxes
to acertain their true natures and use their locations on the HR
diagram to flag them as ``subdwarfs'' (other spectroscopic wavelength
and resolution combinations could also be used).  Because they are not
spectroscopically identified as subdwarfs in our spectra, we use a
conservative notation, [VI], for subdwarfs of types K3.0 through K5.0
to indicate their questionable status.

We have confirmed that spectroscopic indices are useful in separating
late K to late M-type subdwarfs from dwarfs, but that the indices have
limitations when attempting to understand the astrophysical causes
leading to observed subdwarf spectra.  When combined with
trigonometric parallax and photometric information, our results show
that for lower metallicities, subdwarfs are generally bluer and
brighter at optical wavelengths, so they slide up and to the left on
the HR diagram (using axes $V-K_{s}$ vs $M_{Ks}$).  In contrast,
higher gravities make stars redder and less luminous at optical
wavelengths, so subdwarfs generally slide down and to the right on the
same HR diagram.  Because of the complex, and not yet completely
mapped out, interplay of temperatures, metallicities, and gravities,
we conclude that the ``extreme'' and ``ultra'' prefixes {\it only}
outline locations on spectroscopic indices plots, and do not
successfully differentiate the underlying astrophysical causes for
shifts on CaH/TiO5 plots.

Improvements in the subdwarf spectral sequence can be made by
observing wide common proper motion subdwarf binaries, like LHS 2734AB
discussed here.  Assuming identical metallicities, such binaries allow
us to constrain one of the three variables that affects cool dwarf
spectral types.  The ultimate subdwarf spectral sequence will be
three-dimensional, with temperature, metallicity, and gravity as
independent variables (see \citealt{Kirkpatrick2005}, Figure 11).

Previously, gravity effects in cool subdwarfs have been almost
entirely ignored, as metallicity was the factor considered to describe
changing CaH features.  Our results clearly show the importance of
gravity effects.  An additional factor, contamination from an unseen,
evolved, i.e.~white dwarf, companion could also change the slope of
subdwarf spectra.  Thus, comprehensive surveys of subdwarfs for
companions are warranted\footnote{We have an ongoing project to survey
cool subdwarfs for companions.  To date, we have found that cool
subdwarfs have a lower binary fraction than cool dwarfs as reported by
\cite{Riaz2008}, so the contamination assumption from possible unseen
companions appears unlikely.}.  Finally, this work also shows that
current synthetic spectra provide a useful framework in which to
evaluate cool dwarf spectra, but they do not yet provide perfect
matches, indicating that atmospheric models still require fine-tuning
to make additional advances in the future characterization of cool
dwarfs (P. Hauschildt private communication).

\begin{acknowledgements}

We would like to thank the referee, Sandy Leggett, for her very
helpful comments that improved this paper.  We also thank Pat Boeshaar
and Doug Gies for their comments and suggestions.  Peter Hauschildt
was instrumental in helping us use the GAIA model grids, and Vincent
Woolf contributed the old version of synthetic spectra.  We thank
Sebastien Lepine for providing his subdwarf spectra to us before they
were publicly available, and for his comments on section 9.4.  We
appreciate the help of many members of the Georgia State University
(GSU) RECONS team in their data acquisition and reduction efforts,
especially Charlie Finch and Jennifer Winters.  This work has been
supported at GSU by NASA's Space Interferometry Mission, the National
Science Foundation (NSF, grant AST-0507711), and GSU.  Finally, we
wish to thank the other members of the SMARTS Consortium, without whom
the telescopes at CTIO used for this effort might not be available.

This research has made use of the SIMBAD database, operated at CDS,
Strasbourg, France.  This paper has also made use of data from the
SDSS.  Funding for the SDSS and SDSS-II has been provided by the
Alfred P. Sloan Foundation, the Participating Institutions, NSF, the
U.S. Department of Energy, NASA, the Japanese Monbukagakusho, the Max
Planck Society, and the Higher Education Funding Council for England.
This work has used data products from the Two Micron All Sky Survey,
which is a joint project of the University of Massachusetts and the
Infrared Processing and Analysis Center at California Institute of
Technology funded by NASA and NSF.

\end{acknowledgements}



\begin{deluxetable}{lccccc}
\tablewidth{12cm} \tablecaption{Spectroscopic Indices} \tablehead{
\colhead{Object} & \colhead{TiO5} & \colhead{CaH1} & \colhead{CaH2} &
\colhead{CaH3} & \colhead{CaH2$+$CaH3} } 
\startdata 
DEN0515$-$7211& 1.002 &   1.021  &  1.019 &   1.003 &   2.022 \\
G016$-$009AB & 0.973 &   1.018  &  1.021 &   0.998 &   2.019 \\
G022$-$015   & 0.978 &   0.996  &  0.989 &   0.984 &   1.973 \\
G026$-$009ACD& 0.953 &   0.990  &  0.975 &   0.990 &   1.965 \\
GJ0191       & 0.860 &   0.876  &  0.678 &   0.848 &   1.526 \\
GJ0223.1     & 0.958 &   1.001  &  0.981 &   0.971 &   1.952 \\
LEHPM1628    & 0.954 &   0.676  &  0.628 &   0.754 &   1.382 \\
LEHPM3861    & 0.957 &   0.774  &  0.340 &   0.368 &   0.709 \\
LHS0012      & 0.883 &   0.881  &  0.779 &   0.889 &   1.668 \\
LHS0073      & 0.972 &   0.903  &  0.862 &   0.927 &   1.789 \\
LHS0109      & 0.939 &   0.823  &  0.752 &   0.863 &   1.616 \\
LHS0125      & 0.981 &   1.020  &  1.014 &   0.993 &   2.007 \\
LHS0127      & 0.716 &   0.755  &  0.537 &   0.739 &   1.276 \\
LHS0144      & 0.768 &   0.609  &  0.396 &   0.592 &   0.987 \\
LHS0148      & 0.977 &   0.816  &  0.729 &   0.830 &   1.560 \\
LHS0158      & 0.732 &   0.849  &  0.639 &   0.829 &   1.469 \\
LHS0161      & 0.889 &   0.777  &  0.689 &   0.817 &   1.506 \\
LHS0162      & 0.838 &   0.731  &  0.577 &   0.747 &   1.324 \\
LHS0164      & 0.987 &   1.005  &  1.000 &   1.017 &   2.017 \\
LHS0165      & 0.956 &   0.889  &  0.841 &   0.910 &   1.751 \\
LHS0186      & 0.710 &   0.739  &  0.568 &   0.764 &   1.332 \\
LHS0189AB    & 0.630 &   0.733  &  0.492 &   0.719 &   1.211 \\
LHS0193A     & 0.944 &   0.959  &  0.920 &   0.966 &   1.886 \\
LHS0227      & 1.000 &   0.895  &  0.883 &   0.949 &   1.832 \\
LHS0228      & 0.644 &   0.744  &  0.480 &   0.701 &   1.181 \\
LHS0232      & 0.998 &   1.000  &  1.020 &   1.005 &   2.025 \\
LHS0244      & 0.902 &   0.853  &  0.731 &   0.869 &   1.601 \\
LHS0272      & 0.862 &   0.715  &  0.527 &   0.736 &   1.264 \\
LHS0299      & 0.903 &   0.904  &  0.756 &   0.884 &   1.641 \\
LHS0300AB    & 1.031 &   0.933  &  0.886 &   0.943 &   1.830 \\
LHS0318      & 0.934 &   0.747  &  0.617 &   0.797 &   1.413 \\
LHS0326      & 0.948 &   0.782  &  0.689 &   0.841 &   1.530 \\
LHS0327      & 0.995 &   1.004  &  1.025 &   1.005 &   2.030 \\
LHS0334      & 0.475 &   0.492  &  0.277 &   0.470 &   0.747 \\
LHS0335      & 0.960 &   0.721  &  0.640 &   0.791 &   1.431 \\
LHS0360      & 0.986 &   0.840  &  0.778 &   0.886 &   1.664 \\
LHS0367      & 1.011 &   0.880  &  0.854 &   0.929 &   1.783 \\
LHS0375      & 0.879 &   0.604  &  0.414 &   0.583 &   0.997 \\
LHS0381      & 0.883 &   0.728  &  0.555 &   0.730 &   1.285 \\
LHS0385      & 0.982 &   0.787  &  0.712 &   0.841 &   1.553 \\
LHS0398      & 0.910 &   0.767  &  0.644 &   0.808 &   1.452 \\
LHS0401      & 0.997 &   0.959  &  0.911 &   0.958 &   1.869 \\
LHS0406      & 0.686 &   0.801  &  0.576 &   0.789 &   1.365 \\
LHS0418      & 0.913 &   0.924  &  0.858 &   0.924 &   1.782 \\
LHS0424      & 0.951 &   0.917  &  0.852 &   0.927 &   1.778 \\
LHS0440      & 0.763 &   0.789  &  0.608 &   0.796 &   1.404 \\
LHS0507      & 0.991 &   0.958  &  0.969 &   0.980 &   1.949 \\
LHS0515      & 0.813 &   0.524  &  0.355 &   0.508 &   0.863 \\
LHS0518      & 0.966 &   0.933  &  0.903 &   0.951 &   1.854 \\
LHS0521      & 1.006 &   0.937  &  0.950 &   0.969 &   1.919 \\
LHS0541      & 0.795 &   0.678  &  0.538 &   0.701 &   1.239 \\
LHS1490      & 0.301 &   0.778  &  0.296 &   0.569 &   0.865 \\
LHS1970      & 0.878 &   0.433  &  0.476 &   0.686 &   1.162 \\
LHS2067A     & 0.228 &   0.670  &  0.199 &   0.452 &   0.651 \\
LHS2467      & 1.017 &   1.031  &  1.019 &   1.014 &   2.033 \\
LHS2734A     & 1.018 &   0.985  &  0.984 &   0.987 &   1.972 \\
LHS2734B     & 1.086 &   0.720  &  0.669 &   0.828 &   1.497 \\
LHS3620      & 0.836 &   0.684  &  0.544 &   0.727 &   1.271 \\
SCR0242$-$5935 & 0.910 & 0.906  &  0.838 &   0.891 &   1.728 \\
SCR0406$-$6735 & 1.015 & 0.989  &  0.829 &   0.937 &   1.766 \\
SCR0433$-$7740 & 0.929 & 0.887  &  0.760 &   0.864 &   1.624 \\
SCR0529$-$3950 & 0.817 & 0.886 & 0.637 & 0.815 & 1.452 \\
SCR0629$-$6938 & 0.845 & 0.908 & 0.579 & 0.784 & 1.363 \\
SCR0654$-$7358 & 0.938 & 0.835 & 0.653 & 0.825 & 1.478 \\
SCR0701$-$0655 & 0.976 & 0.886 & 0.780 & 0.881 & 1.662 \\
SCR0708$-$4709 & 0.984 & 0.992 & 0.979 & 0.976 & 1.955 \\
SCR0709$-$4648 & 1.013 & 0.899 & 0.846 & 0.915 & 1.761 \\
SCR1107$-$4135 & 0.994 & 0.839 & 0.783 & 0.892 & 1.675 \\
SCR1433$-$3847 & 1.001 & 0.904 & 0.797 & 0.930 & 1.726 \\
SCR1455$-$3914 & 0.915 & 0.788 & 0.707 & 0.849 & 1.555 \\
SCR1457$-$3904 & 0.905 & 0.824 & 0.707 & 0.838 & 1.545 \\
SCR1613$-$3040 & 0.920 & 0.898 & 0.749 & 0.885 & 1.634 \\
SCR1739$-$8222 & 0.945 & 1.009 & 0.797 & 0.882 & 1.679 \\
SCR1740$-$5646 & 0.915 & 0.696 & 0.496 & 0.637 & 1.133 \\
SCR1756$-$5927 & 0.922 & 0.794 & 0.757 & 0.856 & 1.613 \\
SCR1822$-$4542 & 0.817 & 0.818 & 0.624 & 0.809 & 1.433 \\
SCR1843$-$7849 & 0.869 & 0.810 & 0.715 & 0.829 & 1.545 \\
SCR1913$-$1001 & 0.962 & 0.923 & 0.835 & 0.925 & 1.760 \\
SCR1916$-$3638 & 0.931 & 0.647 & 0.493 & 0.666 & 1.159 \\
SCR1958$-$5609 & 0.915 & 0.865 & 0.805 & 0.926 & 1.731 \\
SCR2018$-$6606 & 0.886 & 0.856 & 0.715 & 0.836 & 1.550 \\
SCR2101$-$5437 & 0.929 & 0.969 & 0.726 & 0.904 & 1.630 \\
SCR2104$-$5229 & 0.940 & 0.907 & 0.772 & 0.865 & 1.638 \\
SCR2109$-$5226 & 0.967 & 0.774 & 0.646 & 0.799 & 1.445 \\
SCR2204$-$3347 & 0.874 & 0.711 & 0.569 & 0.762 & 1.331 \\
SIP1342$-$3534 & 0.684 & 0.664 & 0.462 & 0.676 & 1.138 \\ 
WT0135 &0.622 & 0.754 & 0.512 & 0.731 & 1.243 \\ 
WT0233 & 0.939 & 0.933 &0.739 & 0.857 & 1.596 \enddata
\label{tbl.cahn.tio5.index}
\end{deluxetable}


\begin{deluxetable}{cclrccrclllccc}
\rotate
\tabletypesize{\scriptsize}
\tablewidth{23cm}
\tablecaption{Spectral Types}
\tablehead{
\colhead{RA}                   &
\colhead{DEC}                  &
\colhead{Object}               &   
\colhead{$K_{s}$}              &    
\colhead{$V-K_{s}$}            &   
\colhead{$M_{K_{s}}$}          &  
\colhead{$V_{tan}$}            &
\colhead{Old Type}             &
\colhead{New Type}             &
\colhead{metallicity}          &
\colhead{gravity}              &
\multicolumn{3}{c}{Ref}        \\
                               &
                               &
                               &
                               &   
                               &    
                               &   
km/sec                         &   
                               &
                               &
                               &
                               &
$\pi$                          &
$V$                            &
spect                          }

\startdata
12 06 00.9 & $+$14 38 56.8 & G012$-$016        &    7.931$\pm$0.018 &  2.16   &  4.12   &  98.4 & K2         &  GVI     &($-$0.52)\tablenotemark{a} &         & H& 2& 2\\
01 04 26.4 & $-$02 21 59.8 & G070$-$035        &    7.116$\pm$0.020 &  2.04   &  4.15   &  55.8 & G5         &  GVI     &($-$0.67)\tablenotemark{a} &         & H& 4& H\\
02 25 49.8 & $+$05 53 39.5 & G073$-$056        &   10.427$\pm$0.023 &  2.06   &  8.64   &  44.9 &            &  GVI     &($-$1.18)\tablenotemark{b} &         & H& 4&  \\
07 54 34.1 & $-$01 24 44.3 & G112$-$054        &    5.425$\pm$0.023 &  2.00   &  4.01   &  23.7 & K1         &  GVI     &($-$0.94)\tablenotemark{c} &         & H& 1& 2\\
16 13 48.6 & $-$57 34 13.8 & LHS0413           &    5.293$\pm$0.024 &  2.24   &  4.60   & 106.5 & G8/K0V(W)  &  GVI     &($-$1.35)\tablenotemark{c} &         & H& 1&10\\
\tableline
\tableline
15 45 52.4 & $+$05 02 26.6 & G016$-$009AB      &    6.880$\pm$0.024 &  2.27   &  3.46   &  59.2 &  K2V       &  K3.0[VI] &m$-$, ($-$0.77)\tablenotemark{d} &  g      & H& H& 3\\
\tableline										  
\tableline										  
00 50 17.0 & $-$39 30 08.3 & LHS0125           &   11.452$\pm$0.026 &  2.88   &  7.73   & 342.6 &            &  K4.0[VI] &m$-$ &  g      & R& R& ~\\
07 35 46.3 & $+$03 29 36.0 & LHS0232           &   10.841$\pm$0.024 &  2.84   &  6.57   & 346.0 &            &  K4.0[VI] &m$-$ &  g      & Y& R& ~\\
12 25 50.7 & $-$24 33 17.8 & LHS0327           &   10.144$\pm$0.021 &  2.57   &  5.44   & 418.3 &  K0V       &  K4.0[VI] &m$-$ &  g      & R& R& 2\\
11 52 32.0 & $+$27 30 51.3 & LHS2467           &    9.806$\pm$0.017 &  2.43   &  4.94   & 437.0 &  G7V       &  K4.0[VI] &m$-$ &  g      & H& R& 2\\
\tableline										  
\tableline										  
19 07 02.0 & $+$07 36 57.3 & G022$-$015        &    6.469$\pm$0.018 &  2.71   &  4.54   &  97.2 &  K5V       &  K5.0[VI] &m$-$, ($-$0.61)\tablenotemark{e} &  g      & H& H& 2\\
21 32 11.9 & $+$00 13 18.0 & G026$-$009ACD     &    7.082$\pm$0.029 &  2.64   &  3.62   &  96.9 &  K2V       &  K5.0[VI] &m$-$, ($-$1.05)\tablenotemark{f} &  g      & H& H& 2\\
05 54 34.1 & $-$09 23 33.7 & GJ0223.1          &    7.760$\pm$0.020 &  2.96   &  4.84   &  82.1 &  K4V       &  K5.0[VI] &m$-$, ($-$0.62)\tablenotemark{e} &  g      & H& 1& 9\\
\tableline										 
\tableline										 
04 32 36.6 & $-$39 02 03.4 & LHS0193A          &    8.427$\pm$0.023 &  3.23   &  5.80   & 147.3 &            &  K6.0VI  &  m$-$      &  g      & R&16& ~\\
23 43 16.7 & $-$24 11 16.4 & LHS0073           &    9.393$\pm$0.021 &  3.42   &  7.27   & 322.3 &  K5V       &  K6.0VI  &  m$-$      &  g+     & Y&19& 2\\
07 13 40.6 & $-$13 27 57.1 & LHS0227           &   11.036$\pm$0.023 &  3.41   &  7.15   & 362.5 &            &  K6.0VI  &  m$-$      &  g+     & Y& R& ~\\
02 52 45.7 & $+$01 55 49.4 & LHS0161           &   10.995$\pm$0.019 &  3.65   &  8.01   & 272.6 & esdM2      &  K6.0VI  &  m$-$      &  g++    & Y& R& 8\\
\tableline										 
\tableline										 
03 01 40.6 & $-$34 57 56.5 & LHS0164           &   10.641$\pm$0.024 &  2.92   &  8.20   & 192.6 &            &  K7.0VI  &  m$--$     &  g      & R& R& ~\\
05 15 45.1 & $-$72 11 22.2 & DEN0515$-$7211    &   13.211$\pm$0.040 &  3.29   &  9.72   &   4.7 &            &  K7.0VI: &  m$--$     &  g      & 7& 7& ~\\
07 08 32.0 & $-$47 09 30.5 & SCR0708$-$4709    &   10.764$\pm$0.025 &  3.05   & \nodata &\nodata&            &  K7.0VI: &  m$--$     &  g      &  & R&  \\
\tableline										
13 25 14.0 & $-$21 27 06.0 & LHS2734A          &   12.896$\pm$0.037 &  3.23   & too~far &$>$280 &            &  K7.0VI  &  m$-$      &  g      & R& R& ~\\
\tableline										 
\tableline										 
03 06 28.7 & $-$07 40 41.5 & LHS0165           &   11.006$\pm$0.025 &  3.42   &  7.67   & 333.1 &            &  M0.0VI  &  m$---$    &  g+:    & Y& R& ~\\
\tableline										 
11 11 13.7 & $-$41 05 32.7 & LHS0300A          &    9.802$\pm$0.023 &  3.38   &  7.23   & 197.5 &            &  M0.0VI: &  m$--$     &  g      &11&16& ~\\
16 25 14.0 & $+$15 40 54.2 & LHS0418           &   10.072$\pm$0.018 &  3.37   &  6.44   & 301.5 &  K7V       &  M0.0VI  &  m$--$     &  g      & Y& R& 8\\
16 37 05.6 & $-$01 32 01.6 & LHS0424           &   10.803$\pm$0.022 &  3.37   &  7.39   & 279.6 &            &  M0.0VI  &  m$--$     &  g      & Y& R& ~\\
\tableline										 
08 13 27.8 & $-$09 27 56.6 & LHS0244           &   10.729$\pm$0.023 &  3.64   &  6.88   & 412.7 &            &  M0.0VI  &  m$-$      &  g++:   & Y&19& ~\\
\tableline										 
\tableline										 
22 27 59.0 & $-$30 09 30.0 & LHS0521           &   11.463$\pm$0.019 &  3.22   &  8.14   & 222.1 &            &  M0.5VI: &  m$---$    &  g      &11&11& ~\\
21 21 34.8 & $-$19 03 38.6 & LHS0507           &   12.053$\pm$0.026 &  3.15   &  7.81   & 355.2 & K/M sd     &  M0.5VI  &  m$---$    &  g+     & Y& R& Y\\
15 39 39.0 & $-$55 09 10.0 & LHS0401           &    9.407$\pm$0.019 &  3.31   &  7.33   & 142.2 &            &  M0.5VI  &  m$---$    &  g++    & Y& R& ~\\
\tableline										
02 42 26.3 & $-$59 35 01.6 & SCR0242$-$5935    &   12.783$\pm$0.031 &  3.40   & \nodata &\nodata&            &  M0.5VI  &  m$--$     &  g      &  & R&  \\
07 09 37.2 & $-$46 48 58.8 & SCR0709$-$4648    &   11.491$\pm$0.026 &  3.41   & \nodata &\nodata&            &  M0.5VI  &  m$--$     &  g      & ~& R& ~\\
11 07 55.8 & $-$41 35 52.7 & SCR1107$-$4135    &   11.474$\pm$0.019 &  3.49   & \nodata &\nodata&            &  M0.5VI  &  m$--$     &  g      & ~& R& ~\\
17 39 45.4 & $-$82 22 02.2 & SCR1739$-$8222    &   12.190$\pm$0.026 &  3.48   & \nodata &\nodata&            &  M0.5VI  &  m$--$     &  g      &  & R&  \\
\tableline										 
02 02 52.2 & $+$05 42 21.0 & LHS0012           &    8.684$\pm$0.020 &  3.56   &  6.32   & 342.1 & sdM0       &  M0.5VI  &  m$-$      &  g      & Y&13& 8\\
11 11 22.6 & $-$06 31 56.4 & LHS0299           &   11.143$\pm$0.024 &  3.64   &  6.54   & 437.3 &            &  M0.5VI: &  m$-$      &  g      & Y& R& ~\\
14 18 20.4 & $-$52 24 12.6 & LHS0367           &    9.786$\pm$0.019 &  3.41   &  6.26   & 268.6 &            &  M0.5VI: &  m$-$      &  g      & 7& 7& ~\\
04 06 06.7 & $-$67 35 28.8 & SCR0406$-$6735    &   12.804$\pm$0.031 &  3.54   & \nodata &\nodata&            &  M0.5VI  &  m$-$      &  g      &  & R&  \\
04 33 26.5 & $-$77 40 09.7 & SCR0433$-$7740    &   13.361$\pm$0.034 &  3.50   & \nodata &\nodata&            &  M0.5VI  &  m$-$      &  g      &  & R&  \\
14 33 03.3 & $-$38 46 59.6 & SCR1433$-$3847    &   13.592$\pm$0.046 &  3.62   & \nodata &\nodata&            &  M0.5VI  &  m$-$      &  g      &  & R&  \\
18 43 35.7 & $-$78 49 02.5 & SCR1843$-$7849    &   12.591$\pm$0.026 &  3.55   & \nodata &\nodata&            &  M0.5VI  &  m$-$      &  g      & ~& R& ~\\
19 58 31.2 & $-$56 09 10.6 & SCR1958$-$5609    &   12.525$\pm$0.033 &  3.51   & \nodata &\nodata&            &  M0.5VI  &  m$-$      &  g      &  & R&  \\
13 46 55.5 & $+$05 42 56.4 & LHS0360           &   11.662$\pm$0.023 &  3.50   &  6.73   & 524.0 &            &  M0.5VI: &  m$-$      &  g+     & 7& 7& ~\\
\tableline										 
\tableline										 
22 20 27.0 & $-$24 21 49.3 & LHS0518           &   10.393$\pm$0.021 &  3.24   &  7.07   & 230.7 &            &  M1.0VI  &  m$------$ &  g      &11&11& ~\\
07 01 17.7 & $-$06 55 49.3 & SCR0701$-$0655    &   12.996$\pm$0.030 &\nodata  & \nodata &\nodata&            &  M1.0VI  &  m$------$ &  g+     &  &  &  \\
\tableline										 
01 53 09.0 & $-$33 25 02.1 & LHS0148           &   12.832$\pm$0.032 &  3.59   &  8.59   & 374.8 &            &  M1.0VI  &  m$-----$  &  g+:    & 6& 6& ~\\
\tableline										 
01 53 09.0 & $-$33 25 02.1 & SCR1913$-$1001    &   11.929$\pm$0.028 &  3.69   & \nodata &\nodata&            &  M1.0VI  &  m$----$   &  g      &  & R&  \\
00 17 40.0 & $-$10 46 16.9 & LHS0109           &   10.366$\pm$0.021 &  3.51   &  7.63   & 176.1 &  K5        &  M1.0VI  &  m$----$   &  g+     & Y& R& 2\\
21 04 00.5 & $-$52 29 43.5 & SCR2104$-$5229    &   12.763$\pm$0.024 &  3.53   & \nodata &\nodata&            &  M1.0VI  &  m$----$   &  g+     &  & R& ~\\
14 55 35.8 & $-$15 33 44.0 & LHS0385           &   11.062$\pm$0.023 &  3.55   &  7.61   & 403.4 &  M0        &  M1.0VI  &  m$----$   &  g++    & Y& R& Y\\
14 55 51.5 & $-$39 14 33.1 & SCR1455$-$3914    &   11.788$\pm$0.024 &  3.65   & \nodata &\nodata&            &  M1.0VI  &  m$----$   &  g++    &  & R&  \\
12 34 53.1 & $+$05 03 54.1 & LHS0335           &   13.000$\pm$0.030 &  3.60   &  9.74   & 247.2 &            &  M1.0VI  &  m$----$   &  g+++   & Y& R& ~\\
21 09 02.5 & $-$52 26 17.8 & SCR2109$-$5226    &   13.049$\pm$0.036 &  3.66   & \nodata &\nodata&            &  M1.0VI  &  m$----$   &  g+++   & ~& R& ~\\
01 31 04.1 & $-$50 24 54.3 & LEHPM1628         &   13.459$\pm$0.038 &  3.70   & \nodata &\nodata&            &  M1.0VI  &  m$----$   &  g++++  & ~& R& ~\\
\tableline										
06 54 06.3 & $-$73 58 03.6 & SCR0654$-$7358    &   13.285$\pm$0.048 & \nodata & \nodata &\nodata&            &  M1.0VI  &  m$---$    &  g      &  &  &  \\
14 57 49.0 & $-$39 04 51.4 & SCR1457$-$3904    &   12.984$\pm$0.030 &  3.69   & \nodata &\nodata&            &  M1.0VI  &  m$---$    &  g      &  & R&  \\
16 13 53.5 & $-$30 40 58.4 & SCR1613$-$3040    &   12.383$\pm$0.031 &  3.66   & \nodata &\nodata&            &  M1.0VI  &  m$---$    &  g      &  & R&  \\
17 56 27.9 & $-$59 27 18.2 & SCR1756$-$5927    &   12.686$\pm$0.030 &  3.61   & \nodata &\nodata&            &  M1.0VI  &  m$---$    &  g      & ~& R& ~\\
20 18 28.7 & $-$66 06 44.5 & SCR2018$-$6606    &   12.990$\pm$0.030 & \nodata & \nodata &\nodata&            &  M1.0VI  &  m$---$    &  g      &  &  &  \\
\tableline										
05 29 40.9 & $-$39 50 25.6 & SCR0529$-$3950    &   11.651$\pm$0.021 & \nodata & \nodata &\nodata&            &  M1.0VI  &  m$--$     &  g      &  &  &  \\
\tableline										
21 01 45.6 & $-$54 37 31.9 & SCR2101$-$5437    &   12.078$\pm$0.026 &  3.70   & \nodata &\nodata&            &  M1.0VI  &  m$-$      &  g      & ~& R& ~\\
08 01 29.0 & $+$10 43 04.2 & LHS1970           &   13.875$\pm$0.032 &  3.84   &  9.43   & 374.8 & esdM2.5    &  M1.0VI: &  m$-$      &  g+:    & Y& R& 8\\
13 25 14.0 & $-$21 27 06.0 & LHS2734B          &   14.932$\pm$0.136 &  3.90   & too~far &$>$280 &            &  M1.0VI: &  m$-$      &  g+:    & R& R& ~\\
\tableline										
02 42 02.9 & $-$44 30 58.7 & LHS0158           &    9.726$\pm$0.021 &  3.91   &  6.89   & 191.4 &            &  M1.0[VI]&  m         &  g      &11&11& ~\\
17 18 35.0 & $-$43 26 24.0 & LHS0440           &    8.948$\pm$0.023 &  4.03   &  6.78   & 135.6 &  M1V       &  M1.0VI: &  m         &  g+:    &11&11& 2\\
18 22 58.7 & $-$45 42 45.3 & SCR1822$-$4542    &   12.879$\pm$0.027 &  3.92   & \nodata &\nodata&            &  M1.0VI: &  m         &  g+:    &  & R&  \\
\tableline										
\tableline										
07 56 13.4 & $-$67 05 20.6 & WT0233            &   12.627$\pm$0.024 &  3.60   &  7.84   & 326.8 &  M0.0VI    &  M2.0VI  &  m$---$    &  g      & 6& 6& 6\\
11 56 54.8 & $+$26 39 56.3 & LHS0318           &   11.797$\pm$0.018 &  3.65   &  8.77   & 263.8 &            &  M2.0VI: &  m$---$    &  g      &18& R& ~\\
\tableline										
05 11 40.6 & $-$45 01 06.0 & GJ0191            &    5.049$\pm$0.021 &  3.80   &  7.08   & 161.0 &  sdM1.0    &  M2.0VI  &  m$--$, ($-$0.99)\tablenotemark{e} &  g      & Y&13& 8\\
02 56 13.2 & $-$35 08 26.9 & LHS0162           &   11.536$\pm$0.021 &  3.82   &  7.13   & 364.1 &  M1.0VI    &  M2.0VI  &  m$--$     &  g+     & 6& 6& 6\\
21 04 25.5 & $-$27 52 48.5 & LHS3620           &   12.696$\pm$0.027 &  3.93   &  7.94   & 414.6 &            &  M2.0VI  &  m$--$     &  g+     & R& R& ~\\
\tableline										
04 03 38.4 & $-$05 08 05.4 & LHS0186           &   10.854$\pm$0.024 &  4.02   &  7.21   & 295.6 &            &  M2.0VI  &  m$-$      &  g      & Y& R& ~\\
15 43 18.3 & $-$20 15 33.0 & LHS0406           &    9.018$\pm$0.021 &  4.04   &  7.40   & 117.4 &  M1.0V     &  M2.0VI  &  m$-$      &  g      &11&11& 2\\
00 55 43.8 & $-$21 13 05.5 & LHS0127           &   11.733$\pm$0.023 &  4.06   & \nodata &\nodata&            &  M2.0VI  &  m$-$      &  g+     & ~& R& ~\\
06 29 56.4 & $-$69 38 13.4 & SCR0629$-$6938    &   12.901$\pm$0.034 & \nodata & \nodata &\nodata&            &  M2.0VI  &  m$-$      &  g+     &  &  &  \\
\tableline										 
\tableline										 
12 24 26.8 & $-$04 43 36.7 & LHS0326           &   11.234$\pm$0.023 &  3.67   &  7.69   & 304.9 &            &  M3.0VI  &  m$--$     &  g      & R& R& ~\\
15 34 27.7 & $+$02 16 47.5 & LHS0398           &   11.502$\pm$0.025 &  3.74   &  7.52   & 354.3 &            &  M3.0VI  &  m$--$     &  g      & Y&19& ~\\
23 17 05.0 & $-$13 51 04.1 & LHS0541           &   12.414$\pm$0.026 &  4.05   &  8.22   & 420.0 & sdM3.0     &  M3.0VI  &  m$--$     &  g+     & Y& 1& 8\\
22 04 02.2 & $-$33 47 38.9 & SCR2204$-$3347    &   11.601$\pm$0.027 &  3.84   & \nodata &\nodata&            &  M3.0VI  &  m$--$     &  g+     & ~& R& ~\\
17 40 46.9 & $-$56 46 58.0 & SCR1740$-$5646    &   13.195$\pm$0.040 & \nodata & \nodata &\nodata&            &  M3.0VI  &  m$--$     &  g++    &  &  &  \\
19 16 46.5 & $-$36 38 05.8 & SCR1916$-$3638    &   12.947$\pm$0.034 &  3.88   & \nodata &\nodata&            &  M3.0VI  &  m$--$     &  g++    & ~& R& ~\\
\tableline										 
04 25 38.4 & $-$06 52 37.0 & LHS0189AB         &   10.311$\pm$0.037 &  3.94   &  8.59   & 128.3 &            &  M3.0VI: &  m$-$      &  g      & 7& 7& ~\\
07 16 27.7 & $+$23 42 10.4 & LHS0228           &   11.298$\pm$0.018 &  4.20   &  7.54   & 300.6 & sdM2.0     &  M3.0VI  &  m$-$      &  g      & Y& R&14\\
13 42 21.2 & $-$35 34 50.7 & SIP1342$-$3534    &   12.935$\pm$0.030 &  4.37   & \nodata &\nodata&            &  M3.0VI  &  m$-$:     &  g      &  & R&  \\
04 11 27.1 & $-$44 18 09.0 & WT0135            &    9.834$\pm$0.020 &  4.33   & \nodata &\nodata&            &  M3.0VI  &  m$-$:     &  g      & ~& 5& ~\\
09 43 46.2 & $-$17 47 06.2 & LHS0272           &    8.874$\pm$0.021 &  4.29   &  8.34   &  86.3 & sdM3.0     &  M3.0VI  &  m$-$:     &  g+     & Y& R& 8\\
\tableline										  
\tableline										  
14 50 28.8 & $-$08 38 36.8 & LHS0381           &   11.237$\pm$0.021 &  3.90   &  8.43   & 274.9 &  K7.0V     &  M3.5VI  &  m$--$     &  g      & Y&19& Y\\
14 31 38.4 & $-$25 25 33.9 & LHS0375           &   11.507$\pm$0.022 &  4.12   &  9.61   & 157.5 &  esdM4     &  M3.5VI  &  m$--$     &  g+     & Y& R& 8\\
\tableline										  
01 38 49.0 & $+$11 21 36.7 & LHS0144           &   12.080$\pm$0.026 &  4.24   &  8.57   & 386.6 &            &  M3.5VI  &  m$-$      &  g      & Y& R& ~\\
21 55 48.0 & $-$11 21 42.1 & LHS0515           &   12.912$\pm$0.032 &  4.50   &  9.36   & 264.5 &  esdM5     &  M3.5VI  &  m$-$      &  g+     & Y& R&17\\
\tableline										  
\tableline										  
05 00 15.3 & $-$54 06 09.0 & LEHPM3861         &   13.967$\pm$0.059 &  4.47   & \nodata &\nodata&  esdM6     &  M4.0VI  &  m$-$      &  g      & ~& R&15\\
\tableline										  
\tableline										  
03 02 06.3 & $-$39 50 51.8 & LHS1490           &    9.885$\pm$0.023 &  4.37   & \nodata &\nodata&            &  M5.0VI  &  m$-$      &  g      &  & R&  \\
\tableline										  
\tableline										  
12 34 15.7 & $+$20 37 05.7 & LHS0334           &   13.044$\pm$0.029 &  4.98   & \nodata &\nodata&  sdM4.5    &  M6.0VI  &  m$--$     &  g      &  & R&17\\
\tableline										  
08 53 57.0 & $-$24 46 54.0 & LHS2067A          &   11.571$\pm$0.023 &  6.38   &  9.54   &  76.2 &  subdwarf  &  M6.0VI  &  m$-$      &  g      & R& R&12\\

\enddata 

\tablecomments{The last three columns provide references for
parallaxes, $V$ band photometry, and previous spectroscopy
results. ``:'' indicates a questionable sub-type, metallicity, or
gravity.  All $K_{s}$ magnitudes are from the 2MASS all-sky database.
LHS2734A and B have zero parallax so we list them as ``too far'' in
the table.}

\tablenotetext{a}{Metallicity data are from \cite{Carney1994}.}

\tablenotetext{b}{Metallicity data are from \cite{Nordstrom2004}.}

\tablenotetext{c}{Metallicity data are from \cite{Cayrel2001}. There
are multiple metallicities reported for G112-054, so a mean value is
presented.}

\tablenotetext{d}{\cite{Cayrel2001}, \cite{Goldberg2002} and
\cite{Laird1988} report different metallicities, so a mean value is
given.}

\tablenotetext{e}{Metallicity data are from \cite{Woolf2005}}

\tablenotetext{f}{\cite{Morrison2003} and \cite{Allen2000} report
different metallicities, so a mean value is given.}

\tablerefs{ 
 (1) \citealt{Bessel1990}
;(2) \citealt{Bidelman1985}
;(3) \citealt{Cayrel1997}
;(4) \citealt{Carney1994}
;(5) \citealt{Costa2003}
;(6) \citealt{Costa2005}
;(7) \citealt{Costa2006}
;(8) \citealt{Gizis1997}
;(9) \citealt{Hawley1996}
;(10) \citealt{Houk1975}
;(11) \citealt{Jao2005}
;(12) \citealt{Kirkpatrick1995}
;(13) \citealt{Leggett1992}    
;(14)\citealt{Lepine2007}        
;(15) \citealt{Lodieu2005}       
;(16) \citealt{Monteiro2006}     
;(17) \citealt{Reid2005}          
;(18) \citealt{Smart2007}         
;(19) \citealt{Weis1996}          
;(H) \citealt{Hipparcos}
;(R) \citealt{RECONS}
;(Y) \citealt{YPC}}

\label{tbl.spectral.type}
\end{deluxetable}


\begin{deluxetable}{lcccclll}
\tablewidth{18.5cm}
\tablecaption{SDSS Subdwarfs}
\tablehead{
\colhead{Object}            &   
\colhead{$r$}               &    
\colhead{$g-r$}             &   
\colhead{$g-z$}             &   
\colhead{Old Type}          &
\colhead{New Type}          &
\colhead{metallicity}       &
\colhead{gravity}  }
\startdata
SDSS~J085843.89$+$511210.1  &  20.00 & 1.53 & 3.10 & M2.0VI  &   M2.0V   &  m           & \nodata \\
SDSS~J093141.85$+$453914.5  &  19.45 & 1.58 & 3.08 & M2.0VI  &   M2.0V   &  m           & \nodata \\
SDSS~J145447.32$+$011006.8  &  20.35 & 1.44 & 3.45 & M3.0VI  &   M3.0V   &  m           & \nodata \\
\tableline											    
\tableline											    
SDSS~J003755.20$-$002134.2  &  18.70 &      &      & M1.0VI  &   M2.0VI  &  m$-$        &  g      \\
SDSS~J083217.77$+$522408.2  &  19.61 & 1.49 & 3.14 & M2.0VI  &   M2.0VI: &  m           &  g$+$:  \\
\tableline										   	    
\tableline										   	    
SDSS~J033408.64$-$072349.2  &  20.21 & 1.89 & 3.35 & M2.0VI  &   M3.0VI  &  m$----$     &  g$+$:  \\
SDSS~J090434.02$+$513153.9  &  19.40 & 1.78 & 3.29 & M1.0VI  &   M3.0VI  &  m$----$     &  g$+$:  \\
SDSS~J161348.84$+$482016.0  &  18.30 & 1.72 & 3.19 & M1.0VI  &   M3.0VI  &  m$----$     &  g$+$:  \\
\tableline										   	    
SDSS~J112751.35$-$001246.8  &  20.03 &      &      & M1.0VI  &   M3.0VI  &  m$---$      &  g      \\
SDSS~J092708.10$+$561648.1  &  19.37 & 1.62 & 3.02 & M1.0VI  &   M3.0VI  &  m$---$      &  g$+$   \\
SDSS~J100109.54$+$015450.2  &  19.12 & 1.71 & 3.25 & M2.0VI  &   M3.0VI  &  m$---$      &  g$++$  \\
SDSS~J101031.13$+$651327.6  &  19.39 & 1.68 & 3.30 & M1.0VI  &   M3.0VI  &  m$---$      &  g$++$  \\
\tableline										   	    
SDSS~J002228.00$-$091444.8  &  18.92 &-0.89 & 3.03 & M1.0VI  &   M3.0VI  &  m$--$       &  g      \\
SDSS~J024501.77$+$003315.8  &  19.38 & 1.54 & 3.15 & M1.0VI  &   M3.0VI  &  m$--$       &  g      \\
SDSS~J081329.95$+$443945.6  &  19.39 & 1.68 & 3.35 & M2.0VI  &   M3.0VI  &  m$--$       &  g      \\
SDSS~J090238.75$+$471813.6  &  19.97 & 1.73 & 3.21 & M2.0VI  &   M3.0VI  &  m$--$       &  g      \\
SDSS~J092534.16$+$524442.4  &  19.79 & 1.71 & 3.31 & M3.0VI  &   M3.0VI  &  m$--$       &  g      \\
SDSS~J092745.78$+$582122.7  &  20.47 & 1.78 & 3.56 & M3.0VI  &   M3.0VI  &  m$--$       &  g      \\
SDSS~J095147.77$+$003612.0  &  18.27 & 1.58 & 3.15 & M2.0VI  &   M3.0VI  &  m$--$       &  g      \\
SDSS~J115900.70$+$665214.3  &  19.37 & 1.61 & 3.35 & M3.0VI  &   M3.0VI  &  m$--$       &  g      \\
SDSS~J125919.29$-$025402.3  &        &      &      & M2.0VI  &   M3.0VI  &  m$--$       &  g      \\
SDSS~J173452.52$+$603603.1  &  18.78 & 1.71 & 3.36 & M2.0VI  &   M3.0VI  &  m$--$       &  g      \\
SDSS~J215937.69$+$005536.2  &  18.96 & 1.61 & 3.22 & M2.0VI  &   M3.0VI  &  m$--$       &  g      \\
SDSS~J221500.88$+$005217.2  &  19.08 & 1.60 & 3.33 & M2.0VI  &   M3.0VI  &  m$--$       &  g      \\
SDSS~J145547.00$+$602837.3  &  19.18 & 1.70 & 3.39 & M2.0VI  &   M3.0VI  &  m$--$       &  g$+$   \\
SDSS~J104320.47$+$010439.4  &  19.16 & 1.66 & 3.37 & M2.0VI  &   M3.0VI  &  m$--$       &  g$++$  \\
\tableline										   	    
SDSS~J084105.39$+$032109.6  &  20.05 & 1.45 & 3.18 & M3.0VI  &   M3.0VI  &  m$-$        &  g      \\
SDSS~J091451.98$+$453152.8  &  19.02 & 1.64 & 3.49 & M3.0VI  &   M3.0VI  &  m$-$        &  g      \\
SDSS~J093024.66$+$554447.7  &  19.14 & 1.56 & 2.96 & M3.0VI  &   M3.0VI  &  m$-$        &  g      \\
SDSS~J094306.37$+$465701.4  &  19.74 & 1.59 & 3.29 & M3.0VI  &   M3.0VI  &  m$-$        &  g      \\
SDSS~J224854.83$-$091723.2  &  19.85 & 1.57 & 3.39 & M3.0VI  &   M3.0VI  &  m$-$        &  g      \\
SDSS~J235830.60$-$011413.2  &  19.99 & 1.54 & 3.33 & M2.0VI  &   M3.0VI  &  m$-$        &  g      \\
SDSS~J113501.76$+$033720.3  &        &      &      & M3.0VI  &   M3.0VI: &  m           &  g$+$:  \\
\tableline										   	    
\tableline										   	    
SDSS~J105122.43$+$603844.8  &  17.15 & 1.65 & 3.21 & M2.0VI  &   M4.0VI  &  m$---$      &  g      \\
SDSS~J092429.76$+$523410.7  &  18.70 & 1.67 & 3.21 & M2.0VI  &   M4.0VI  &  m$---$      &  g      \\
\tableline										   	    
SDSS~J031314.28$-$000619.8  &  20.33 & 1.48 & 3.39 & M2.0VI  &   M4.0VI  &  m$--$       &  g      \\
SDSS~J082230.00$+$471645.8  &  19.39 & 1.67 & 3.38 & M2.0VI  &   M4.0VI  &  m$--$       &  g      \\
SDSS~J143930.77$+$033317.3  &  19.38 & 1.80 & 3.58 & M3.0VI  &   M4.0VI  &  m$--$       &  g$+$   \\
\tableline										   	    
SDSS~J003541.84$+$003210.1  &  20.15 & 1.64 & 3.37 & M3.0VI  &   M4.0VI  &  m$-$        &  g      \\
SDSS~J003701.37$-$003248.3  &  20.15 & 1.60 & 3.50 & M3.0VI  &   M4.0VI  &  m$-$        &  g      \\
SDSS~J010811.89$+$003042.4  &  17.34 & 1.61 & 3.43 & M3.0VI  &   M4.0VI  &  m$-$        &  g      \\
SDSS~J083002.73$+$483251.6  &  19.92 & 1.90 & 3.69 & M3.0VI  &   M4.0VI  &  m$-$        &  g      \\
SDSS~J171745.22$+$625337.0  &  18.66 & 1.65 & 3.57 & M3.0VI  &   M4.0VI  &  m$-$        &  g      \\
SDSS~J221625.03$-$003122.5  &  19.28 & 1.58 & 3.37 & M3.0VI  &   M4.0VI  &  m$-$        &  g      \\
SDSS~J223802.82$-$082532.4  &  19.71 & 1.78 & 3.59 & M3.0VI  &   M4.0VI  &  m$-$        &  g      \\
SDSS~J224605.41$+$141640.6  &  17.00 & 1.57 & 3.48 & M3.0VI  &   M4.0VI  &  m$-$        &  g      \\
SDSS~J230303.49$-$010656.7  &  18.95 & 1.62 & 3.42 & M3.0VI  &   M4.0VI  &  m$-$        &  g      \\
SDSS~J235116.25$-$003104.8  &  19.48 & 1.49 & 3.31 & M3.0VI  &   M4.0VI  &  m$-$        &  g      \\
\tableline										   	    
\tableline										   	    
SDSS~J150511.33$+$620926.3  &  18.61 & 1.81 & 3.45 & M3.0VI  &   M5.0VI  &  m$--$       &  g      \\
\tableline										   	    
SDSS~J230805.24$+$001812.7  &  19.36 & 1.70 & 3.72 & M3.0VI  &   M5.0VI  &  m$-$        &  g      \\  
\enddata

\tablecomments{Column definitions are the same as in
Table~\ref{tbl.spectral.type}.  Double lines separate each type.}

\label{tbl.sdss.subdwarf}
\end{deluxetable}




\clearpage


\clearpage


\clearpage


\clearpage


\clearpage


\clearpage


\clearpage


\clearpage


\clearpage


\clearpage

\clearpage


\clearpage


\clearpage


\clearpage


\clearpage


\clearpage


\clearpage


\clearpage


\clearpage


\clearpage


\clearpage


\clearpage


\clearpage


\clearpage


\clearpage


\clearpage


\clearpage


\clearpage


\clearpage


\clearpage


\end{document}